%% file: master.tex
\documentclass[11pt,a4paper]{article}

\usepackage{xcolor}
\usepackage[colorlinks=true, linkcolor=black!50!blue, urlcolor=blue, citecolor=blue, anchorcolor=blue]{hyperref}
\usepackage[font=small,labelfont=bf,margin=0mm,labelsep=period,tableposition=top]{caption}
\usepackage[a4paper,top=3cm,bottom=2.5cm,left=2.5cm,right=2.5cm,bindingoffset=0mm]{geometry}
\usepackage{subcaption}
\usepackage{amsmath}
\usepackage{amsfonts}
\usepackage{amssymb}
\usepackage{authblk}
\usepackage{dsfont}
\usepackage{pifont}
\usepackage{booktabs}
\usepackage{tabularx}
\usepackage{siunitx}
\usepackage{graphicx} \graphicspath{{figs/}}
\usepackage{epstopdf}
\usepackage{epsfig}
\usepackage{framed}
\usepackage{makeidx}
\usepackage{simplewick}
\usepackage{placeins}
\usepackage{bbold}
\usepackage{braket}
\usepackage{amsthm}
\usepackage{cleveref}
\usepackage{cite}
\crefname{equation}{eq.}{eqs.}
\creflabelformat{equation}{(#2#1#3)}
\crefname{figure}{Fig.}{Figs.}
\crefname{table}{Table}{Tables}
\crefname{section}{Section}{Sections}
\crefname{subsection}{Section}{Sections}
\crefname{appendix}{Appendix}{Appendices}
\crefname{subappendix}{Appendix}{Appendices}
\theoremstyle{definition}

\newtheorem*{example*}{Example}
\crefname{example}{Example}{Examples}
\input{commands}

\begin{document}
\newgeometry{top=1.5cm,bottom=1.5cm,left=1.5cm,right=1.5cm,bindingoffset=0mm}
\vspace{-2.0cm}
\vspace{0.3cm}

\begin{center}
  {\Large \bf The proton structure one mode at a time:\\[0.2cm]
    a Neural Tangent Kernel analysis of a DIS fit}
  
  \vspace{1.0cm}

\vspace{1.1cm}

   Amedeo Chiefa, Luigi Del Debbio, and Richard Kenway

  \vspace{0.2cm}

  \it{ \small
    The Higgs Centre for Theoretical Physics, \\ 
    School of Physics and Astronomy, The University of Edinburgh,\\
    Peter Guthrie Tait Road, Edinburgh EH9 3FD, United Kingdom\\[0.1cm]
  }
  \vspace{0.7cm}
\end{center}

\begin{abstract}
  We study the training dynamics of a feed-forward neural network (NN) through
  the analysis of the neural tangent kernel (NTK) in a phenomenologically
  relevant fit of parton distribution functions (PDFs) from deep inelastic
  scattering (DIS) data. The exact integration of the gradient flow derived in a
  previous single-flavour analysis is generalised to the multi-flavour case,
  yielding an analytical perspective on the relationship between data and
  methodology during training. The Shapley value recently used in the context of
  PDFs is derived within our statistical framework for observables that depend
  linearly on the PDFs, and we outline a geometric interpretation of training.
  We expose the role of gradient-based optimisation algorithms and of the
  parametrisation in the regularisation of the ill-posed inverse problem that
  affects the determination of PDFs from experimental data, demonstrating that
  the regularisation amounts to the incremental inclusion of NTK modes, each
  learned at the timescale set by the corresponding eigenvalue. We find that
  only a few leading modes of the NTK are necessary to reconstruct the target
  PDFs, while the noise of the data is carried by the sub-leading ones learned
  at later stages of training. Finally, we consider the impact of the
  initialisation on the training dynamics, and recover a transient behaviour
  observed in the previous analysis and in the machine learning (ML) literature.
  Beyond sharpening our understanding of the training dynamics, our framework
  can be applied to other parametrisations (such as fixed functional forms) and
  ML applications in physics, as well as to stress tests in the presence of
  inconsistent data.
\end{abstract}

\newpage

\tableofcontents
\vspace{5mm}

\clearpage

\include{sections/introduction}

\include{sections/framework}
\include{sections/geostat}
\include{sections/results}
\include{sections/init_dependence}
\include{sections/conclusions}

\section*{Code and data availability}
The fits presented in this paper were performed with the open-source NNPDF
code~\cite{NNPDF:2021uiq,nnpdf_zenodo}, using the \texttt{ntk\_colibri} branch
(commit \texttt{16786e8e7e}). The NTK analysis was carried out with
\texttt{ntkpdf}~\cite{ntkpdf} (commit \texttt{90bda63b35}), a specialised
interface built on top of Colibri~\cite{Costantini:2025agd} (\texttt{ntk}
branch, commit \texttt{fb82175704}). The fit outputs used in this work are
available from the authors upon request.

\section*{Acknowledgements}
We are very grateful to M Ubiali, M N Costantini, V S Sánchez, and E Cole for
helpful discussions during the early stages of this work and for support with
the Colibri framework. LDD is supported by an STFC Consolidated Grant
(ST/T000600/1, ST/X000494/1). AC used the AI assistant Claude (Anthropic) for
manuscript editing and for ensuring consistency throughout the text, and for
assistance with coding and the production of the figures. All scientific
content, including the methodology, results and their interpretation, is the
sole responsibility of the authors.

\newpage

\appendix
\include{sections/dln}
\include{sections/dln_dynamics}

\bibliography{bib/ntk_wide_nets,bib/training_dynamics,bib/geometry_statistics,bib/pdfs,bib/ml_hep,bib/collider}

\end{document}

%% file: commands.tex
\newcommand{\cov}{C}
\newcommand{\covy}{\cov_Y}
\newcommand{\real}{\mathbb{R}}

\newcommand{\gauss}{\mathcal{N}}
\newcommand{\expct}{\mathbb{E}}

\newcommand{\covar}{\mathrm{Cov}}
\newcommand{\manfd}{\mathcal{M}}

\newcommand{\id}{\mathbb{I}}
\newcommand{\cl}{c.l.} 
\renewcommand{\id}{\mathbb{1}}
\newcommand{\diag}{\mathrm{diag}}

\newcommand{\ndata}{N_{\rm data}}
\newcommand{\ngrid}{N_{\rm grid}}
\newcommand{\nvec}{N_{\rm vec}}
\newcommand{\nf}{N_{\rm flav}}   
\newcommand{\obs}{Y}

\newcommand{\covdata}{\cov_Y}

\newcommand{\loglhd}{\mathcal{L}}

\newcommand{\nreps}{N_{\rm rep}}
\newcommand{\nout}{N_{\rm out}}   
\newcommand{\nparams}{N_{\rm params}}
\newcommand{\vlin}{\mathcal{V}_{\rm lin}}
\newcommand{\parameters}{763}

\newcommand{\ie}{\textit{i.e.}}
\newcommand{\eg}{\textit{e.g.}}

\newcommand{\cfr}{\textit{cfr.}}

  \newcommand{\fk}{(\mathrm{FK})}

\newcommand{\fin}{f^{(\rm in)}}

\newcommand{\mker}{\mathrm{Ker}(M)}

\newcommand{\xa}{x_{\alpha}}
\newcommand{\rfk}{R^{(\rm FK)}}

\newcommand{\ntk}{\Theta}
\newcommand{\fker}{f^{\parallel}}

\newcommand{\fperp}{f^{\perp}}

\newcommand{\zperp}{Z_{\perp}}
\newcommand{\zpar}{Z_{\parallel}}
\newcommand{\NN}{\textrm{NN}}
\newcommand{\ntkimg}{\mathrm{Im}(\ntk)}
\newcommand{\ntkker}{\mathrm{Ker}(\ntk)}
\newcommand{\ppar}{P_{\parallel}}
\newcommand{\pperp}{P_{\perp}}
\newcommand{\dperp}{d^{\perp}}

\newcommand{\tcntk}{t_{\rm NTK}}
\newcommand{\tntk}{\tau_{\rm NTK}}
\newcommand{\tesc}{\tau_{\rm esc}}
\newcommand{\tend}{\tau_{\rm end}}
\newcommand{\wght}[1]{W^{(#1)}}
\newcommand{\uwght}[1]{\Omega^{(#1)}}
\newcommand{\uchain}[2]{\Omega_{#1#2}}
\newcommand{\bias}[1]{b^{(#1)}}
\newcommand{\preact}[1]{\phi^{(#1)}}
\newcommand{\act}[1]{\rho^{(#1)}}

\newcommand{\coeff}{\mathcal{C}}
\newcommand{\emb}{\mathcal{R}}
\newcommand{\spec}[1]{\mathrm{spec}\left(#1\right)}

\renewcommand{\it}{\textit}

\newcommand{\tgnt}[1]{T_{#1}\manfd}
\newcommand{\fisher}{\mathcal{F}}

\newcommand{\fb}{f}
\newcommand{\pcond}[2]{p\left(#1 \, | \, #2 \right)}
\newcommand{\kldiv}[2]{D_{\rm KL} \left( #1 \, || \, #2 \right)}
\newcommand{\pyf}{p_f(Y)}
\newcommand{\pydf}{p_{f+\delta f}(Y)}

%% file: sections/introduction.tex
\section{Introduction}
\label{sec:intro}

The data collected at the Large Hadron Collider (LHC) up to the now-completed
third data-taking period (Run 3) have brought the field of particle physics into
a regime where sub-percent accuracy is becoming the benchmark for both
experimental measurements and theoretical predictions. These data have
underpinned the most recent validations of the Standard Model (SM) through the
measurement of its
parameters~\cite{LHCb:2021bjt,ATLAS:2024erm,CMS:2024lrd,LHCb:2025nob,ATLAS:2015ihy,LHCb:2015jyu,CMS:2018ktx,LHCb:2024ygc,CMS:2024ony,ATLAS:2023lhg},
making precision itself a non-negotiable condition to probe physics beyond the
SM (BSM)~\cite{Armadillo:2026mvp}. 

These results, however, are rarely read off the raw data straight away. The
quantities of interest have to be reconstructed by solving an \emph{inverse
problem}~\cite{DelDebbio:2026abr}, so that the reliability of the determination
is bounded not only by the data but also by how well the methodology -- the
inference procedure -- is kept under control. Many methodologies rest on machine
learning (ML) tools~\cite{krause_2026_21626667}, and their reliability --
uncertainty quantification, calibration, and validation -- has itself drawn
significant
attention~\cite{Haussmann:2026gbi,Araz:2025vuw,Araz:2026img,Cruz-Martinez:2026apu,Kramer:2026eoq}.
With the High-Luminosity LHC (HL-LHC) era~\cite{ZurbanoFernandez:2020cco} and
promising future collider
facilities~\cite{Proceedings:2026xrb,Armadillo:2026mvp} on the horizon, the
requirements on the methodology will become even more stringent. 

Meeting these requirements is, however, inherently difficult. Most of the
inverse problems that challenge particle physics are indeed plagued by their
ill-posed nature~\cite{DelDebbio:2026abr}; the determination of parton
distribution functions (PDFs) from global analyses of experimental data is no
exception. PDFs are fundamental quantities that encode the inner structure of a
hadron -- the proton, in our case -- and their extraction from data relies on an
intricate interplay between three main ingredients: the data that constrain the
model, the theory that predicts the data, and the methodology that dictates how
the two combine. The problem is ill-posed: a finite amount of data cannot pin
down the continuous functional form of the PDFs, leaving the methodology to fix
the unconstrained directions. As in the case of
unfolding~\cite{Brenner:2019lmf,Brenner:2025wwd}, any methodology therefore
specifies a set of \emph{implicit} and \emph{explicit} assumptions that
regularise the problem, the former being more often than not hidden within the
methodology itself, and therefore difficult to spell out precisely.

That the choice of methodology has a significant impact on the result is now
well established. Comparisons across different methodologies reveal sizeable
differences both at the level of
PDFs~\cite{Harland-Lang:2024kvt,Cruz-Martinez:2026rct} and of the theoretical
predictions~\cite{Chiefa:2025loi}, even though the data entering the fits are
largely the same. Understanding the origin of these deviations is crucial
because PDFs are the primary input to essentially every theoretical prediction
that involves a proton in the initial state. Moreover, they are used in the
inference of SM parameters~\cite{Ball:2025xgq,Ball:2026qno}, and are at the core
of indirect searches for new physics, where an inaccurate determination of the
PDFs can absorb a BSM signal~\cite{Hammou:2023heg,Cole:2026eex}. 

It is thus not surprising that, together with the remarkable achievements in the
theoretical calculations now approaching N$^3$LO
accuracy~\cite{McGowan:2022nag,Cridge:2024icl,NNPDF:2024nan}, the faithfulness
of delivered PDF uncertainties is becoming the subject of active
investigations~\cite{Costantini:2024wby,Barontini:2025lnl,Costantini:2026sel}.
Pinning down which features of the final results are dictated by the data, and
which are inherited from the methodology, is key to distinguishing genuine
physical signals from uncontrolled artefacts of the inference procedure.

Efforts to disentangle the role of the methodology in PDF determinations are not
lacking. The Colibri framework~\cite{Costantini:2025agd} allows different
methodologies and inference strategies to be compared within the same
infrastructure. Bayesian
approaches~\cite{DelDebbio:2021whr,Candido:2024hjt,Costantini:2025wxp,Medrano:2025cmg}
have proven successful in laying out a framework where the assumptions are
explicit -- and so are their implications -- while tools originally developed in
the context of explainable AI have been applied to PDFs either to expose the
inner features of the PDFs~\cite{Kriesten:2023uoi,Kriesten:2024are} or to make
the role of the data more transparent~\cite{Bonnet-Guerrini:2026thq}. Most of
these studies, however, propose alternative ways to approach the problem, while
a systematic study of the methodologies that set the current standard in PDF
determination is still lacking.

We began to address this point in Ref.~\cite{Chiefa:2025cap} where we focused on
the NNPDF4.0 methodology~\cite{NNPDF:2021njg,NNPDF:2021uiq}, which rests on the
propagation of the experimental uncertainties into PDFs through Monte Carlo
sampling of the data and on the parametrisation of PDFs by means of a neural
network (NN). Using a single-flavour toy model, we studied the training dynamics
of the NN by inspecting the evolution of the neural tangent kernel
(NTK)~\cite{jacot2018neural,lee:2020,huang2019dynamics}. We showed that, under
the assumption of a \emph{lazy} training regime~\cite{LazyNTKChizat:2019},
characterised by a slowly varying NTK, the gradient flow can be integrated
exactly, providing an analytical description of the training from the onset of
the lazy regime onwards. Whether these results hold in a more realistic fit
remained, however, an open question.

\subsubsection*{Our contribution}
In this work we extend the previous analysis to a realistic fit of PDFs, based
on a comprehensive dataset of deep inelastic scattering (DIS) measurements that
allows for the simultaneous fit of eight independent flavours. We adopt the
NNPDF4.0 methodology with plain gradient descent, with no sum rules and
positivity constraints, and generate synthetic data from a known set of PDFs, so
that the results can be validated against the underlying law. Within these
settings, we aim at quantifying the impact of the implicit assumptions discussed
above and at understanding the training dynamics of the NN. Our contribution can
be summarised as follows.

We first generalise the analytical solution of the gradient flow to the
multi-flavour case and validate it against the empirically trained network. The
implicit regularisation of the inverse problem is thus made manifest as the
projection of the training flow onto the sub-space of non-vanishing modes of the
NTK. In this picture, training is viewed as an incremental inclusion of NTK
modes, each learned at the timescale set by the corresponding eigenvalue. We
show that only a few of the leading modes of the NTK spectrum are needed to
reconstruct the underlying law, while the sub-leading ones contribute to the
noise of the solution. The choice of when to stop the training -- usually
delegated to a stopping criterion -- thus becomes a trade-off between the signal
and noise of the data. The empirical analysis is supplemented with a derivation
of the Fisher information matrix of the PDFs, from which we recover, for DIS
observables, the Shapley value of Ref.~\cite{Bonnet-Guerrini:2026thq}. We then
outline a geometric interpretation of the NTK and of the analytical solution.

Using the analytical solution, we demonstrate that the central value and most of
the error budget of the PDFs are driven by the data. The initial condition,
although it does not bias the central value, contributes to the uncertainty by
an amount that depends on the flavour and on the region in $x$, and remains
comparable to that of the data in the extrapolation regions. We then investigate
the dependence on the width of the initialisation distribution of the
parameters, and find that it can have a considerable impact on the training
dynamics. In particular, we show that for small initialisations the network
operates in the linear regime, which slows training. This phenomenon can be
detected by inspecting the NTK.

\subsubsection*{Related works}
The NTK was first introduced in Ref.~\cite{jacot2018neural}, which showed that,
in the infinite-width limit in which the network at initialisation is a Gaussian
process~\cite{Neal1996,Lee:2017qzq,matthews2018gaus}, the NTK is deterministic
and constant during training, and gradient descent reduces to kernel learning.
This was then recast in weight space~\cite{lee:2020}, showing that wide NNs
train like their linearisation around the parameters at initialisation. The
properties of the NTK during training have been studied in
Refs.~\cite{huang2019dynamics,Dyer:2019uzd}, while examples of methods borrowed
from statistical and quantum field theory used to describe wide networks and
their finite-width corrections can be found in
Refs.~\cite{Halverson:2020trp,Demirtas:2023fir,PhysRevE.104.064301}. The
connection with the Fisher information matrix and the geometric interpretation
of \cref{sec:geostat} were inspired by
Refs.~\cite{sun_geometric_2025,LY201740,pmlr-v89-liang19a}. Early stopping as a
regularisation has been studied in linear models in
Refs.~\cite{pmlr-v89-ali19a,coeurdoux2026rmt}. The study of the dependence on
the initialisation was inspired by
Refs.~\cite{jacot2022sts,FUKUMIZU2000317,zhang2026sts}.
\par\bigskip\noindent The remainder of this paper is organised as follows. In
\cref{sec:framework} we review the dataset and the methodology, generalise the
analytical solution of the gradient flow to the multi-flavour case, and use it
to interpret early stopping as a bias-variance trade-off. \cref{sec:geostat} is
devoted to the derivation of the Fisher information matrix of the PDFs and to
the geometric interpretation of the NTK. In \cref{sec:results} we present the
NTK analysis of the DIS-only fit and compare the analytical solution with the
trained network, while in \cref{sec:init_dependence} we study how the training
dynamics depends on the initialisation. We summarise our findings in
\cref{sec:conclusions}, where we also propose future improvements and
applications. A closed-form expression of the NTK for deep linear networks is
derived in \cref{app:dln_ntk}, while in \cref{app:init_dynamics} we collect
supplementary material to complement the discussion of
\cref{sec:init_dependence}.

%% file: sections/framework.tex
\section{Methodological and theoretical framework}
\label{sec:framework}

In this section, we provide a brief overview of the methodological and
experimental ingredients that enter this analysis. We then summarise in
\cref{sec:gradient_flow} the main results of Ref.~\cite{Chiefa:2025cap} and show
how the analytical solution of the gradient flow generalises to the case of a
multi-flavour PDF fit. In doing so, we introduce the NTK and the operator $M$,
both fundamental objects in the learning process, for which we provide a
statistical and geometric interpretation in \cref{sec:geostat}. Finally, in
\cref{sec:early_stopping} we interpret early stopping through the lens of the
NTK, thus exposing the role of each NTK mode in the bias-variance trade-off
along the training dynamics.

\subsection{Dataset and methodology}
\label{sec:dataset_method}

The experimental input comprises all the DIS measurements included in the
NNPDF4.0 global analysis, both neutral- and charged-current, restricted to
observables whose theoretical predictions depend linearly on the PDFs. After
kinematic cuts, essential for ensuring accuracy in the theoretical predictions,
the dataset amounts to $\ndata = 2968$ points. Here we assume the data to be
distributed according to a multivariate Gaussian distribution centred on the
experimental value and with variance given by the experimental covariance matrix
$\covy$. We refer the reader to the original NNPDF4.0
release~\cite{NNPDF:2021njg} and references therein for a detailed description
of each experimental input entering the dataset. In the following, we provide a
concise account of the methodological aspects involved in this analysis,
summarised in \cref{tab:dis_fit_settings}.

\paragraph{Theoretical predictions.}
These are computed as in the previous study, although the multi-flavour
structure introduced here adds a layer of complexity to the NTK analysis. In
fact, the FK tables are now rank-3 tensors that approximate the convolution
integral of the factorisation theorem by way of a Riemann sum over a grid of
points $\xa$, where $\alpha = 1,\dots,\ngrid$.\footnote{Here we use the same
grid of $\ngrid=50$ points used in Ref.~\cite{Chiefa:2025cap}. Note that
different datasets may require different subsections of this grid for their
predictions to be computed, depending on the physical constraints of the
experiment. Whenever this is the case, the FK table is padded in order to have a
uniform grid dimension across all datasets.} The predictions are then computed
by convoluting the PDFs $f$, now including $\nf=14$ flavours each discretised on
the grid in $x$, with the FK table
\begin{equation}
  T_I 
  = \sum_{\alpha=1}^{\ngrid} \sum_{i=1}^{\nf} \fk_{I, \alpha}^{i} f_{i\alpha},
  \label{eq:theory_prediction}
\end{equation}
where $I = 1,\dots,\ndata$ and $i=1,\dots,\nf$ run over the data points and
flavours, respectively. Note that, with the present dataset, only a fraction of
the $\nf$ flavours are independent and therefore fitted, as we show below.
Following the NNPDF convention, FK tables act on the momentum densities $xf$,
and throughout this work $f$ denotes the grid of these values, $f_{i\alpha}
\equiv x_\alpha f_i(x_\alpha, Q_0)$. In order to ease the notation, and avoid
carrying along many indices in the discussion, we use a vectorisation convention
to condense flavour and grid indices into a single \emph{flavour-major} ordered
index $a=1, \dots, \nvec$, with $\nvec = \nf \times \ngrid$. We then write the
convolution in a compact matrix notation as
\begin{equation}
  \label{eq:convolution_vectorised}
  T_I = \sum_{a=1}^{\nvec} \fk_{Ia} f_a\,,
  \quad
  \rm{or}
  \quad
  T = \fk f\,,
\end{equation}
where the FK table is reshaped as a matrix
\begin{equation}
  \fk = \begin{bmatrix}
    \mathrm{FK}^{\Sigma} & \mathrm{FK}^{g} & \cdots & \mathrm{FK}^{V_{35}}
  \end{bmatrix}
  \in\real^{\ndata \times \nvec}\,,
  \quad
  \mathrm{FK}^{i}\in\real^{\ndata \times \ngrid}\quad\forall i \in \left\{\Sigma,\,g,\,\dots,\,V_{35}\right\}\,,
  \label{eq:FK_vectorised}
\end{equation}
and PDFs are cast into a single column vector
\begin{equation}
  \fb = \begin{bmatrix}
    \fb_{\Sigma} \\
    \fb_{g} \\
    \vdots \\
    \fb_{V_{35}}
  \end{bmatrix}\,,
  \quad
  \fb_i \in \real^{\ngrid}\quad\forall i \in \left\{ \Sigma,\,g,\,\dots,\,V_{35} \right\}\,.
  \label{eq:PDFVecExplicit}
\end{equation}

\begin{table}[t]
    \centering
    \begin{minipage}[t]{0.48\textwidth}
        \centering
        \begin{tabular}{l|l}
            \textbf{Parameter} & \textbf{Value} \\
            \hline
            Architecture & [25, 20, 8], $\nparams = \parameters$ \\
            Fitted flavours & $\Sigma$, $g$, $V$, $V_3$, $V_8$, $T_3$, $T_8$,
            $T_{15}$ \\
            Activation functions & [$\tanh$, $\tanh$, linear] \\
            Initialisation & Glorot normal \\
            Optimiser & GD \\
            Learning rate & $1.0 \times 10^{-6}$ \\
            Clipnorm & $0.0$\\
            Max epochs & $2\times 10^4$ \\
            \hline
        \end{tabular}
        \caption{Settings of the baseline fit.}
        \label{tab:dis_fit_settings}
    \end{minipage}
    \hfill
    \begin{minipage}[t]{0.48\textwidth}
        \centering
        \begin{tabular}{c|c|c}
            \textbf{Flavours} & \textbf{Small-$x$} ($\delta$) & \textbf{Large-$x$} ($\epsilon$) \\
            \hline
            $\Sigma$ &   $[1.089,\;  1.119]$   & $[1.475,\; 3.119]$ \\
            $g$      &   $[0.7504,\; 1.098]$  & $[2.814,\; 5.669]$ \\
            $V$      &   $[0.479,\;  0.7384]$  & $[1.549,\; 3.532]$ \\
            $V_3$    &   $[0.1073,\; 0.4397]$ & $[1.733,\; 3.458]$ \\
            $V_8$    &   $[0.5507,\; 0.7837]$ & $[1.516,\; 3.356]$ \\
            $T_3$    &   $[-0.4506,\; 0.9305]$& $[1.745,\; 3.424]$ \\
            $T_8$    &   $[0.5877,\; 0.8687]$ & $[1.522,\; 3.515]$ \\
            $T_{15}$ &   $[1.089,\; 1.141]$   & $[1.492,\; 3.222]$ \\
            \hline
        \end{tabular}
        \caption{Ranges of the preprocessing exponents.}
        \label{tab:dis_preprocessing_ranges}
    \end{minipage}
\end{table}

\paragraph{The NNPDF4.0 parametrisation.}
The PDFs are parametrised by means of a feed-forward NN with two hidden layers
having 25 and 20 nodes, respectively. The activation function is the hyperbolic
tangent in the hidden layers, while a linear function is used in the output
layer. We parametrise $\nout=8$ independent flavours, which correspond to the
output nodes of the NN. These are, in order, $\Sigma$, $g$, $V$, $V_3$, $V_8$,
$T_3$, $T_8$, and $T_{15}$. The PDFs are parametrised at the initial scale $Q_0
= 1.65$ GeV, which is also the scale at which they are shown throughout this
work. Each flavour is supplemented by preprocessing factors of the form
$x^{1-\delta}(1-x)^{\epsilon}$ that encode the expected behaviour of the PDFs at
small- and large-$x$. Note also that since the grid in $x$ on which the PDFs are
evaluated has both logarithmic and linear scales, the input layer is
supplemented with a transformation from the bare input $x$ to the doublet $(x,
\log x)$, which is then fed as input to the first hidden layer. This ensures
that both the linear and the logarithmic points in $x$ are treated on an equal
footing by the model. Denoting by $\NN_{\alpha j} = \NN(x_{\alpha})_j$ the
$j$-th flavour of the NN evaluated at point $x_{\alpha}$, the model can be
written as
\begin{equation}
  f_{i\alpha}
  = A_i \sum_{j=1}^{\nout} x_{\alpha} ^{1-\delta_j} (1-x_{\alpha})^{\epsilon_j}\NN_{\alpha j} R_{ji}^{(\rm FK)},
  \label{eq:nnpdf_model}
\end{equation}
where $R^{(\rm FK)}$ is the rotation matrix that rotates the $\nout$-dimensional
output space of the NN to the $\nf=14$ basis of the FK tables. The
normalisation factors $A_i$ are computed by imposing the momentum sum rule for
the gluon and the valence sum rules for $V$, $V_3$, and $V_8$, defined as
\begin{equation}
  A_g = \frac{1 - \int_0^1 dx \, x\widehat{\Sigma}(x)}{\int_0^1 dx \, x \, \widehat{g}(x)}, \quad
  A_{V} = \frac{3}{\int_0^1 dx \, \widehat{V}(x)}, \quad
  A_{V_3} = \frac{1}{\int_0^1 dx \, \widehat{V}_3(x)}, \quad
  A_{V_8} = \frac{3}{\int_0^1 dx \, \widehat{V}_8(x)}.
\end{equation}
Here a hat denotes the un-normalised distribution, \ie\ $x\widehat{f}_i(x)$ is
the right-hand side of \cref{eq:nnpdf_model} with $A_i = 1$. Note that the
momentum sum rule constrains a momentum fraction, and hence involves the
integrals of $x\Sigma$ and $xg$, whereas the valence sum rules count valence
quarks and therefore involve the integrals of the distributions themselves. The
normalisation factors depend indirectly on the weights of the NN and on the
preprocessing exponents, and therefore they change accordingly during training.
However, the dependence of the normalisation factors on the parameters makes the
model non-local in $x$, and it is not yet included in the analytical treatment
of the training dynamics that we show in \cref{sec:gradient_flow}. For this
reason, in the following we do not impose the sum rules and set $A_i = 1$
throughout this work.

\paragraph{Monte Carlo replicas.}
Experimental uncertainties are propagated to the fitted PDFs by means of the
Monte Carlo replica method~\cite{Costantini:2024wby}. We adopt the closure-test
procedure and train an ensemble of neural networks on a set of synthetic data
generated using a known underlying PDF set $\fin$, here taken to be the central
replica of the NNPDF4.0 parton set~\cite{NNPDF:2021njg}. Working with synthetic
data, for which we know the underlying \emph{true law}, allows us to assess the
performance of the model without introducing spurious effects of potentially
inconsistent data~\cite{Barontini:2025lnl}. We work throughout with level-2 (L2)
pseudodata. These are generated by first smearing the predictions computed with
the true law $\fin$ with random noise $\eta$ sampled from the data distribution,
and then adding an extra noise $\xi^{(k)}$ sampled independently for each
replica $k$ from the same distribution
\begin{equation}
  \label{eq:l2_data}
  Y^{(k)} = \fk \fin + \eta + \xi^{(k)},
  \quad
  \eta, \xi^{(k)} \sim \gauss(0, \covy),
\end{equation}
where $k$ labels the replica. We refer to the previous work in
Ref.~\cite{Chiefa:2025cap} and to Refs.~\cite{DelDebbio:2021whr,NNPDF:2021njg}
for a detailed treatment of the closure-test methodology.

\paragraph{Initialisation of the NN parameters.}
At initialisation, the exponents of the small- and large-$x$ factors are drawn
from a uniform distribution whose ranges, different for each flavour, are shown
in \cref{tab:dis_preprocessing_ranges}. These ranges were chosen in the
hyperoptimisation run of Ref.~\cite{NNPDF:2021njg}, to which we refer for a
detailed discussion. For the NN, the weights are drawn from the Glorot normal
distribution~\cite{glorot2010understanding}, while the biases are deliberately
set to zero. The Glorot normal distribution is a zero-mean Gaussian distribution
with a layer-dependent variance such that
\begin{equation}
  \expct\left[\wght{\ell}_{ij}\right] = 0\,,
  \quad
  \expct\left[\wght{\ell}_{ij}\wght{\ell'}_{i'j'}\right] 
  = \delta_{ii'}\delta_{jj'} \delta_{\ell\ell'} \sigma^2_{\ell}\,,
  \quad
  \sigma_{\ell}^2 = \frac{2}{n_{\ell-1} + n_{\ell}}\,,
  \label{eq:glorot_normal}
\end{equation}
where $\wght{\ell}_{ij}$ is the weight matrix of layer $\ell$ with
$i=1,\cdots,n_{\ell}$ and $j=1,\cdots,n_{\ell-1}$, the two integers $n_{\ell}$
and $n_{\ell-1}$ being the output and input dimensions of the hidden layer
$\ell$ of the NN.\footnote{The Keras implementation of the Glorot normal
initialisation truncates the Gaussian at two standard deviations. The fits
presented here use the untruncated Gaussian, with the variance of
\cref{eq:glorot_normal} unchanged.} Although this is the conventional choice in
the NNPDF methodology, we consider some variations -- and their impact on the
training dynamics -- in \cref{sec:init_dependence}.

\paragraph{Loss function and diagonal basis.}
Each replica of the model is trained to the corresponding replica of the data by
minimising an appropriate loss function. Unlike the NNPDF4.0 methodology,
here the fit is performed in the diagonal basis of the experimental covariance
matrix $\covy$. The experimental covariance matrix is diagonalised through the
orthogonal transformation $\covy = R D  R^T$, where $D$ is a diagonal matrix of
the eigenvalues of $\covy$, and $R$ is an orthogonal matrix whose columns are
the eigenvectors of $\covy$. Data and theoretical predictions are then rotated
to the diagonal basis as
\begin{equation}
  \label{eq:diagonal_basis}
  \widetilde{Y} = R^T Y, \quad
  \widetilde{T} = R^T T.
\end{equation}
For a given replica $k$, the loss function is then defined as the mean-squared
error weighted by the inverse of the diagonal matrix $D$, supplemented by
positivity and integrability penalties
\begin{equation}
  \label{eq:loss}
  \loglhd^{(k)}
  = \frac{1}{N_{\rm dat}} \lVert\widetilde{Y}^{(k)} - \widetilde{T}^{(k)}\rVert_{D^{-1}}^2
  + \mathcal{L}_{\rm pos} 
  + \mathcal{L}_{\rm int},
\end{equation}
where $\lVert \cdot \rVert_{D^{-1}}^2$ is the squared norm weighted by the
inverse of the diagonal matrix $D$. The additional terms $\mathcal{L}_{\rm pos}$
and $\mathcal{L}_{\rm int}$ in \cref{eq:loss} are Lagrange multipliers that
constrain the PDFs to satisfy the positivity and integrability conditions,
respectively. However, because the analytical treatment of the training dynamics
has not yet been extended to include these constraints, and because we also want to
keep the fitting procedure as close as possible to the previous study, these
terms are neglected in this analysis.

\paragraph{Minimisation and early stopping.}
The loss is minimised using vanilla gradient descent (GD), and no stochasticity
is introduced in the training. This is done to keep the training steps as close
as possible to the analytical solution of the gradient flow, which is derived
without any stochasticity, momentum terms, or adaptive learning rates. Only
the parameters of the neural network, namely weights and biases, are updated in
the minimisation, while the preprocessing factors are kept fixed after
initialisation. In order to prevent over-fitting, we split the dataset into a
training and a validation set -- the former is minimised during training while
the latter is monitored. The splitting is performed at the level of the
diagonalised dataset, where the data points are uncorrelated, so that the split
does not spoil the correlation structure of the original data. The training is
left to run for a fixed number of epochs, $\tend$.

\subsection{The analytical solution of the gradient flow}
\label{sec:gradient_flow}

The training of the network amounts to minimising the loss function in
\cref{eq:loss} by iterative steps of GD. For the purpose of the analytical
treatment, it is convenient to describe the discrete parameter updates as a
continuous \emph{gradient flow}, in which the parameters $\theta_{\mu}$ --
collectively the weights and biases of the network -- evolve in the continuous
training time $t$ along the negative gradient of the loss,
\begin{equation}
  \frac{d\theta_{t,\mu}}{dt} = -\partial_{\mu}\loglhd_t\,,
  \label{eq:gradient_flow}
\end{equation}
where $\partial_{\mu} = \partial/\partial\theta_{\mu}$ and the subscript $t$
denotes the derivative with respect to the parameter $\theta_{\mu}$ at training
time $t$. Note that the evolving parameters are those of the NN, while the
preprocessing factors are kept fixed after initialisation. For a sufficiently
small learning rate $\eta$, this continuous flow approximates the discrete GD
trajectory~\cite{barrett2021implicit}, with the training time $t$ related to the
number of epochs $\tau$ by the linear relation $t = \eta\tau$. For the sake of
the discussion, in the following we work in the original space of the data,
noting that the rotation to the diagonalised basis is obtained by applying
\cref{eq:diagonal_basis}.

\subsubsection*{The continuous gradient flow}

Because the theory predictions are linear in the PDFs, the derivative of the
predictions $T$ with respect to the PDFs is the time-independent FK table,
$\partial T/\partial f = \fk$. A few algebraic steps then translate the flow of
the parameters into a flow for the PDFs themselves,
\begin{equation}
  \frac{d f_t}{dt}
  = \ntk_t\, \fk^T \covdata^{-1}\left(\obs - \fk f_t\right)\,,
  \label{eq:field_flow}
\end{equation}
where the NTK $\ntk$~\cite{jacot2018neural,Chiefa:2025cap} is
defined as
\begin{equation}
  \ntk_{t,\,i_1\alpha_1 i_2\alpha_2}=
  \sum_{j_1,j_2=1}^{\nout}
  R^{(\rm FK)}_{j_1 i_1} R^{(\rm FK)}_{j_2 i_2}\,
  x^{1-\delta_{j_1}}_{\alpha_1} (1-x_{\alpha_1})^{\epsilon_{j_1}}
  x^{1-\delta_{j_2}}_{\alpha_2}(1-x_{\alpha_2})^{\epsilon_{j_2}}
  \sum_{\mu}
  \partial_{\mu} \NN_{t,\alpha_1 j_1}\,\partial_{\mu} \NN_{t,\alpha_2 j_2}\,.
  \label{eq:ntk_def}
\end{equation}
The NTK carries two pairs of flavour and grid indices. With the same
flavour-major vectorisation used for the FK tables and the PDFs, each pair is
collapsed into a single composite index, so that $\ntk_t$ is represented as a
symmetric, positive semi-definite matrix
\begin{equation}
  \ntk_t \in \real^{\nvec \times \nvec}\,,
\end{equation}
acting on the same space as the vectorised PDFs $f$. It is then convenient to
recast \cref{eq:field_flow} in the compact form
\begin{equation}
  \frac{d f_t}{dt} = -\ntk_t\left( M f_t - b \right)\,, \qquad
  b = \fk^T \covdata^{-1} \obs\,,
  \label{eq:flow_equation}
\end{equation}
where $b$ is the \emph{pull} that collects the data-driven source term, and the
metric $M$ is defined as
\begin{equation}
  M = \fk^T \covdata^{-1} \fk \in \real^{\nvec \times \nvec}\,.
  \label{eq:M_operator}
\end{equation}
Since the parameters are updated by minimising the \emph{training} loss alone,
the data $\obs$, the FK tables, and the covariance $\covdata$ appearing in
\cref{eq:flow_equation} are those of the training set.

\begin{figure}
    \centering
    \includegraphics[width=1.0\textwidth]{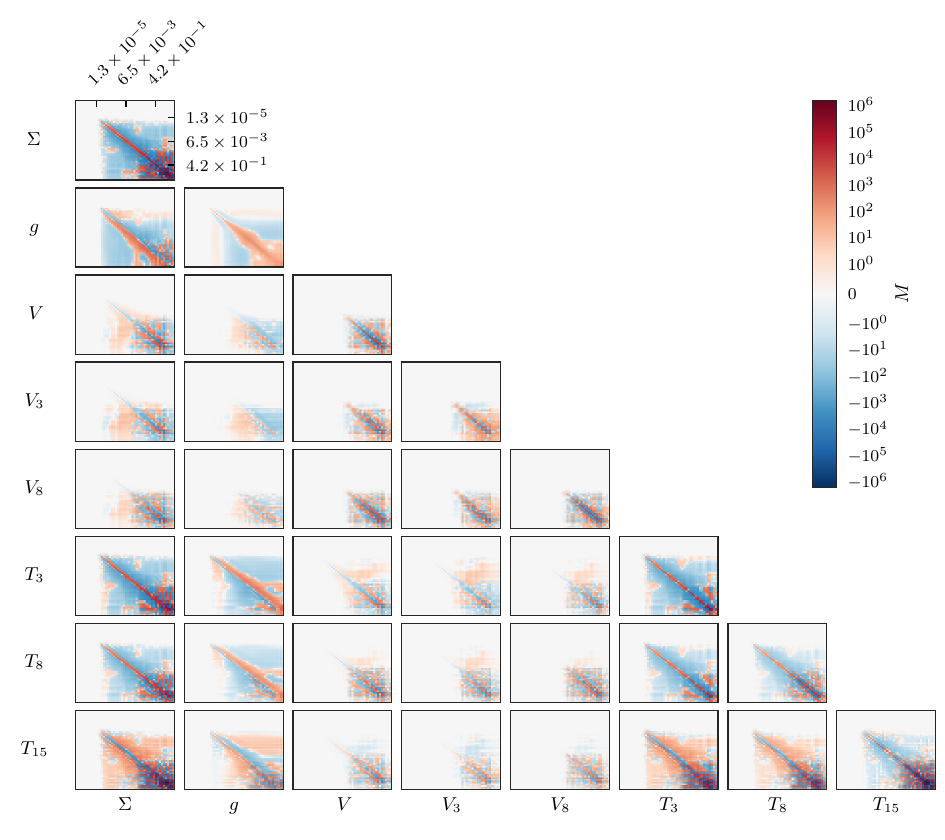}
    \caption{The block matrices of the operator $M$ for the eight independent
    fitted flavours. Each block represents the correlations between a pair of
    flavours on the $x$-grid as seen through the FK tables and weighted by the experimental
    uncertainties. The colour bar on the right encodes the size of these
    correlations: red for the correlated points, blue for the anti-correlated
    ones, white for points not constrained by the data. The operator $M$ is
    symmetric by construction, so that only the lower triangular block is
    shown.}
    \label{fig:fkspace}
\end{figure}

The flow is thus driven by the interplay of two matrices: the NTK $\ntk_t$,
which encodes the dependence on the architecture of the network, and the
operator $M$. The latter carries the physical correlations between flavours
induced by the underlying theory, then weighted by the experimental
uncertainties. In \cref{fig:fkspace} we show the matrix $M$ for the DIS dataset
used in this work. The matrix is presented in a block structure, where each
block represents the correlations between a pair of flavours. Being symmetric by
construction, the matrix is shown only in its lower-triangular part. Applying an
SVD to this matrix reveals that it is affected by many vanishing modes -- flat
directions in PDF space that the data are not able to constrain. Therefore, any
contribution lying in its kernel is left unchanged by the flow. In
\cref{sec:res_power_data} we give a statistical interpretation of the operator
$M$, while its connection to the Shapley value~\cite{Bonnet-Guerrini:2026thq} is
discussed in \cref{sec:shapley}.

\subsubsection*{The frozen NTK and the training Hamiltonian}

The NTK in \cref{eq:ntk_def} depends on the training time $t$ in a non-trivial
way that prevents a direct integration of \cref{eq:flow_equation}. However, as
discussed at length in Ref.~\cite{Chiefa:2025cap}, we expect the training
dynamics to exhibit two distinct regimes: an early \emph{rich} phase, in which
the NTK evolves rapidly and aligns its spectrum with the physically relevant
directions, followed by a \emph{lazy} phase, in which the NTK becomes
approximately constant. In the lazy regime, the NTK can be \emph{frozen} to its
value at a reference time $\tcntk$,
\begin{equation}
  \ntk_t = \ntk_{\tcntk} \equiv \ntk\,, \qquad \forall\,t \geq \tcntk\,,
  \label{eq:frozen_ntk}
\end{equation}
so that \cref{eq:flow_equation} reduces to a linear differential equation with
constant coefficients. Note that in the infinite-width
limit~\cite{jacot2018neural,lee:2020} the NTK is constant from initialisation,
and the linearised dynamics holds along the whole trajectory. This is not
necessarily true in our context, as also previously shown in
Ref.~\cite{Chiefa:2025cap}, so that \cref{eq:frozen_ntk} holds from $\tcntk$
onwards. Of course, the two notions coincide only when the rich phase is absent.

The solution to the flow equation is obtained by projecting the flow onto the
eigenbasis of the frozen NTK. Being symmetric and positive semi-definite, the
NTK admits the spectral decomposition
\begin{equation}
  \ntk = Z\Lambda Z^T\,,
  \qquad
  \Lambda\,,Z\,\in\real^{\nvec\times\nvec}\,,
  \label{eq:NTK_diagonalisation}
\end{equation}
where $\Lambda$ is the diagonal matrix of the eigenvalues and $Z$ is the
orthogonal matrix of the eigenvectors, whose columns are defined as
\begin{equation}
  Z_{ak} \equiv z^{(k)}_{a}\,,
  \qquad 
  a,\,k = 1,\dots,\nvec\,.
  \label{eq:NTK_eigenvecs_def}
\end{equation}
Note that each eigenvector can have non-vanishing entries for multiple flavours
at the same time, so that the label $k$ cannot be associated with a single
flavour. Although not yet discussed, we expect the spectrum of the NTK to be
populated by many vanishing modes that make up its kernel, $\ntkker$. Denoting
by $\dperp$ the number of non-vanishing modes that define the image of the NTK,
$\ntkimg$, we can decompose the eigenspace as follows
\begin{equation}
  \Lambda = \begin{bmatrix}
    \Lambda_{\perp} & 0\\
    0 & 0
  \end{bmatrix}\,,
  \qquad
  \Lambda_{\perp}\in\real^{\dperp\times\dperp}\,,
  \label{eq:eigval_decomposition}
\end{equation}
which induces a similar decomposition of the eigenvectors
\begin{equation}
  Z = \left[ \zperp \; \zpar \right]\,,
  \qquad
  \zperp\in\real^{\nvec\times\dperp}\,,
  \quad
  \zpar\in\real^{\nvec\times(\nvec-\dperp)}\,,
  \label{eq:eigvec_decomposition}
\end{equation}
satisfying the following identities:
\begin{align}
    &Z_{s}^T Z_{s} = \id_s\,,
    \quad
    P_s = Z_{s} Z_{s}^T \neq \id_s\,,
    \label{eq:x_proj}
    \quad
    s\in\left\{\parallel,\, \perp\right\}\\[2pt]
    &\zperp\zperp^T + \zpar\zpar^T = \pperp + \ppar = \id_{\nvec}\,,
    \label{eq:Z_completeness}
\end{align}
where $\pperp$ and $\ppar$ are the projectors onto $\ntkimg$ and $\ntkker$,
respectively. Following the definitions of Ref.~\cite{Chiefa:2025cap}, the
vector of PDFs $f$ can be decomposed accordingly
\begin{equation}
  \fker = \zpar^T f\,, \qquad
  \fperp = \Lambda_{\perp}^{-1/2}\zperp^T f\,.
\end{equation}
By construction, the components that live in the kernel do not evolve during
lazy training and remain at their initial condition, $\fker_t = \fker_0$.
Instead, the components of the image evolve according to the
\emph{training Hamiltonian}
\begin{equation}
  H^{\perp} = \Lambda_{\perp}^{1/2}\zperp^T M \zperp \Lambda_{\perp}^{1/2}\,,
  \label{eq:training_hamiltonian}
\end{equation}
which is symmetric and positive definite under the mild assumption $\mker \cap
\ntkimg = \{ 0 \}$. 

We emphasise that \cref{eq:training_hamiltonian} is where the regularisation of
the inverse problem is manifest. The matrix $M$, singular by construction, is
projected onto the image of the NTK, so that only the physical directions that
the parametrisation can reconstruct -- the eigenvectors of the NTK -- enter the
dynamics. The intersection above dictates whether this is sufficient to
regularise the problem. If the intersection is non-trivial, there are some
directions explored by the parametrisation that are not constrained by the data:
the training Hamiltonian would then inherit such flat directions of $M$, and the
chosen functional parametrisation would fail in regularising the problem. If,
however, as happens with NNs, the intersection is trivial, all the flat
directions of $M$ are cured by the projection and the training Hamiltonian is
invertible.

With this consideration in mind, we can write
\begin{equation}
  H^{\perp}\, w^{(i)} = h^{(i)}\, w^{(i)}\,,
  \qquad i = 1,\dots,\dperp\,,
  \qquad w^{(i)} \in \real^{\dperp}\,,
  \label{eq:hamiltonian_eigen}
\end{equation}
and it is convenient to map the eigenvectors $w^{(i)}$ back to the PDF space through
\begin{equation}
  Q = \zperp \Lambda_{\perp}^{1/2} W\,,
  \quad \mathrm{or} \quad
  Q_{a i} \equiv q^{(i)}_{a} = \sum_{k=1}^{\dperp} \sqrt{\lambda^{(k)}}\, z^{(k)}_{a}\, w^{(i)}_{k}
  \in\real^{\nvec}\,,
  \label{eq:hamiltonian_eigvec_pdf}
\end{equation}
where $k$ runs over the non-vanishing modes of the NTK. As we now show, the
vectors $q^{(i)}$ form the natural basis in which the trained solution is
expressed, each mode being learned on a timescale set by the inverse eigenvalue
$1/h^{(i)}$. One can verify the following properties
\begin{equation}
  q^{(i)T} M\, q^{(j)} = h^{(i)}\, \delta_{ij}\,,
  \qquad
  \sum_{i=1}^{\dperp} q^{(i)} q^{(i)T}
  = \zperp \Lambda_{\perp} \zperp^T = \ntk\,.
  \label{eq:modes_properties}
\end{equation}
The first equality implies that the vectors $q^{(i)}$ are not orthogonal in the
Euclidean sense but rather with respect to $M$; the second one provides an
alternative decomposition of the NTK.

We finally observe that if we write $q^{(i)} = \lVert q^{(i)} \rVert\, n^{(i)}$,
with $n^{(i)}$ a unit vector in PDF space and $\lVert q^{(i)} \rVert$ the
standard Euclidean norm, then we have
\begin{equation}
  h^{(i)} = \lVert q^{(i)} \rVert^2\, n^{(i)T} M\, n^{(i)}\,,
  \label{eq:hval_decomposition}
\end{equation}
which shows that each eigenvalue of the training Hamiltonian is the product of
two terms. The first one is the weight in the mode expansion of the NTK, as in
the second expression of \cref{eq:modes_properties}, and can be small if the
direction is poorly covered by the parametrisation. The second term is discussed
more extensively in \cref{sec:res_power_data}, and quantifies the resolution
power of the data along a given direction $n$. Both contribute to the size of
$h^{(i)}$, and therefore to the speed at which each mode is resolved by the
analytical solution (see below).

\subsubsection*{The analytical solution}
Integrating the flow in the two subspaces (see Ref.~\cite{Chiefa:2025cap} for
the explicit derivation) yields the central result
\begin{equation}
  f_{t} = U(t)\, f_{0}
    + V(t)\,\obs\,,
  \label{eq:analytic_sol}
\end{equation}
which expresses the trained solution at any time $t$ as the sum of a linear
transformation $U(t)$ of the initial condition $f_0$ and a linear transformation
$V(t)$ of the data, provided that the lazy regime is satisfied. Following
Ref.~\cite{Chiefa:2025cap}, the evolution operator $U(t)$ is the sum of three
contributions,
\begin{equation}
  U(t) = \hat{U}^{\perp}(t) + \check{U}^{\perp}(t) + U^{\parallel}\,,
  \label{eq:U_operator}
\end{equation}
where
\begin{equation}
  \hat{U}^{\perp}(t) = \zperp \Lambda_{\perp}^{1/2}\, e^{-H^{\perp} t}\,
    \Lambda_{\perp}^{-1/2} \zperp^T\,,
  \qquad
  U^{\parallel} = \zpar \zpar^T\,.
  \label{eq:U_hat}
\end{equation}
The contributions from $\check{U}^{\perp}(t)$ and $V(t)$ are more easily
expressed by introducing the operator
\begin{equation}
  \begin{split}
    \mathcal{M}(t)
    &= \zperp \Lambda_{\perp}^{1/2} \left( \id - e^{-H^{\perp} t} \right)
      \left( H^{\perp} \right)^{-1} \Lambda_{\perp}^{1/2} \zperp^T \\
    &= \sum_{i=1}^{\dperp} \frac{1 - e^{-h^{(i)} t}}{h^{(i)}}\;
      q^{(i)} q^{(i)T}\,,
  \end{split}
  \label{eq:M_t_operator}
\end{equation}
in terms of which we can write
\begin{equation}
  \check{U}^{\perp}(t) = -\mathcal{M}(t)\, M\, U^{\parallel}\,,
  \qquad
  V(t) = \mathcal{M}(t)\, \fk^T \covdata^{-1}\,.
  \label{eq:evolution_operators}
\end{equation}
As observed in Ref.~\cite{Chiefa:2025cap}, the operator $\mathcal{M}(t)$ can be
seen as a regularised inverse of $M$ restricted to the image of the NTK, with
the training time $t$ being the regularisation parameter. In fact, it can be verified
that
\begin{equation}
  \mathcal{M}_{\infty} 
  = \lim_{t\rightarrow\infty}\mathcal{M}(t) 
  = \zperp \left(\zperp^T M \zperp\right)^{-1} \zperp^T\,,
  \label{eq:M_infty}
\end{equation}
which satisfies $\mathcal{M}_{\infty} M \mathcal{M}_{\infty} =
\mathcal{M}_{\infty}$. 

The regularisation induced by the training time $t$ is enforced by the filter
factor $(1-e^{-h^{(i)} t})$: the modes with $h^{(i)} t \gg 1$ have been
completely resolved and are thus fully inverted, while those with $h^{(i)} t \ll
1$ have covered only a fraction of $h^{(i)} t$ and are suppressed by $t$. We
refer the reader to Ref.~\cite{Chiefa:2025cap} for an extensive discussion of
these operators. Here we provide a second reading of the analytical expression,
which ultimately facilitates our discussion in \cref{sec:early_stopping}.

\paragraph{An alternative form.}
The three contributions to $U(t)$ can be summed explicitly. Using the definition
of the training Hamiltonian in \cref{eq:training_hamiltonian}, one can prove
that $\mathcal{M}(t)\, M \zperp\zperp^T = \zperp\zperp^T - \hat{U}^{\perp}(t)$,
which combined with \cref{eq:Z_completeness} yields
\begin{equation}
  U(t) = \id_{\nvec} - \mathcal{M}(t)\, M\,.
  \label{eq:U_compact}
\end{equation}
The analytical solution in \cref{eq:analytic_sol} then takes the compact form
\begin{equation}
  f_t = f_0 - \mathcal{M}(t) \left( M f_0 - b \right)\,,
  \label{eq:analytic_sol_compact}
\end{equation}
where $b$ is the pull defined in \cref{eq:flow_equation}. This form offers a
second reading of the solution. The combination $M f_0 - b$ is proportional to
the gradient of the loss with respect to $f$, evaluated at the initial condition
$f_0$. The trained solution is thus obtained by displacing the initial condition
by the operator $\mathcal{M}(t)$ applied to this gradient. For small $t$ one has
$\mathcal{M}(t) \simeq t\,\ntk$, and \cref{eq:analytic_sol_compact} reduces to
the first discrete step of the flow in \cref{eq:flow_equation}. 

\subsection{Early stopping as a bias-variance trade-off}
\label{sec:early_stopping}

We conclude this section by showing how the analytical solution disentangles the
role of early stopping in the training process. To do so, we use the L2
pseudodata of \cref{eq:l2_data} to recast \cref{eq:analytic_sol_compact} into
the \emph{teacher-student} form, which reads
\begin{equation}
  f_t
  = \underbrace{f_0 + \mathcal{M}(t)\, M \left(\fin - f_0 \right)}_{\textrm{teacher part}}
  + \underbrace{\mathcal{M}(t)\, \fk^T \covdata^{-1} \left( \eta + \xi^{(k)} \right)}_{\textrm{noise part}}\,.
  \label{eq:analytic_sol_compact_l2}
\end{equation}
If we neglect the fluctuations of the NTK, and treat $f_0$ and $\mathcal{M}$ as
deterministic objects, the two terms appearing in this expression have a clear
interpretation.

The first term -- the \emph{teacher} in the language of teacher-student settings
-- depends on $\fin$ and corrects the initial condition $f_0$ by the regularised
inversion of the residual $\fin - f_0$. It carries no noise of the data, and its
role is to move the central value closer to the target. By using
\cref{eq:M_t_operator}, we see that this is achieved by incrementally including
new modes as they are resolved by the filter factor. For $t\to\infty$, all
available modes are used and $\mathcal{M}_{\infty}$ is the inverse of $M$
restricted to $\ntkimg$. The teacher part thus reduces to the solution of the
least-squares error with the noiseless data $\fk\fin$ within $f_0 + \ntkimg$.

The second term, on the other hand, injects the noise of the data and carries no
information about $\fin$. Note that $\eta$ is common across replicas, shifting
the central value away from the target, while $\xi^{(k)}$ is centred at zero and
fluctuates with the replicas, contributing to the variance of the solution. The
interplay with the teacher term becomes manifest if these two sources of noise
are decomposed along the modes $q^{(i)}$. Using \cref{eq:M_t_operator}, the
shift to the central value becomes
\begin{equation}
  \mathcal{M}(t)\,\fk^T\covdata^{-1}\eta
  = \sum_{i=1}^{\dperp}
    \frac{1-e^{-h^{(i)}t}}{n^{(i)T} M\, n^{(i)}}\,
    n^{(i)}\, n^{(i)T}\fk^T\covdata^{-1}\eta\,,
  \label{eq:cv_noise}
\end{equation}
while the contribution to the variance is
\begin{equation}
  \covar\left[\mathcal{M}(t)\,\fk^T\covdata^{-1}\xi^{(k)}\right]
  = \mathcal{M}(t)\, M\, \mathcal{M}(t)
  = \sum_{i=1}^{\dperp}
    \frac{\left(1-e^{-h^{(i)}t}\right)^2}{n^{(i)T} M\, n^{(i)}}\,
    n^{(i)} n^{(i)T}\,,
  \label{eq:cov_noise}
\end{equation}
where in both expressions we used \cref{eq:hval_decomposition} to express the
weights in terms of the inverse of the resolution power of the data along the
$i$-th direction, $n^{(i)T} M n^{(i)}$.\footnote{This quadratic form quantifies
how well a given vector is constrained by the data, and it is extensively
discussed in \cref{sec:res_power_data}.} When $t\gg1/h^{(i)}$, the weights of
the $i$-th mode in \cref{eq:cv_noise,eq:cov_noise} reduce to the same factor.
Therefore, those directions that are poorly constrained by the data are also
those with the largest impact on the central value and on the variance. By
\cref{eq:hval_decomposition}, we also see that these poorly constrained
directions yield small $h^{(i)}$, and thus correspond to the slower modes in the
spectrum.\footnote{The converse is not true in general: a small $h^{(i)}$ can
also result from a small norm $\lVert q^{(i)} \rVert$, namely from a direction
that is well constrained by the data but poorly covered by the parametrisation.
Such a mode would be slow without injecting a large amount of noise. However, we
expect a NN to be flexible enough to cover the directions constrained by the
data, so that the slowest modes are mainly those poorly constrained by the
data.}

We conclude that the inclusion of new modes, while improving the teacher term,
can actually lead to a poor performance of the overall solution, as we shall
explicitly see in \cref{sec:analytical_training}
(\cref{fig:AnltSolInit,fig:UVContributions}). Early stopping can thus be seen as
a prescription to set the trade-off between the bias (the teacher plus the noise
shift) and the variance (injected by $\xi^{(k)}$). We finally note that a
similar result is also found in Ref.~\cite{coeurdoux2026rmt} for a simplified,
solvable setting.

%% file: sections/geostat.tex
\section{Information content and geometry of a PDF fit}
\label{sec:geostat}

In the previous section, we introduced the two main objects of the present
analysis, namely the NTK and the operator $M$. In the following, we explore the
statistical and geometric interpretation of these two, which allows us to
clarify their role in the solution of the inverse problem. In
\cref{sec:res_power_data} we show that $M$ is the Fisher information matrix
(FIM) of a statistical model whose parameters are the PDFs themselves. In
\cref{sec:shapley} we recover the Shapley value of
Ref.~\cite{Bonnet-Guerrini:2026thq} from the operator $M$ for predictions that
are linear in the PDFs. Finally, in \cref{sec:geometric_interpretation} we
present a novel geometric interpretation of a PDF fit. This section is
complementary to the numerical analysis presented in this work. The reader
interested in the main results can jump to \cref{sec:results}.

\subsection{Resolving power of the data}
\label{sec:res_power_data}

We show that the matrix $M$, defined in \cref{eq:M_operator}, is the FIM that
measures the amount of information carried by the data $Y$ about the PDFs $f$.
Let $\pcond{Y}{f} \equiv \pyf$ denote the likelihood of the data $Y$ given a set
of PDFs $f$. In our setup, its negative log-likelihood is
\begin{equation}
  -2 \log \pyf \propto \left(Y - T[f] \right)^T \covdata^{-1} \left(Y - T[f] \right)\,,
  \label{eq:NLL} 
\end{equation}
which is the usual quadratic form that compares predictions against data in
units of the experimental uncertainties. Note that \cref{eq:NLL} is our
deliberate choice in defining a notion of distance between two points in the
space of the data, with the metric given by the inverse of the experimental
covariance matrix $\covdata$. As our ultimate goal in our analysis is the PDF,
we want to find a similar notion of distance, induced by \cref{eq:NLL}, in the
space of PDFs.\footnote{In geometric terms, \cref{eq:NLL} sets the notion of
distance on a statistical manifold whose points live in the space of the data.
The distance in PDF space is the \emph{pull-back} metric induced by
\cref{eq:NLL} obtained by reversing the forward map of
\cref{eq:convolution_vectorised}.} 

To do so, we use a variational approach and study how much a small perturbation
$\delta f$ in the PDFs affects the original likelihood $\pyf$. We use the
Kullback--Leibler (KL) divergence to compare the two distributions,
\begin{equation}
  \kldiv{\pyf}{\pydf} \equiv
  \int_Y \pyf \log \frac{\pyf}{\pydf}\,,
  \label{eq:KLDiv}
\end{equation}
where $\pydf$ is the likelihood given the perturbation $\delta f \in
\real^{\nvec}$. We introduce the \emph{score function} $s \equiv
s\left(Y;f\right) \in \real^{\nvec}$, whose entries are defined as
\begin{equation}
  s_{i\alpha} = \frac{1}{\pyf} \frac{\partial \pyf}{\partial f_{i\alpha}} = \frac{\partial \log \pyf}{\partial f_{i\alpha}}\,,
  \label{eq:ScoreFunc}
\end{equation}
which measures how sensitive the probability of the observed data is to a change
in the PDFs, relative to the original distribution $\pyf$. We expand
\cref{eq:KLDiv} in powers of $\delta f$ and, after some algebraic manipulations,
we can write
\begin{equation}
  \kldiv{\pyf}{\pydf} 
  = \delta f^T \, \expct\left[ s\right] 
  + \frac{1}{2} \delta f^T \, \fisher \, \delta f
  + \mathcal{O} \left( \delta f^3 \right) \,,
  \label{eq:KLDivExp}
\end{equation}
where the expectation values are evaluated with respect to $\pyf$ and $\fisher$
is the FIM defined as
\begin{equation}
  \fisher = - \expct\left[ \frac{\partial^2 \log \pyf}{\partial f \partial f^T} \right]\,,
  \label{eq:FIMDef}
\end{equation}
which is the expected local curvature of the log-likelihood. Assuming mild
regularity conditions on $\pyf$~\cite{Lehmann1998}, one can show that:
\begin{enumerate}
  \item The score has zero mean, and therefore the first term in
  \cref{eq:KLDivExp} disappears;
  \item The FIM admits an alternative form, which is $\fisher = \expct\left[ s
        s^T \right]$. 
\end{enumerate}
From these two properties, it follows that the FIM is the variance of the score
function: a large fluctuation of the score means that the data are very
sensitive to $f$, implying that the data strongly constrain the PDFs. We can
finally compute the FIM explicitly using either of the two definitions and show
that $\fisher = M$. Therefore, \cref{eq:KLDivExp} can finally be written as
\begin{equation}
  \kldiv{\pyf}{\pydf} 
  = \frac{1}{2} \delta f^T \, M \, \delta f
  + \mathcal{O} \left( \delta f^3 \right) \,.
  \label{eq:KLDivExp2}
\end{equation}
This result shows that the matrix $M$ is the positive semi-definite metric on
the space of PDFs endowed with the statistical structure inherited from
\cref{eq:NLL}. A different prescription for comparing the data with the theory
would imply a different notion of distance for the PDFs.

We emphasise that \cref{eq:KLDivExp2} can be read in two complementary ways.
Read forward, it is a statement on the distribution of the data given $f$: it
measures how much a perturbation of the PDFs along a given direction changes the
likelihood of the observations. Directions along which the quadratic form is
large produce predictions that are statistically distinguishable from the
original ones, while directions along which the quadratic form is small leave
the distribution of the data essentially unchanged. Read backward, the same
equation becomes a statement on the constraining power of the data about the
PDFs. A perturbation that the data would clearly resolve is a perturbation for
which the fit has little freedom, and the corresponding direction in PDF
space is therefore tightly constrained. Conversely, the flat directions of
$M$, and in particular those belonging to $\mker$, can be varied at no cost in
the likelihood: they are left undetermined by the data, and are fixed only by
the parametrisation. The spectrum of $M$ thus provides a direct map of which
regions of the PDF space are informed by the experiments and which ones are not.

\subsection{The Shapley value in the DIS case}
\label{sec:shapley}

In game theory, the Shapley value is an estimator that quantifies the
contribution to the total payoff of each player among a collaborating group.
Adapted to the context of PDFs, it has been recently used in
Ref.~\cite{Bonnet-Guerrini:2026thq} to measure how much each individual flavour
and kinematic region in $x$ contributes to the description of the data. Read
backward, in the sense of \cref{sec:res_power_data}, this statement thus
provides a direct probe of the constraining power of the data on each flavour.

Here we show that the quadratic form of \cref{eq:KLDivExp2} is exactly the
object that is computed, albeit in a different language, by the Shapley-value
analysis: there the perturbations $\delta f$ are localised bumps acting on a
given flavour in a narrow region of $x$, and the associated payoff is the change
they induce in the likelihood. The matrix $M$ is therefore at the core of this
estimator, as far as DIS observables are concerned.

In the referenced work, a localised Gaussian bump centred at a chosen $\hat{x}$,
calibrated to the PDF uncertainty, is applied with both signs to the PDFs. A
coalition configuration is defined as a subset $S$ of flavours for which the
perturbation is applied. In our vectorised convention, a PDF vector of this
coalition is expressed as
\begin{equation}
  f^{(S\pm)} = \fb + \delta f^{(S\pm)}\,,
\end{equation}
where $\fb$ denotes the vector of unperturbed PDFs as in
\cref{eq:PDFVecExplicit}, and $\delta f^{(S\pm)}$ is the vector of either
positive or negative Gaussian bumps with non-zero entries only on the flavours
belonging to $S$. For a flavour $j\in S$, the Gaussian perturbation centred on
$\hat{x}$ is defined as~\cite{Bonnet-Guerrini:2026thq}
\begin{equation}
  \delta f_{j\alpha}^{(\pm)} = \pm \Delta_j^{(\pm)}(\hat{x}) \exp \left[
    -\frac{(\log_{10} x_{\alpha} - \log_{10} \hat{x})^2}{2w^2}
  \right]\,,
  \label{eq:ShapleyPerturbation}
\end{equation}
where $\Delta_j^{(\pm)}(\hat{x})$ are the upper and lower bounds of the 68\%
confidence interval of the PDF ensemble at $\hat{x}$. For a given coalition $S$,
the resulting change in the $\chi^2$ for a chosen dataset $\mathcal{D}$ is
measured by the value function $v(S)$, defined as
\begin{equation}
  v(S) = \frac{1}{2}\left[
            \chi^2\left( f^{(S+)}; \mathcal{D} \right) + \chi^2\left( f^{(S-)}; \mathcal{D} \right)
          \right]\,,
  \label{eq:ValueFunction}
\end{equation}
where $\chi^2\left( f^{(S\pm)}; \mathcal{D} \right)$ is evaluated using the
perturbed PDFs in the coalition $S$ for a chosen dataset $\mathcal{D}$. Denoting
by $\delta T^{(S\pm)}$ the change in the predictions induced by the Gaussian
bump $\delta f^{(S\pm)}$, \cref{eq:ValueFunction} can be decomposed into four
components
\begin{equation}
  \begin{split}
    v(S) = &\chi^2\left(f; \mathcal{D}\right) + \left(Y-T\right)^T\covy^{-1} \left( \delta T^{(S+)} + \delta T^{(S-)} \right) \\
           & + \frac{1}{2}\left[ \left(\delta T^{(S+)} \right)^T \covy^{-1}\,\delta T^{(S+)} + \left(\delta T^{(S-)} \right)^T \covy^{-1}\,\delta T^{(S-)}\right]\,,
  \end{split}
  \label{eq:ValueFunction2}
\end{equation}
where $T$ denotes the unperturbed prediction. In the case of DIS-like
observables, where predictions are computed as in
\cref{eq:convolution_vectorised}, this expression simplifies drastically if we
further assume symmetric bounds in \cref{eq:ShapleyPerturbation}, which implies
$\delta f^{(S)} \equiv \delta f^{(S+)} = - \delta f^{(S-)}$. We
can then write
\begin{equation}
    v(S) 
    = \chi^2\left(f; \mathcal{D}\right) + \left(\delta T^{(S)} \right)^T \covy^{-1}\,\delta T^{(S)}
    = \chi^2\left(f; \mathcal{D}\right) + \left(\delta f^{(S)} \right)^T M\,\delta f^{(S)} \,,
  \label{eq:ValueFunction3}
\end{equation}
where in the second equality we used \cref{eq:convolution_vectorised} to express
the second term in PDF space. For a given coalition $S$, we define the marginal
contribution of flavour $i$ as
\begin{equation}
  \phi_i(S) = v(S \cup \{i\} ) -  v(S)\,,
  \label{eq:ShapleyMarginal}
\end{equation} 
that is, the difference in the value functions with and without $i$ perturbed.
The Shapley value $\Phi_i(\hat{x})$ is the combinatorially weighted average of
this marginal contribution over all possible $2^{\nf-1}$ coalitions
$\mathcal{B}\setminus\{i\}$ that can be formed from the remaining $\nf-1$
flavours\footnote{The coalitions can be easily visualised with $u$, $d$, and $s$
($\nf=3$) and the Shapley value evaluated for $u$. In this case, all possible
coalitions that exclude $u$ are $\{d\}$, $\{s\}$, $\{d,s\}$ and
$\varnothing$ with weights $1/6$, $1/6$, $2/6$ and $2/6$.}
\begin{equation}
  \Phi_i = \sum_{S \subseteq \mathcal{B}\setminus\{i\}} w(S)\, \phi_i(S)\,,
  \label{eq:ShapleyValue}
\end{equation}
where $w(S)$ is the weight factor for a given coalition $S$ defined as
\begin{equation}
  w(S) = \frac{|S|!\,\left(\nf - |S| - 1\right)!}{\nf!}\,,
  \label{eq:ShapleyWeights}
\end{equation}
where $|S|$ denotes the number of elements in the coalition.

Recall that $M$ is a symmetric matrix whose blocks $M_{ij}$ are specified by a
pair of flavour indices, see \cref{fig:fkspace}. Since $\delta f^{(S)}$ has
support only on the flavours belonging to $S$, and only in a narrow window of
$x$ around $\hat{x}$, the quadratic form probes the sub-block of $M$ with both
flavour indices in $S$, localised in that same window. After some simple
algebra, one can then show that in the DIS-like case the Shapley value of
\cref{eq:ShapleyValue} reduces to the following expression
\begin{equation}
  \Phi_i 
  = \delta f_i^T M_{ii} \delta f_i + \sum_{j\neq i} \delta f_i^T M_{ij} \delta f_j\,,
  \label{eq:ShapleyValueFinal}
\end{equation}
which is the sum of quadratic forms that measure the actual contribution of the
perturbation as seen by the FK tables and weighted by the experimental
uncertainty. The diagonal entries $M_{ii}$ therefore quantify the sensitivity
of the data to each perturbed flavour on its own, while the off-diagonal
$M_{ij}$ terms encode the interference between simultaneously perturbed
flavours. The weighted average over all possible coalitions is then a sum over
all the correlations between flavours as encoded in the matrix $M$.

\subsection{A geometric interpretation of training}
\label{sec:geometric_interpretation}

We now propose a geometric interpretation of the training dynamics and of the
analytical solution of \cref{eq:analytic_sol}. Information Geometry
(IG)~\cite{e22101100,amari2016information} has long been proposed as a
theoretical framework that aims at describing statistical models by using the
language of differential geometry. For instance, in
Ref.~\cite{pmlr-v89-liang19a} IG is used to provide generalisation bounds in
ReLU networks that are invariant under reparametrisations. Natural gradient
descent~\cite{Amari:1998jzd} in machine learning follows from geometric
considerations applied to statistical manifolds. Finally, in collider physics IG
has been used to analyse the latent geometry of jet taggers~\cite{Kuntz:2026kuv}
and to construct bases of correlator observables~\cite{Bal:2026pzw}. Yet, to the
best of our knowledge, applications in the context of PDF determinations are
still missing. Here we do not aim at a rigorous treatment of the geometric
structure of a parametric fit to PDFs, which we leave to future work, but rather
focus on a more intuitive visualisation of these ideas.

\subsubsection*{The role of the model}
The functional model that parametrises the PDFs is a \emph{forward map} from the
parameter space $\mathcal{P} \subseteq \real^{\nparams}$ to the vectorised space
of PDFs, $\real^{\nvec}$. As parameters vary, this map traces a manifold
$\manfd_f$ in the ambient space $\real^{\nvec}$. A point $f$ on $\manfd_f$ is a
set of PDFs evaluated on the $x$-grid, and the training process can be
visualised as a trajectory on this manifold. At each point, the tangent space
$\tgnt{f}$ is spanned by the columns of the Jacobian, \ie\ by the \emph{tangent
vectors}
\begin{equation}
  J_{\mu} = \partial_{\mu} f \in \real^{\nvec}\,,
  \qquad \mu = 1, \dots, \nparams\,,
  \label{eq:tangent_features_def}
\end{equation}
each encoding the sensitivity of the function to a displacement of the
corresponding parameter. Note that the tangent vectors are not necessarily
orthogonal, nor linearly independent. We use the word \emph{frame} to denote
this overcomplete set of vectors that live in the tangent plane. In the NN
parametrisation of \cref{eq:nnpdf_model}, $\nparams=\parameters$ tangent vectors
live in a space of dimension $\nvec=\ngrid\cdot\nf=700$, and thus are
necessarily linearly dependent. Note also that the manifold $\manfd_f$ and the
related tangent bundle induced by the forward map are \emph{intrinsic} geometric
objects, and therefore invariant under reparametrisations.

The choice of the parametric model not only determines the manifold $\manfd_f$
-- the set of functional forms reachable by the parametrisation -- but also
specifies how the tangent space is covered. In fact, while the span of the
tangent space is fixed by the geometry, the way this space is populated by the
tangent vectors is entirely dictated by the parametrisation. Two different
models can then share the same tangent plane at a common point on
$\manfd_f$, but define different frames that explore the neighbourhood with
different sensitivity. A simplified example can help clarify this point.

\begin{example*}
\label{ex:frames}
Consider the forward map $f(x; \theta_1, \theta_2) = \theta_1 \theta_2 x$, which
has tangent vectors $J_1 = \theta_2 x$ and $J_2 = \theta_1 x$. Here both vectors
lie in the one-dimensional tangent space spanned by $x$, $\mathrm{span}\{x\}$,
and therefore the two tangent vectors are linearly dependent. A displacement in
the parameters provokes a displacement in the function along the tangent space
\begin{equation}
  \delta f(x) 
  = \partial_1 f(x)\, \delta\theta_1 + \partial_2 f(x)\, \delta\theta_2
  = \left( \delta\theta_1\theta_2 + \theta_1\delta\theta_2 \right)x\,.
  \label{eq:OverParmExample}
\end{equation}
Any displacement along the direction $(\delta\theta_1, \delta\theta_2) \propto
(\theta_1, -\theta_2)$ leaves the function unchanged at first order. This is not
a coincidence, but a consequence of the rescaling symmetry $(\theta_1, \theta_2)
\to (c\,\theta_1, \theta_2/c)$ that leaves the function untouched. Compare now
with the forward map $f(x; \theta) = \theta x$. This model induces the same
functional manifold and the same tangent plane (line spanned by $x$), but now
with a single tangent vector $J = x$. The two parametrisations thus select
different frames on the same tangent space, and the NTK of the two cases reads
\begin{equation}
  \ntk = \left( \theta_1^2 + \theta_2^2 \right) x\, x^T
  \qquad \textrm{and} \qquad
  \ntk = x\, x^T\,,
  \label{eq:NTKExample}
\end{equation}
respectively. These have the same eigenvector, but different eigenvalues. In
fact, in the first case the strength with which the direction $x$ is covered
depends on the point in parameter space, while in the second case it is
constant.
\end{example*}

\noindent
From this viewpoint, the NTK is nothing but the Gram matrix of the tangent
vectors,
\begin{equation}
  \ntk = \sum_{\mu} J_{\mu}\, J_{\mu}^{T}\,,
  \label{eq:NTKfeatures}
\end{equation}
and as such it encodes the covering capability of the frame. We can understand
how tangent vectors make up the eigenvectors of the NTK by taking the quadratic
form (also known as the Rayleigh quotient) with any unit direction $n$ in PDF space,
that is
\begin{equation}
  n^{T} \ntk\, n = \sum_{\mu} \left( J_{\mu},\, n \right)^{2}\,,
  \label{eq:coverage}
\end{equation}
where the spectrum is given by the stationary points of \cref{eq:coverage}.
Therefore, if $n$ is an eigenvector of $\ntk$, the quadratic form receives a
major contribution from all tangent vectors $J_{\mu}$ with a large component
along $n$, while tangent vectors that are mildly aligned with $n$ give
negligible contributions. An eigenvector is thus the collection of tangent
vectors with a common principal component. We finally note that, in contrast to
the manifold, the NTK is a property of the parametrisation because it depends on
the point $\theta$ in parameter space and it is not invariant under
reparametrisations.

\subsubsection*{Training as a flow on the manifold}
A gradient flow on the manifold of parameters $\manfd_{\theta}$ is generally
defined as
\begin{equation}
  \dot{\theta} = - G^{-1} \nabla_{\theta}\loglhd\,,
  \label{eq:GFManifold}
\end{equation}
where $G$ is the chosen metric on $\manfd_{\theta}$, and $\nabla_{\theta}$ is
the covariant derivative of the loss with respect to the parameters. The flow in
functional space, as in \cref{eq:flow_equation}, is thus the push-forward of
\cref{eq:GFManifold} through the Jacobian $J$ of the forward map 
\begin{equation}
  \dot{f} = J \dot{\theta} = - J G^{-1} J^T \nabla_{f}\loglhd\,,
  \qquad
  J\in\real^{\nvec\times\nparams}\,,
  \label{eq:GFManifoldpush-forward}
\end{equation}
where $\nabla_{\theta}\loglhd = J^T \nabla_{f}\loglhd$ and $\nabla_f \loglhd = M
f - b$. In other words, the push-forward converts a ``velocity'' vector in
parameter space to a velocity vector in PDF space.
\Cref{eq:NTKfeatures,eq:flow_equation} then follow from choosing the metric in
\cref{eq:gradient_flow} to be the Euclidean one, $G = \id$.\footnote{Note that
$G^{-1}$ is the co-metric and in index notation the gradient flow in parameter
space reads $\dot{\theta}^{\mu} = - G^{\mu\nu} \partial_{\nu} \loglhd$. The
push-forward of the flow in \cref{eq:GFManifoldpush-forward} then reads
$\dot{f}^{\alpha} = - J^{\alpha}_{\mu} G^{\mu\nu} J^{\beta}_{\nu}
\partial_{\beta} \loglhd$. As a consequence, the NTK in \cref{eq:NTKfeatures} is
the push-forward of the co-metric $G^{-1}$ and, since it is not injective in the
over-parametrised case, it is not invertible and therefore cannot be used to
define a metric in the Riemannian sense.} The role of the tangent vectors is
then made explicit if we expose them in the flow equation. By setting $G$ to be
the Euclidean metric, \cref{eq:GFManifoldpush-forward} becomes
\begin{equation}
  \dot{f} = - \sum_{\mu} J_{\mu}\, \left(J_\mu,\, \nabla_f \loglhd \right)\,,
  \label{eq:FlowTangentFeatures}
\end{equation}
from which we conclude that each vector $J_{\mu}$ contributes to the trajectory
independently and with the same form for the weight. A different choice of the
metric $G$ in parameter space would lead to a flow where each direction is
weighted accordingly. When $G$ is the FIM of the parametric model,
\cref{eq:GFManifoldpush-forward} is known as natural gradient
flow~\cite{Amari:1998jzd}. 

The velocity of the flow along any direction $n$ is bounded by the frame at a
point $f$. If we project the flow, \cref{eq:GFManifoldpush-forward}, along $n$
and use the NTK as the metric of the semi-inner product, the Schwarz inequality
yields
\begin{equation}
  n^T \dot{f} 
  = n^T \ntk\, \nabla_f \loglhd
  \leq \sqrt{n^T \ntk\, n}\, \sqrt{\nabla_f \loglhd^T \ntk\, \nabla_f \loglhd}\,,
\end{equation}
where the right-hand side factorises into two contributions. The first term is
the coverage of \cref{eq:coverage}, which measures the alignment between $n$ and
the frame of the parametrisation. The second term represents the pull of the
data as seen by the eigenvectors of the NTK, similar to
\cref{eq:training_hamiltonian}. If the latter is small, then the data constrain
directions that are not accessible to the model, and therefore the rate of
change of $f$ along $n$ is tightly bounded for any direction $n$. On the other hand,
assuming the model is flexible enough so that the second term is non-vanishing,
any direction $n$ that is covered by the frame, \cref{eq:coverage}, results in a
larger bound indicating that this direction can be learned fast. Finally, if
$n\in\ntkker$, the first term vanishes and the dynamics along this direction
stops, no matter how strong the pull of the data is. This implies that two
models that realise the same function at some point of their training could in
principle depart along different trajectories depending on their frames. A
minimal example can again help visualise this point.

\begin{example*}
  Consider the one-dimensional manifold of linear functions $f(x) = a\,x$, and
  three parametrisations of it: $a = \theta$, $a = \theta^{3}$, and $a =
  \theta_{1}\theta_{2}$. The manifold is the same in the three cases -- a
  straight line through the origin with the same tangent plane
  $\mathrm{span}\{x\}$ at every point. In the first case, there is one tangent
  vector $x$ with constant NTK, $\ntk = x\,x^{T}$. In the second
  parametrisation, the tangent vector is $3\theta^{2} x$ and $\ntk =
  9\theta^{4}\, x\,x^{T}$, which varies along the trajectory and vanishes at
  $\theta = 0$, where the flow stalls. Finally, the third parametrisation
  induces a pair of tangent vectors $\{\theta_{2}x, \theta_{1}x\}$, with $\ntk =
  (\theta_{1}^{2}+\theta_{2}^{2})\, x\,x^{T}$. So although they share the same
  functional form, the three models experience different dynamics as dictated by
  the three NTKs.
\end{example*} 

\begin{figure}[t]
  \centering
  \includegraphics[width=\linewidth]{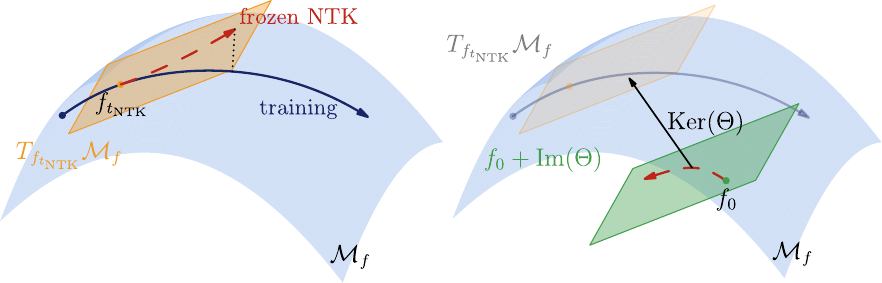}
  \caption{Schematic representation of the training dynamics in the functional
  manifold $\manfd_f$. (Left) Training is a curve on $\manfd_f$ (solid blue).
  Freezing the NTK at $\tcntk$ selects the tangent plane $\tgnt{f_{\tcntk}}$, on
  which the flow continues (dashed red). The curvature of the manifold separates
  the two trajectories (dotted). (Right) In this work the frozen flow starts
  from an independent initialisation $f_0$, and lives on the affine plane $f_0 +
  \ntkimg$ parallel to the tangent plane. Their separation along $\ntkker$ is
  conserved by the flow.}
  \label{fig:ntk_geometry_cartoon}
\end{figure}

\subsubsection*{The frozen NTK dynamics as a tangent plane}

Freezing the NTK at some epoch is thus equivalent to selecting the tangent plane
-- and its frame -- at one point along the trajectory. From this point onwards,
the analytical solution in \cref{eq:analytic_sol_compact} then continues the
training on this plane rather than on the manifold itself. If, over the region
explored by the fit, the manifold can be approximated with its tangent plane,
the approximation is faithful. Otherwise, the curvature of the manifold
introduces an error that scales with the learning rate. This is the geometric
visualisation of the core idea underlying the NTK dynamics first introduced in
Refs.~\cite{jacot2018neural,lee:2020}.

Note that in the next section we freeze the NTK at some point along the
trajectory, but then evolve the flow from an initial condition $f_0$ independent
of the training trajectory along which the NTK was frozen. The compact form of
the analytical solution, \cref{eq:analytic_sol_compact}, can help us visualise
this geometric picture. Since $\mathcal{M}(t)$ maps onto $\ntkimg$ by
construction, the displacement $f_t - f_0$ lies in $\ntkimg$ at any time, so
that the trajectory generated from $f_0$ is confined to the affine plane
\begin{equation}
  f_0 + \ntkimg\,,
  \label{eq:affine_plane}
\end{equation}
which is the plane through the initial condition $f_0$ and parallel to the
tangent plane of the frozen dynamics, $\tgnt{f_{\tcntk}}$. The kernel component
of $f_0$ is left untouched by the flow, and $\lVert \ppar \left(f_0 - f_{t_{\rm
NTK}} \right)\rVert$ measures the distance between the two tangent planes. The
second term of \cref{eq:analytic_sol_compact}, which is driven by the gradient
of the loss and by the pull of the data at the frozen point, traces the trajectory
on the affine space.

\subsubsection*{The physical directions}
We emphasise that what really drives the analytic solution of a PDF fit is the
training Hamiltonian $H^{\perp}$ of \cref{eq:training_hamiltonian}. In fact,
$H^{\perp}$ plays the same role as the NTK in \cref{eq:coverage}, but with the
Euclidean metric replaced by $M$.\footnote{To be more precise, $M$ is not a
metric as it is not invertible. Nonetheless, it can still be used to compute a
semi-inner product.} Therefore, the directions are then also weighted by how
well they resolve the data: a direction that is strongly covered by the frame
but poorly constrained by the data will be given less weight, and \textit{vice
versa}. The training Hamiltonian is thus the combination of two interplaying
components: on the one hand the spectrum of the NTK dictating what the
\emph{parametrisation} sees, on the other the matrix $M$ giving the resolvable
physical directions.

%% file: sections/results.tex
\section{The NTK analysis of a DIS-only fit}
\label{sec:results}

The methodology outlined in \cref{sec:framework} forms our baseline for the NTK
analysis of a DIS-only fit. We begin this section with a brief discussion of the
distribution of PDFs at initialisation in \cref{sec:results_initialisation},
followed by a comprehensive study of the training dynamics in terms of the NTK,
including the evolution of its spectrum and the corresponding eigenvectors, in
\cref{sec:results_ntk_training}. Finally, in \cref{sec:analytical_training} we
compare the analytical solution of the training dynamics with the empirical
training of the network.

\begin{figure}[t!]
  \centering
  \includegraphics{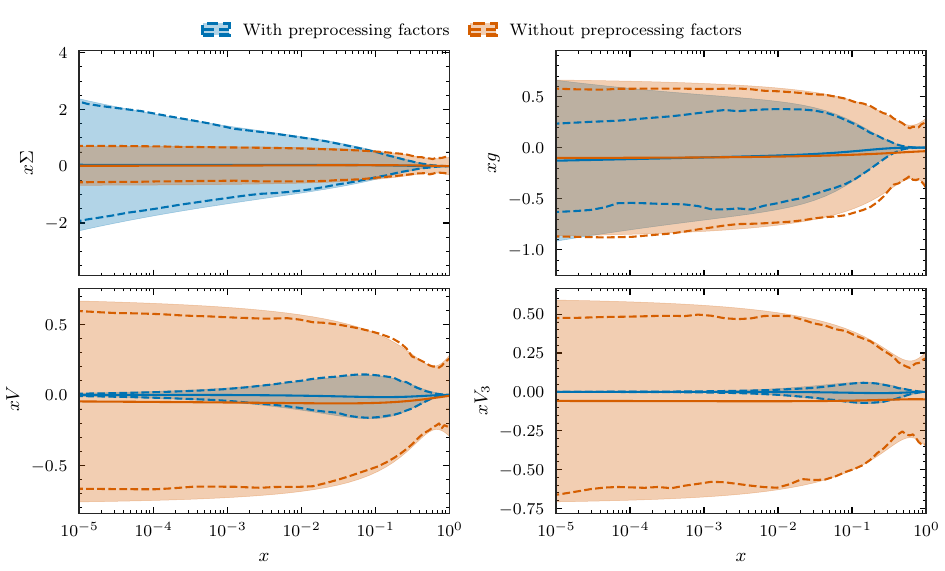}
  \caption{The output of the ensemble of models at initialisation at the initial
  scale $Q_0$, here for $\Sigma$, $g$, $V$, and $V_3$. We compare the case of
  the model with preprocessing factors (blue) and without preprocessing factors
  (orange), the latter corresponding to the contribution of the neural network
  alone. The solid lines represent the mean value computed over an ensemble of
  120 replicas, while the shaded bands represent the one-sigma uncertainty
  computed as the variance over the same ensemble. The 68$\%$ \cl\ is also shown
  as dashed lines.}
  \label{fig:dis_initialisation}
\end{figure}

\subsection{The model at initialisation}
\label{sec:results_initialisation}

The probability distribution at initialisation of the model parameters --
weights and biases of the NN and the preprocessing factors of the PDFs --
induces a corresponding distribution in the functional space of the PDFs, which
we show in \cref{fig:dis_initialisation,fig:dis_initialisation_2}. In the eight
panels, one for each fitted flavour, we show the mean (solid line), variance
(shaded bands), and 68$\%$ \cl~(dashed lines) of the output at initialisation
across all values in $x$ for an ensemble of $\nreps = 120$ replicas. We show
both the cases with and without the preprocessing factors in order to
disentangle the contribution of the NN alone from the contribution of the large-
and small-$x$ preprocessing factors. A few comments are in order.

First, we note that when the preprocessing factors are removed altogether, the
distributions at initialisation appear very similar irrespective of the flavour.
This is indeed expected, as the NN alone treats all flavours on the same footing
at initialisation. Furthermore, the inclusion of $\log x$ in the input layer
prevents the network from collapsing to zero for small values of $x$, resulting
in a homogeneous behaviour in this region, as already observed in
Ref.~\cite{Chiefa:2025cap}. On the other hand, the preprocessing factors act as
a flavour-dependent deformation of the shapes of the PDFs and, with the
exception of $g$ and $\Sigma$, the inclusion of the preprocessing factors tends
to narrow the spread of the distributions, with a significant reduction of the
variance towards small-$x$. 

Note that in a Bayesian approach to the solution of the inverse problem, the
prior distribution of the parameters at initialisation induces a corresponding
prior distribution in the functional space of the PDFs. The prior distribution
at initialisation is therefore a crucial ingredient in the Bayesian inference,
as it encodes our prior knowledge and assumptions about the PDFs before any data
are taken into account. In particular, the choice of the prior distribution can
have a significant impact on the posterior distribution of the PDFs, especially
in regions where data are scarce or absent. The analytical solution discussed in
\cref{sec:analytical_training} allows us to study the impact of the prior on the
final result. 

\begin{figure}[t!]
  \centering
  \includegraphics{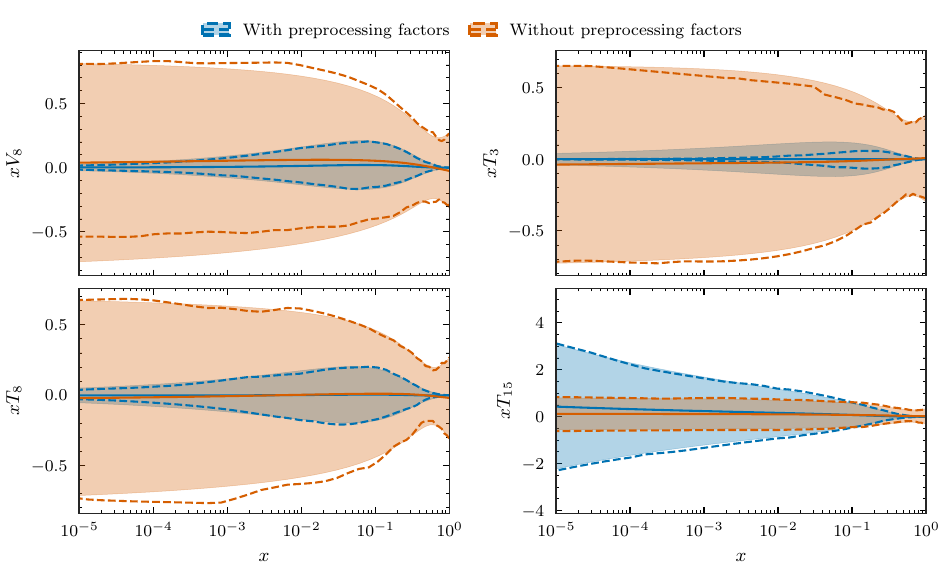}
  \caption{Continuation of \cref{fig:dis_initialisation} with $V_8$, $T_3$,
  $T_8$, and $T_{15}$.}
  \label{fig:dis_initialisation_2}
\end{figure}
\begin{figure}[t!]
  \centering
  \includegraphics[width=0.7\linewidth]{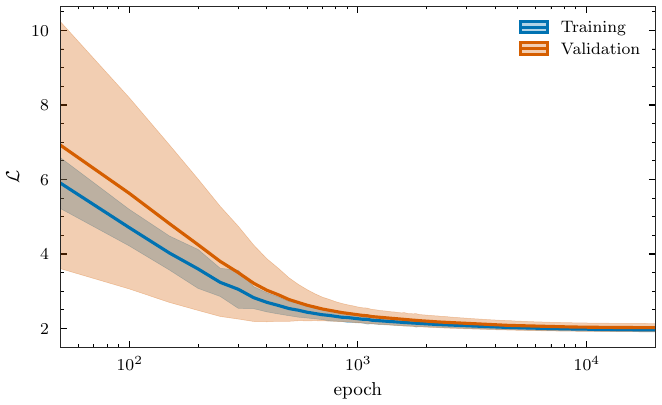}
  \caption{The evolution of the loss function during training for both the
  training (blue) and validation (orange) sets of data. The loss is the first
  term of \cref{eq:loss}, normalised to the number of data points of the
  corresponding set. The lines represent the mean value computed over an
  ensemble of 120 replicas, while the shaded bands represent the one-sigma
  uncertainty computed as the variance over the same ensemble. Note the
  logarithmic scale in the horizontal axis.}
  \label{fig:loss-functions}
\end{figure}

\begin{figure}[p]
  \centering
  \includegraphics[width=\linewidth]{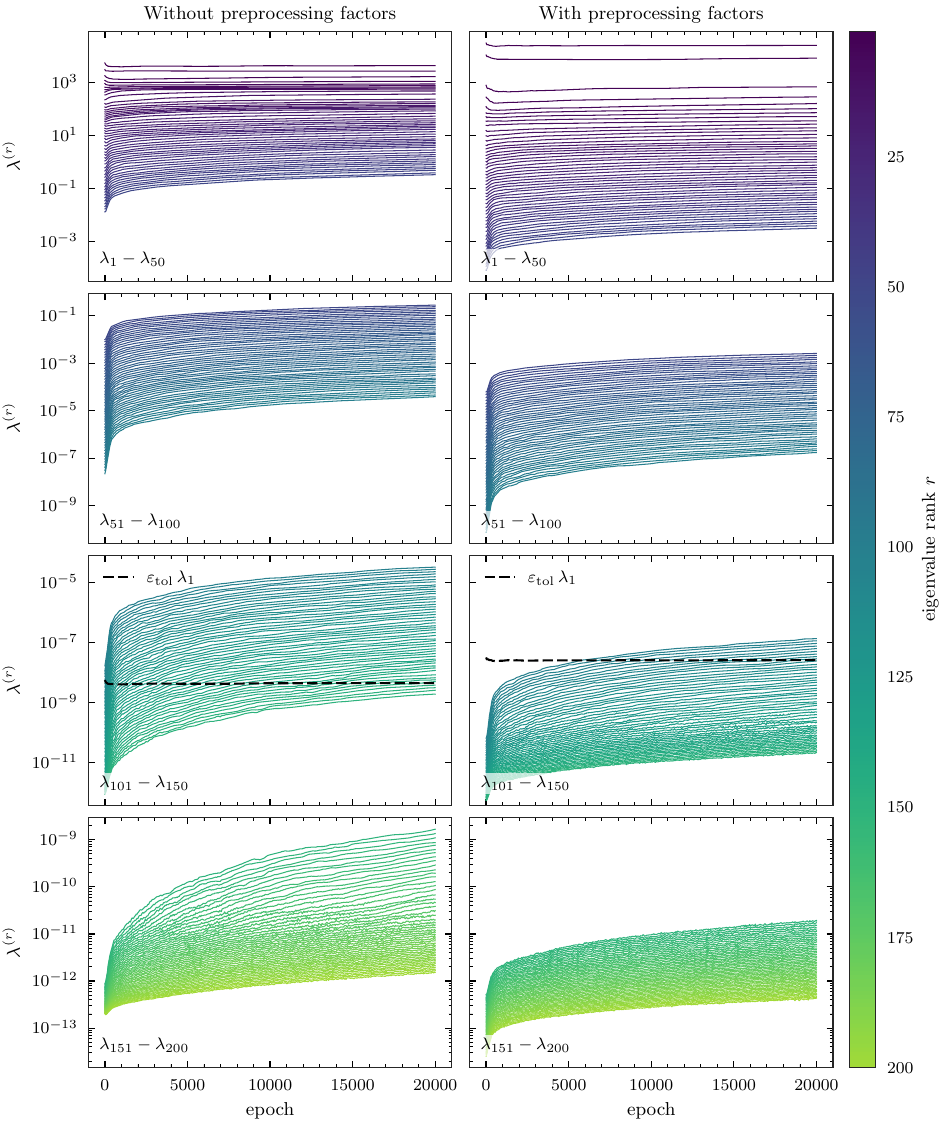}
  \caption{Eigenvalues of the NTK during training for both cases without (left
   column) and with (right column) preprocessing factors. The figure displays
   the first 200 ranks of the NTK spectrum, distributed evenly across the four
   rows. The colour of each curve encodes the rank $i$ of the eigenvalues, as
   indicated by the colour bar. The lines represent the median value computed
   over an ensemble of 120 replicas. The dashed black line marks the separation
   between the image and the kernel of the NTK, \ie\ the numerical-zero relative
   threshold $\varepsilon_{\rm tol}\,\lambda_1(t)$, with $\varepsilon_{\rm tol}
   = 10^{-12}$. The dimension of \cref{fig:DimNTKSpace} is precisely the number
   of eigenvalues lying above that line at each training time.}
  \label{fig:EigvalsNTK}
\end{figure}

\subsection{Diagnostic of the NTK during training}
\label{sec:results_ntk_training}

Our analysis of the training dynamics starts by inspecting the training (blue)
and validation (orange) loss curves along the training trajectory, as shown in
\cref{fig:loss-functions}. Here we use the methodology outlined in
\cref{sec:framework} to train the NNs, and we recall that momentum and valence
sum rules, as well as positivity and integrability constraints, are \emph{not}
imposed in any of the fits here considered. 

The behaviour of the two loss functions is significantly different from the one
observed in the previous study~\cite{Chiefa:2025cap}, where the loss function
was characterised by a steady plateau followed by a sudden drop. In contrast,
here the loss functions decay smoothly, with a rapid decrease in the first few
epochs followed by a slower variation until reaching a plateau. In the
following, we show how this behaviour is reflected in the internal structure of
the NTK. We then devote \cref{sec:init_dependence} to exploring how the training
dynamics is affected by the initialisation prescription, recovering the
behaviour originally observed in Ref.~\cite{Chiefa:2025cap}.

\begin{figure}[t!]
  \centering
  \includegraphics[width=\linewidth]{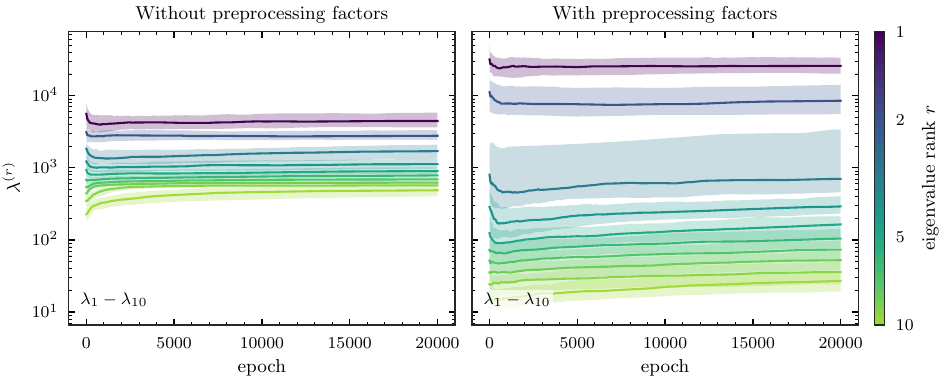}
  \caption{Close-up view of the leading ten eigenvalues of the NTK during training for
   both cases without (left column) and with (right column) preprocessing
   factors. As in \cref{fig:EigvalsNTK}, the colour of each curve encodes the
   rank $i$ of the eigenvalues, as indicated by the colour bar. Here the solid
   lines represent the median value while the shaded bands represent the 68\%
   \cl, both computed over an ensemble of 120 replicas.}
  \label{fig:EigvalsNTKzoom}
\end{figure}

\subsubsection*{Spectrum of the NTK}
\begin{figure}[h!]
  \centering
  \includegraphics[width=0.7\linewidth]{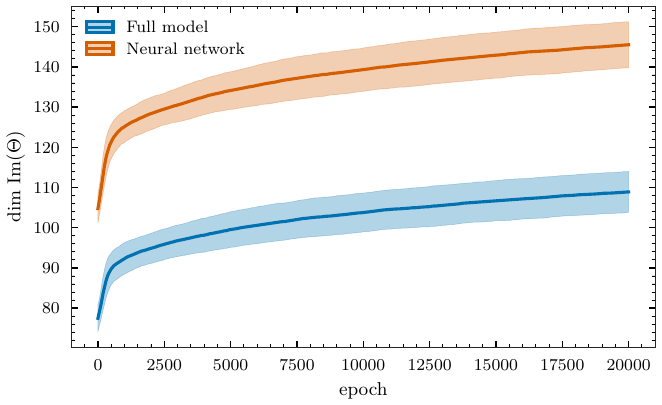}
  \caption{Dimension of the image of the NTK during training as a function of
  the training time. We show both the case of the full model with preprocessing
  factors (blue) and the case of the NN alone (orange). The solid line
  represents the mean value computed over an ensemble of 120 replicas, while the
  shaded bands represent the one-sigma uncertainty computed as the variance over
  the same ensemble. At each training time, this dimension is the number of
  eigenvalues lying above the image/kernel threshold $\varepsilon_{\rm
  tol}\,\lambda_1$ drawn in \cref{fig:EigvalsNTK}.}
  \label{fig:DimNTKSpace}
\end{figure}

In \cref{fig:EigvalsNTK} we show the first 200 eigenvalues of the NTK as
functions of the training time, distributed evenly across the four rows. The
left and right columns display the eigenvalues of the NTK computed without and
with the preprocessing factors, respectively.\footnote{We remark that the two
cases do not correspond to different fits. The NTK without preprocessing factors
is obtained from the same trained model by setting the preprocessing factors to
one in \cref{eq:ntk_def}. The two columns therefore differ only by the factors
$x^{1-\delta_j}(1-x)^{\epsilon_j}$ that multiply the NTK of the NN.} The curves
are medians over the 120 fitted replicas, and the rank is encoded by the colour
as indicated by the colour bar. Error bars are not shown for the sake of
clarity, but we support our discussion by means of
\cref{fig:EigvalsNTKzoom,fig:DimNTKSpace}, the first showing a close-up view of
the leading ten eigenvalues where the 68\% \cl~is also shown, the second
tracking the dimension of the NTK, with and without preprocessing factors, over
the entire training process.\footnote{The dimension of the image of the NTK is
defined as the number of eigenvalues that are above the tolerance
$\varepsilon_{\rm tol}= 10^{-12}$ relative to the largest eigenvalue.}

From inspecting these figures, we observe that the main results of
Ref.~\cite{Chiefa:2025cap} remain qualitatively true in a DIS-only fit: the
spectrum of the NTK is characterised by a hierarchical structure that spans
several orders of magnitude and, as training progresses, the NN discovers new
modes represented by eigenvalues that cross the kernel threshold and enter the
image of the NTK. Unlike the single-flavour toy model, however, such behaviour
is localised at the beginning of the training, which is reminiscent of the rapid
decrease of the loss function observed in \cref{fig:loss-functions}. After this
discovery phase, which we refer to as the \emph{rich
regime}~\cite{Chiefa:2025cap,pmlr-v125-woodworth20a}, the spectrum of the NTK
stabilises with the leading eigenvalues remaining approximately constant and the
lower ones increasing slowly. We identify this latter phase as the \emph{lazy
regime}, where we assume the NTK to be approximately constant and the training
dynamics to be well described by the analytical solution, as discussed in
\cref{sec:analytical_training}.

The comparison between the two columns of \cref{fig:EigvalsNTK} reveals that the
inclusion of the preprocessing factors does not affect the qualitative behaviour
of the spectrum, although it contributes to the magnitude of the eigenvalues. In
fact, a closer inspection of \cref{fig:EigvalsNTKzoom} reveals that when
preprocessing factors are excluded, the first ten eigenvalues are all confined
within one order of magnitude and the gap between consecutive curves is reduced.
On the other hand, upon inclusion of the preprocessing factors, the eigenvalues
span about three orders of magnitude, with increasing spacing between
consecutive curves. Note also that the inclusion of the preprocessing factors
increases the spread over replicas too, as shown by the 68\% \cl~bands.

The same pattern is also reflected in the dimension of $\ntkimg$ shown in
\cref{fig:DimNTKSpace}. Here we see that the number of modes that populate
$\ntkimg$ is systematically larger for the NN alone than for the full model. We
therefore conclude that the dynamics of the full model is controlled by fewer
modes than those actually available to the bare network. This can be understood
by noting that the modes with support in the regions where the preprocessing
factors are small have their eigenvalues suppressed, thus falling below the
threshold. In these regions, the behaviour of the PDFs remains the one imposed
by the preprocessing factors, resulting in a smaller $\ntkimg$.

\subsubsection*{Modes of the training Hamiltonian}
\begin{figure}[p]
  \centering
  \begin{subfigure}{\linewidth}
    \centering
    \includegraphics[width=0.83\linewidth]{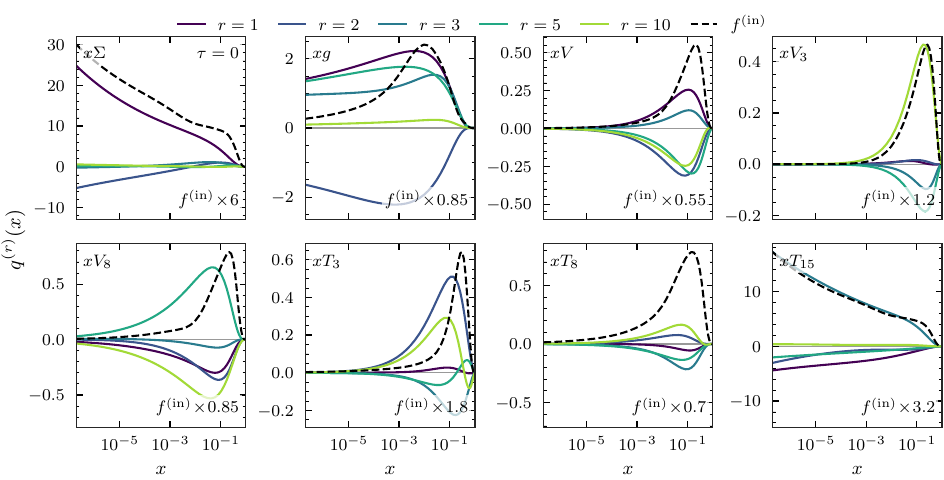}
  \end{subfigure}
  \begin{subfigure}{\linewidth}
    \centering
    \includegraphics[width=0.83\linewidth]{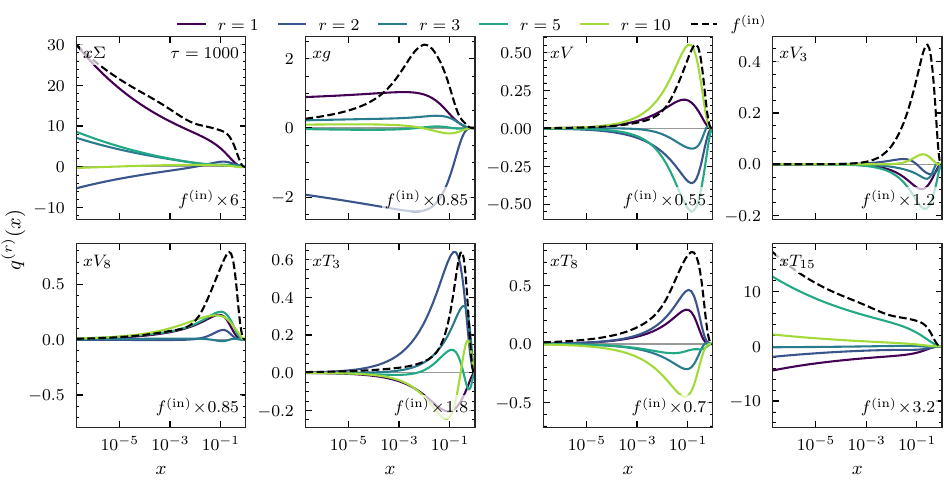}
  \end{subfigure}
  \begin{subfigure}{\linewidth}
    \centering
    \includegraphics[width=0.83\linewidth]{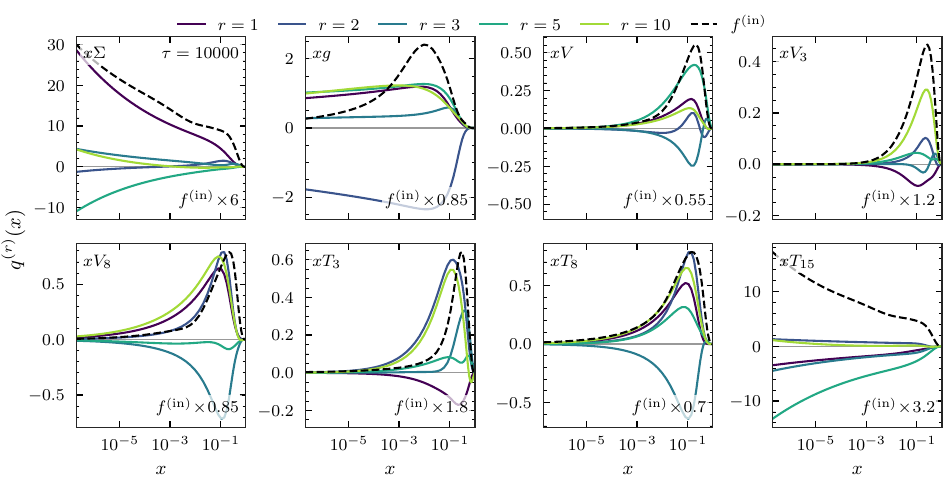}
  \end{subfigure}
  \caption{The modes $q^{(r)}$ of the training Hamiltonian,
  \cref{eq:hamiltonian_eigvec_pdf}, as functions of $x$ and shown at three
  training epochs: $\tau=0$ (top block), $\tau=1000$ (middle block), and
  $\tau=10000$ (bottom block). The eight panels of each block show, for each
  fitted flavour, the ranks $r=1,2,3,5,10$ coloured as indicated in the legend.
  The modes are those of a single replica. The vertical scale differs between
  flavours but is the same across the three epochs. The overall sign of each
  mode is chosen so that its coefficient in the expansion of $\fin$ on the modes
  is positive. The dashed black line is the central replica of the underlying
  PDF $\fin$ used to generate the data, rescaled for each flavour by the factor
  indicated, common to the three epochs.}
  \label{fig:QmodesEpch1}
\end{figure}

We now turn to discussing the modes $q^{(i)}$ defined in
\cref{eq:hamiltonian_eigvec_pdf}, whose role in this context is twofold. First,
they expose the functional forms that the parametrisation can represent when
solving the inverse problem. Second, they shape the analytical solution through
the mode expansion of \cref{eq:analytic_sol_compact}. We show their components
for each of the eight fitted flavours in \cref{fig:QmodesEpch1} for a
representative subset of ranks, here chosen to be $r=1,2,3,5,10$. In order to
investigate their evolution along the training trajectory, we evaluate the modes
at $\tau=0$, 1000, and 10000, shown in the three blocks of panels.

Starting from the upper block, we see that the modes exhibit a non-trivial
structure already at initialisation, before any optimisation step is taken. An
inspection by rank reveals some associations between ranks and flavours. For
instance, the first mode makes up most of $\Sigma$ and contributes to $g$, while
being smaller for all other flavours. The second mode contributes mostly to
$T_3$ and $g$ (with a negative sign), while the tenth almost entirely recovers the
shape of $V_3$. Still, these examples do not point to any systematic pattern
between ranks and flavours. It is worth noting that even at initialisation the
modes capture much of the shape of $\fin$, and in \cref{sec:analytical_training}
we explore how this is reflected in the analytical solution.

We do not observe significant patterns emerging when looking across subsequent
epochs, but rather a reorganisation of the flavours covered by each rank.
However, we note that the first mode remains tied to $\Sigma$ across the other
two blocks, while the contribution of the other ranks to each flavour changes
with time without any visible correlation. In general, the modes display
functional forms similar to the underlying law at later stages.

In conclusion, the inspection of these modes is a useful tool to gain insight
into the kind of primitive functions the parametrisation relies on. We remark
that these modes depend on the parametrisation and, as discussed extensively in
\cref{sec:init_dependence}, they change drastically if a different
initialisation is adopted.

\subsection{Analytical training}
\label{sec:analytical_training}
\begin{figure}[t!]
  \centering
  \includegraphics[width=0.7\linewidth]{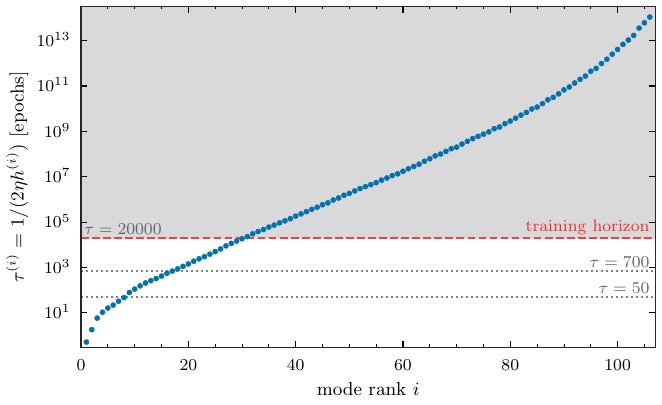}
  \caption{Characteristic resolution times $\tau^{(i)} = 1/(2\eta h^{(i)})$, in
  epochs, of every mode $i$ in $\ntkimg$ at the frozen epoch $15000$ (ordered by
  decreasing $h^{(i)}$). The threshold separating the image from the kernel is
  obtained with $\varepsilon_{\rm tol} = 10^{-12}$, as in
  \cref{fig:DimNTKSpace}, giving $\dperp\simeq106$ median modes at this epoch. The red
  dashed line marks the training horizon ($\tau_{\rm max}=20000$ epochs). The
  dotted lines are the three evolution times $\tau$ of \cref{fig:AnltSolInit}.
  Points are medians over the ensemble of replicas.}
  \label{fig:ResolutionTimes}
\end{figure}

The analytical solution reviewed in \cref{sec:gradient_flow} can be seen as a
linear combination of the Hamiltonian modes $q^{(i)}$, each weighted by a
time-dependent factor that decays exponentially with the training time. This
decay sets the resolution timescale at which each mode is incrementally included
in the solution, and it is defined as
\begin{equation}
  \tau^{(i)} = \frac{1}{2\eta h^{(i)}}\,,
  \label{eq:tau}
\end{equation}
where $h^{(i)}$ is the corresponding eigenvalue of the training Hamiltonian
$H^{\perp}$ and $\eta$ is the learning rate.\footnote{The factor of $2$ in the
denominator arises from the missing factor of $1/2$ in the definition of the
loss function.}  Note that the characteristic timescale $\tau^{(i)}$ is
expressed in units of epochs, and the analogous continuous version is obtained
by multiplying it by the learning rate $\eta$. As training proceeds, the modes
associated with the largest eigenvalues are learned first, while those
associated with smaller eigenvalues are learned later.

Freezing the NTK at a specific epoch $\tntk$ is thus equivalent to selecting a
set of modes with their corresponding \emph{learning timescales}. In the
remainder of this paper we work with the reference epoch $\tntk=15000$, but
comment on other possible choices later. The corresponding characteristic
resolution times are shown in \cref{fig:ResolutionTimes} as medians over the 120
replicas (leaving about 106 modes), ordered by decreasing eigenvalue $h^{(i)}$.
The red dashed line marks the number of modes learned by the analytical solution
when evolved up to $\tau=20000$ (the largest epoch of the numerical training),
which we denote as the \emph{training horizon}. Everything above this line (grey
shaded area) is excluded from the expansion.

The hierarchical structure is not surprising, as it is a direct consequence of
the spectrum of the NTK shown in \cref{fig:EigvalsNTK}. However, we observe that
the points cover a significant range of timescales, spanning about 14 orders of
magnitude. For instance, if the analytical solution is evolved up to the
training horizon, we retain less than a third of the available modes. This is
another point of contrast with the previous study~\cite{Chiefa:2025cap}, where
within the training horizon the analytical solution encompassed all the
available modes. Such a difference is however expected, as it reflects the
increased complexity of the current setting. We now see how these modes shape
the PDFs in the analytical solution.

\subsubsection*{Convergence of the analytical solution}

\begin{figure}[tp]
  \centering
  \includegraphics[width=\linewidth]{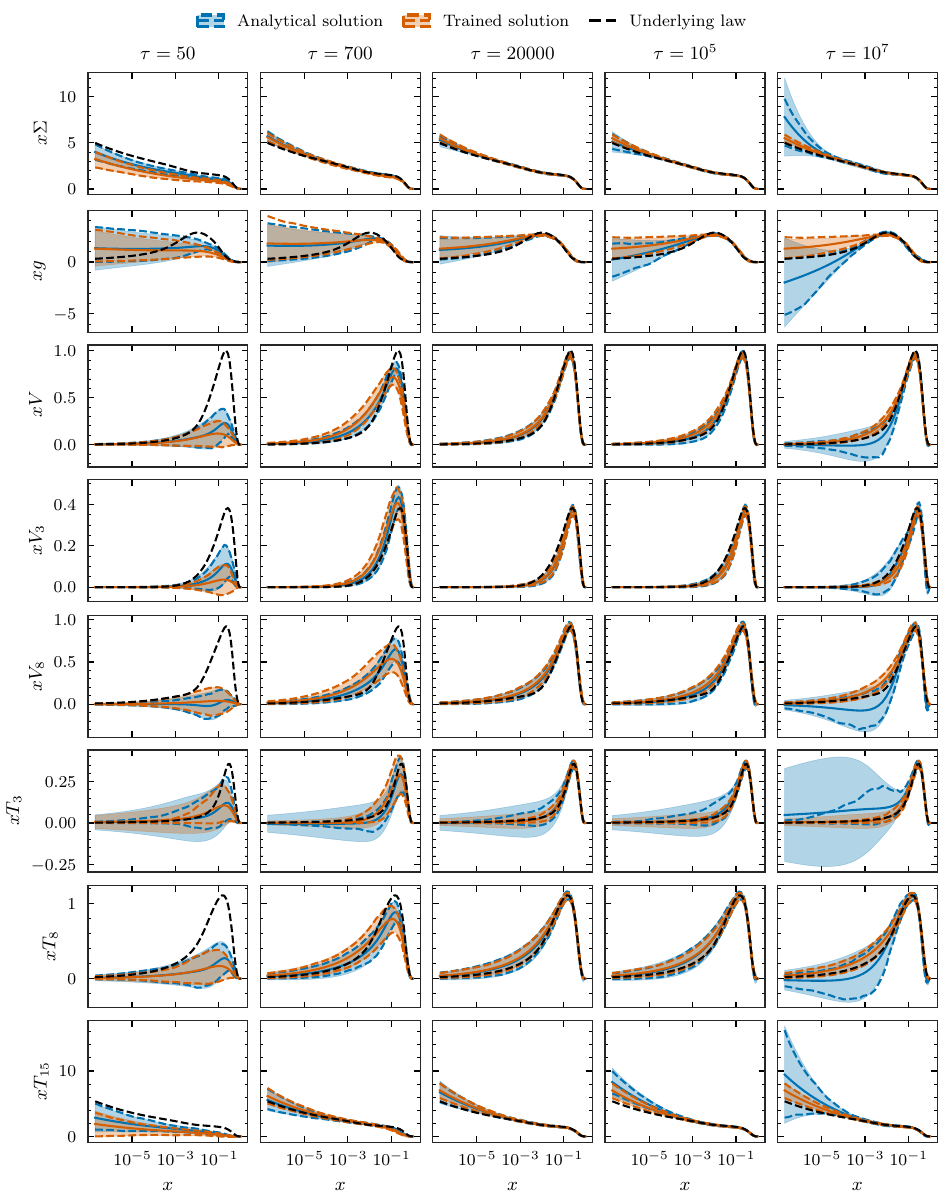}
  \caption{The analytical solution (blue) of the eight fitted flavours (rows)
  computed with the NTK frozen at $\tntk=15000$ and for five evolution times
  (columns). The initial condition $f_0$ of the analytical solution is an
  independent ensemble of PDFs at initialisation for all panels. In the first
  three columns the analytical solution is compared against the trained network
  (orange) evaluated at the same training time, while in the last two columns
  the trained network is kept fixed at the last available epoch, $\tau_{\rm
  max}=20000$. Curves show the mean, one-sigma band and $68\%$ \cl\ over the
  ensemble of replicas. The underlying law $\fin$ is also drawn as a dashed
  black line.}
  \label{fig:AnltSolInit}
\end{figure}

In order to quantify the amount of information encoded in the NTK at $\tntk$, we
apply the analytical solution to an ensemble of models at initialisation. The
initial condition is thus trained -- or \emph{evolved} -- according to
\cref{eq:analytic_sol} and compared against the empirically trained network.
With the help of \cref{fig:ResolutionTimes}, we select three representative
evolution times which are $\tau=50$, 700, and 20000. These are indicated by the
dotted grey lines in \cref{fig:ResolutionTimes}, and on average they amount to
8, 17, and 30 resolved modes, respectively. The resulting PDFs are collected in
\cref{fig:AnltSolInit}, where the analytical solution is computed for each of
the eight fitted flavours (rows) at the three representative evolution times
(columns). In addition, we show two additional columns where the analytical
evolution is computed at $\tau=10^5$ and $\tau=10^7$, respectively -- both
beyond the training horizon such that the remaining modes are resolved. The
analytical solution is compared to the empirically trained network at the same
evolution times, except for the last two columns where the trained solution is
held at its last available epoch $\tau_{\rm max} = 20000$. 

The agreement between the analytical and trained solutions is generally good
across all flavours and for the first three evolution times. Moreover, the two
solutions converge onto the underlying law at the same pace and, at
$\tau=20000$, both reconstruct the underlying law within errors. The analytical
solution is thus a good approximation of the discrete gradient steps used to
optimise the network, up to some exceptions that we now discuss. Most
importantly, we also conclude that only the first 30 modes or so are actually
needed to obtain a good reconstruction of the underlying law.

These results are consistent with the slowly varying trajectories of the leading
NTK eigenvalues (\cref{fig:EigvalsNTK}) and the monotonic, rapid decay of the
loss function (\cref{fig:loss-functions}), the latter well approximated by the
exponential behaviour of the analytical solution, \cref{eq:analytic_sol}. This
is significantly different from our previous work~\cite{Chiefa:2025cap}, where
the analytical solution converged to the underlying law much faster than the
trained network. This, however, was due to the inconsistency between the smooth
behaviour of the loss function induced by the analytical solution and the abrupt
learning phase that characterised the trained solution. In
\cref{sec:init_dependence} we explore what happens to the validity of the
analytical solution when this sudden learning phase is induced in the training
dynamics.

The only notable exception across the eight flavours is the small-$x$ region of
$T_3$. Although the analytical solution is compatible with the underlying law
within errors, the uncertainty bands at small-$x$ deviate significantly from
those of the trained network, despite the central value being almost identical.

We show in the last two columns of \cref{fig:AnltSolInit} that this deviation
does not improve upon inclusion of additional modes in the analytical solution.
At $\tau=10^5$, the uncertainties are slightly reduced, at the cost of a shift
in the central value of the analytical solution around $x\sim10^{-2}$. When the
solution is pushed up to $\tau=10^7$, the PDFs drastically worsen across all
flavours both in terms of central value and uncertainties. This is consistent
with \cref{sec:early_stopping}, and demonstrates that the modes learned at later
stages during training  -- those further down in the spectrum -- are poorly
constrained by the data. Their inclusion thus introduces noise in both central
values and uncertainties with a noise that depends on the matrix $M$.

\subsubsection*{Impact of the choice of $\tntk$}
\begin{figure}[t!]
  \centering
  \includegraphics[width=0.85\linewidth]{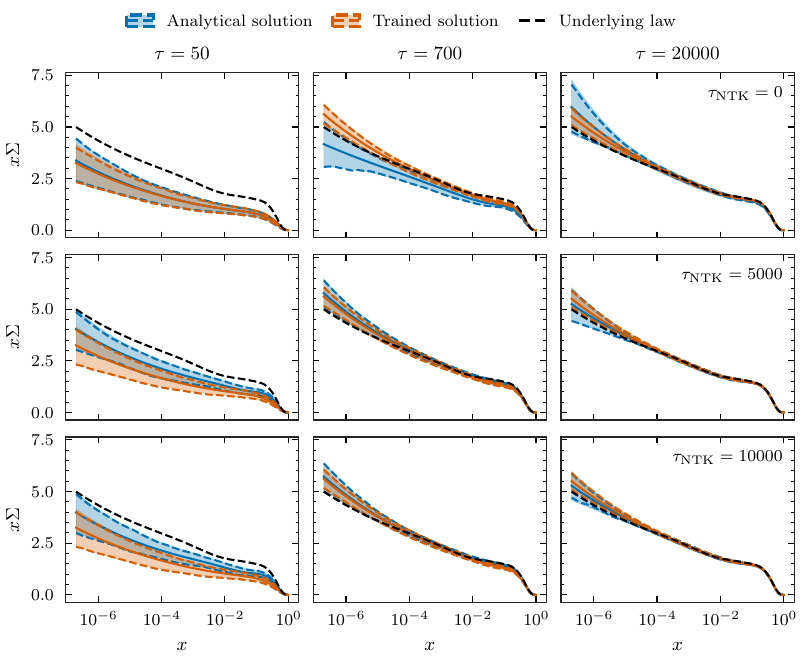}
  \caption{Analytical (blue) versus empirically trained (orange) evolution of
  the singlet $\Sigma$, for three choices of the epoch at which the NTK is
  frozen ($\tntk=0,5000,10000$, rows) and the same three evolution times $\tau$
  as in \cref{fig:AnltSolInit} (columns). Statistics and the underlying law
  (dashed black) are as in \cref{fig:AnltSolInit}.}
  \label{fig:AnltSolSigma}
\end{figure}
One might wonder how the choice of $\tntk$ affects the analytical solution, and
in particular if it retains its validity at initialisation. We address these
questions in \cref{fig:AnltSolSigma}, where we show the analytical solution of
the singlet $\Sigma$ for three different choices of $\tntk$, namely $\tntk=0$,
5000, and 10000, laid out in rows. The solution is compared against the trained
network at the first three evolution epochs $\tau$ of \cref{fig:AnltSolInit},
represented by the columns. The underlying solution is also shown.

Surprisingly, the analytical solution performs well even when the NTK is frozen
at initialisation, as it is capable of tracking the trained network and
reconstructing the underlying law. This was already anticipated when discussing
the modes $q^{(i)}$ in \cref{fig:QmodesEpch1}, where we observed that these have
non-trivial shapes even at initialisation. We remark that this result is
different from the behaviour observed in our previous
study~\cite{Chiefa:2025cap}, and more similar to the notion of lazy training of
Refs.~\cite{lee:2020,jacot2018neural}. The mild deviations that are observed are
mostly ascribed to the lack of sub-leading eigenvalues that are not yet
discovered at initialisation (see \cref{fig:EigvalsNTK}). In fact, the agreement
with the trained solution improves as $\tntk$ increases, namely when more modes
are included in the analytical solution (\cref{fig:ResolutionTimes}). Later in
\cref{sec:init_dependence} we show to what extent this behaviour is tied to the
choice of initialisation.

\subsubsection*{The initial condition and the data}

\begin{figure}[p]
  \centering
  \includegraphics[width=\linewidth]{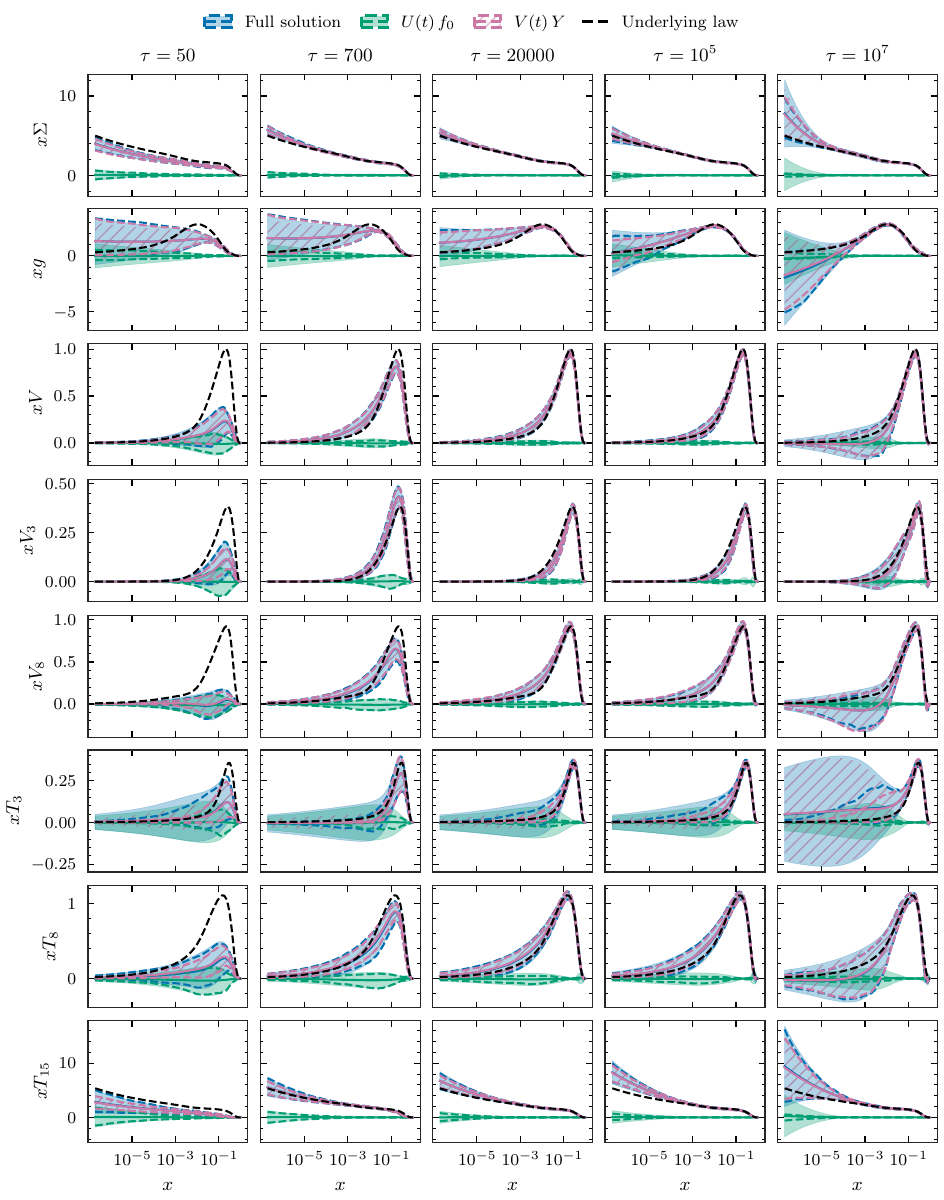}
  \caption{Decomposition of the analytical solution of \cref{eq:analytic_sol}
  into the contribution from the initial condition, $U(t)f_0$ (green), and the
  contribution from the data, $V(t)\obs$ (purple, hatched), together with their
  sum (blue), for the eight fitted flavours (rows) and the five evolution times
  $\tau$ of \cref{fig:AnltSolInit} (columns). The NTK is frozen at epoch
  $15000$. The initial condition and the statistics over the ensemble of
  replicas are as in \cref{fig:AnltSolInit}. The dashed black line is the
  underlying law $\fin$.}
  \label{fig:UVContributions}
\end{figure}

\begin{figure}[tp]
  \centering
  \includegraphics[width=\linewidth]{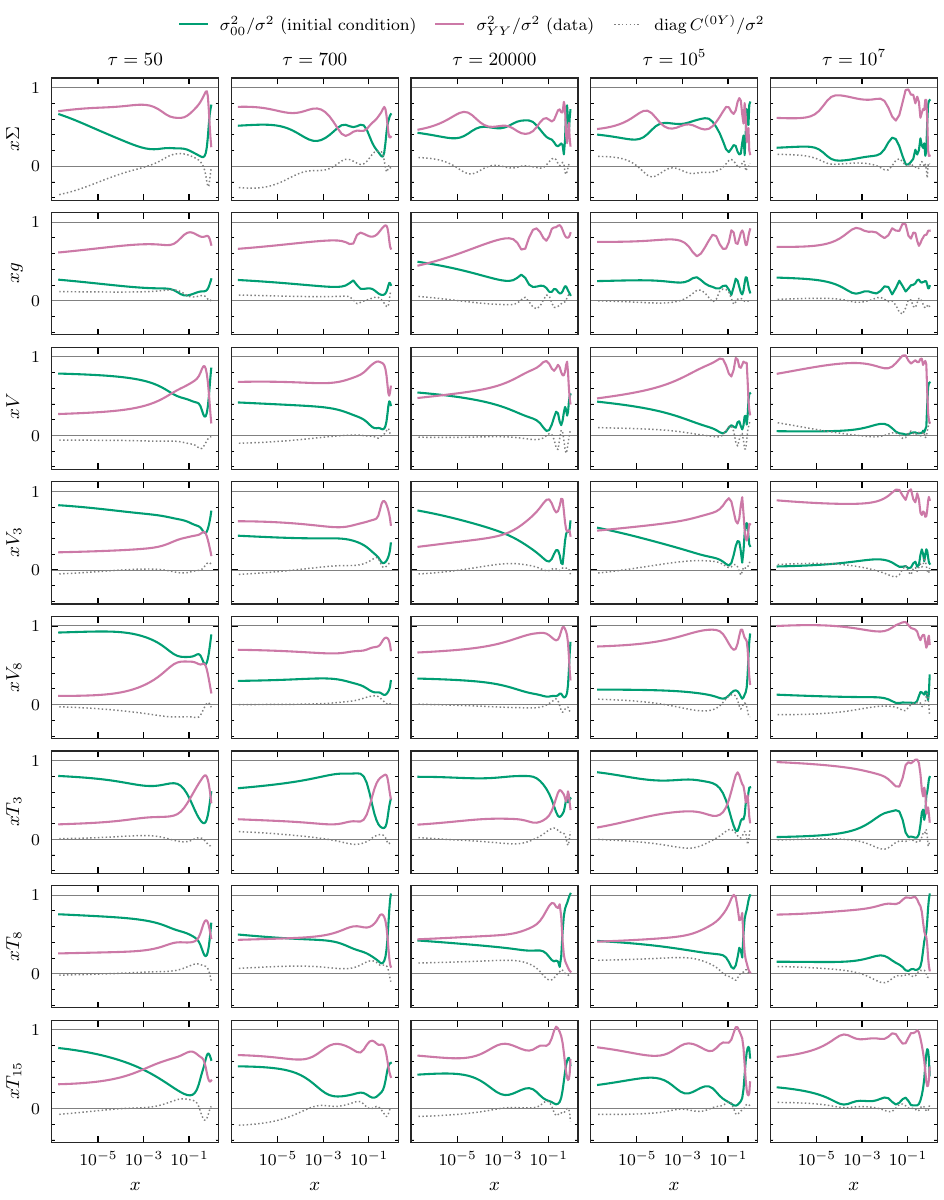}
  \caption{Fractions of the variance of the analytical solution carried by the
  initial condition (green), by the data (purple), and by their correlation
  (dotted grey). Each curve is the ratio between the diagonal entries of the
  corresponding terms in \cref{eq:error_budget} and the total covariance, so
  that all fractions sum to one at each point in $x$. The layout and settings
  are the same as in \cref{fig:AnltSolInit}. The two horizontal grey lines mark
  zero and one of the y-scale. The point at $x=1$, where PDFs vanish by
  construction, is omitted.}
  \label{fig:ErrorBudget}
\end{figure}

The analytical solution in \cref{eq:analytic_sol} splits into two contributions:
the first one accounting for the initial condition, $U(t)f_0$, and the second
one accounting for the pull of the data, $V(t)\obs$. For the sake of the
discussion, we denote these two terms by $f_U$ and $f_V$, respectively. We also
remark that the two terms are stochastic quantities, in that the evolution
operators are built from the NTK of each replica, $f_0$ is drawn from the
distribution at initialisation, and $\obs$ fluctuates with the pseudodata. The
covariance of the solution then splits into three terms~\cite{Chiefa:2025cap},
\begin{equation}
  \covar\left[f_t, f_t^T\right]
  = \underbrace{\covar\left[f_U, f_U^T\right]}_{\cov^{(00)}_t}
  + \underbrace{\covar\left[f_V, f_V^T\right]}_{\cov^{(\obs\obs)}_t}
  + \underbrace{\covar\left[f_U, f_V^T\right]
    + \covar\left[f_V, f_U^T\right]}_{\cov^{(0\obs)}_t}\,,
  \label{eq:error_budget}
\end{equation}
namely the uncertainty inherited from the initial condition, the one propagated
from the data, and their correlation. In \cref{fig:UVContributions} we show
$f_U$ and $f_V$ separately, together with their sum and the underlying law
$\fin$, while in \cref{fig:ErrorBudget} we show the diagonal components of the
three terms of \cref{eq:error_budget} as fractions of the total variance. The
layout, the reference $\tntk$, and the evolution epochs $\tau$ are those of
\cref{fig:AnltSolInit}.

A first inspection of \cref{fig:UVContributions} reveals that $f_V$ is the
dominant contribution to the central value across most of the range in $x$ and
for all flavours, moving the solution closer to the underlying law as the
evolution epoch increases. The term $f_U$ is instead centred around zero for all
flavours and at all epochs, and thus it mainly contributes to the total error
budget rather than to the central value. Notably, the $f_V$ term for $\Sigma$ is
able to capture most of the functional form of the underlying law already at
$\tau=50$.

The relative weight of the uncertainties of the two contributions -- and their
correlation -- is better appreciated in \cref{fig:ErrorBudget}, and it varies
across panels. At $\tau=50$, most of the variance in the extrapolation region at
small- and large-$x$ of $V$, $V_3$, $V_8$, $T_3$, and $T_8$ is carried by $f_U$,
the latter decreasing towards the region covered by data, where $f_V$ becomes
comparable or even larger. For $\Sigma$ and $T_{15}$ the contribution of $f_U$
is mostly confined to $x\lesssim10^{-4}$, where it is dominant, and at
$x\gtrsim0.5$, where it displays a peak. Note that in $\Sigma$ the two
contributions are significantly anti-correlated at $x\lesssim10^{-5}$, as shown
by the dotted grey line. For $g$, the contribution of $f_U$ is systematically
sub-dominant over the whole range in $x$, $f_V$ contributing as much as 80\% of
the total error budget.

As training proceeds, the analytical solution suppresses the contribution of
$f_U$ in the region covered by the data, and by $\tau=20000$ it accounts for
about 10\% of the total variance of most of the flavours around $x\sim10^{-1}$.
In the two extrapolation regions the two terms remain roughly comparable, with a
balance that depends on the flavour. The two notable exceptions to this rule are
$\Sigma$ and $T_3$. In the former, the two terms contribute in equal parts at
small-$x$, although in this region both uncertainties are very small, as shown
in \cref{fig:UVContributions}. In the latter, the contribution of $f_U$ remains
the leading source of uncertainty until $x\lesssim10^{-1}$, accounting for up to
80\% of the total error budget, consistent with the corresponding panel of
\cref{fig:UVContributions}.

We already observed in \cref{fig:AnltSolInit} that pushing the evolution beyond
the training horizon does not help reduce the uncertainties of $T_3$ in the
extrapolation region, and it actually worsens the other flavours as well. From
\cref{fig:UVContributions,fig:ErrorBudget} we now see that much of the worsening
previously noted is not due to an amplification of the initial condition, but
rather to an inflation of the term driven by the data. Indeed, we see from the
last column of \cref{fig:ErrorBudget} that while the relative size of $f_U$
drops for all flavours down to 10\%, the uncertainties of $f_V$ dominate the
entire small-$x$ region. As already noted for \cref{fig:AnltSolInit}, this
agrees with the interpretation given in \cref{sec:early_stopping}, confirming
that sub-leading modes can lead to undesirable effects.\footnote{At present, we
cannot explain precisely why the inflation of $f_V$ is more pronounced in $T_3$
when more modes are included in the solution. We leave this question for future
investigation.}

Taken together, \cref{fig:UVContributions,fig:ErrorBudget} show that the initial
condition, and therefore the dependence on the initialisation, does not yield a
significant bias with the initialisation shown in
\cref{fig:dis_initialisation,fig:dis_initialisation_2}, but it does contribute
to the total uncertainty with a relative weight that depends on the flavour and
on the training time. We stress that the validity of this conclusion is not
confined to the analytical solution, but extends to the trained network to the
extent that the former is a good approximation of the training dynamics.
Furthermore, small corrections due to the discrete updates and to the deviations
from \cref{eq:frozen_ntk} do not change the qualitative picture of this result.
How the latter changes with the initialisation is discussed in the next section.

%% file: sections/init_dependence.tex
\section{Dependence on the initialisation}
\label{sec:init_dependence}

The initialisation of the parameters of the NN defines the initial condition
from which every replica starts its training course. Inspired by results found
with Deep Linear Networks
(DLNs)~\cite{jacot2022sts,FUKUMIZU2000317,zhang2026sts} and more complicated
architectures such as those used in Large Language
Models~\cite{hang2026LLMSmallInit}, in this section we explore the dependence of
the training dynamics upon variations of the initialisation distribution of the
NN parameters.

To do so, we introduce to the setup of \cref{sec:framework} a power-law scaling
of the variance of the Glorot normal distribution of the NN weights, defined as
\begin{equation}
  \sigma^2_{\ell} = \left(
    \frac{2}{n_{\ell} + n_{\ell-1}}
  \right)^{\gamma}\,,
  \label{eq:init_variance}
\end{equation}
where $\gamma$ is a free parameter that controls the width of the initialisation
distribution. The standard Glorot normal initialisation of the baseline analysis
is then recovered by the case $\gamma = 1$. By fixing the architecture as in
\cref{tab:dis_fit_settings}, \cref{eq:init_variance} allows us to explore two
different regimes of initialisation widths.\footnote{The fits presented in this
section share the same random seeds for both the sampling of the parameters and
the generation of the data, so that differences between fits can be ascribed to
a different value of $\gamma$ replica by replica and not to spurious statistical
fluctuations.} For $\gamma \lesssim 1$, the initialisation width becomes larger
than the standard initialisation, while for $\gamma \gtrsim 1$ the
variance is suppressed, and the initialisation width tends to zero in the limit
$\gamma \to \infty$. 

In \cref{sec:init_gamma} we look at the PDFs at initialisation with different
values of $\gamma$, while in \cref{sec:init_speed} we inspect the impact on the
training dynamics. We validate the analytical solution at different values of
$\gamma$ in \cref{sec:init_frozen}, and in \cref{sec:init_optimal} we explore
which figures of merit can be used to identify an optimal value of $\gamma$. We
conclude in \cref{sec:init_width} where we clarify the connection between the
limit $\gamma\rightarrow\infty$ and wide NNs.

\begin{figure}[t]
  \centering
  \includegraphics[width=\linewidth]{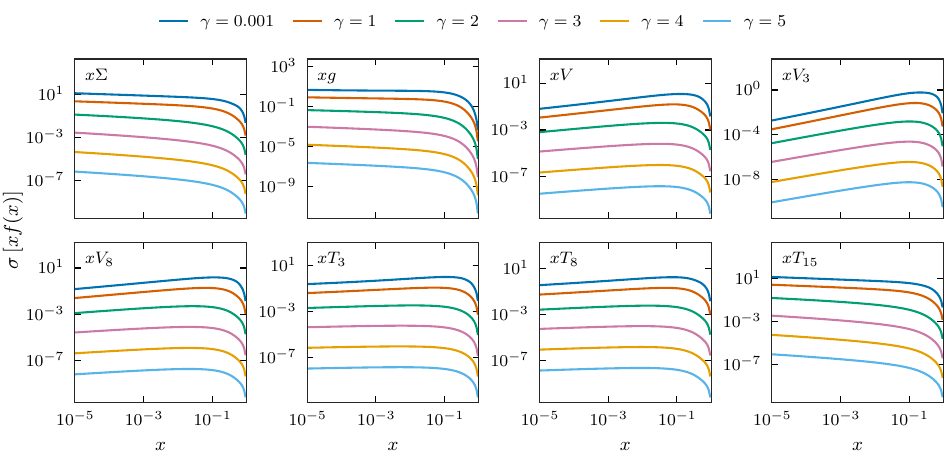}
  \caption{One-sigma width $\sigma[xf(x)]$ of the PDFs at initialisation as a
  function of $x$ at the initial scale $Q_0$, computed as the standard deviation
  over $\nreps = 120$ replicas of the full model, preprocessing factors
  included. Each panel shows one of the eight fitted flavours, and each curve
  one of the six values of the initialisation exponent $\gamma$ of
  \cref{eq:init_variance}. Both axes are logarithmic.}
  \label{fig:InitPDFsGamma}
\end{figure}

\begin{figure}[t]
  \centering
  \includegraphics[width=\linewidth]{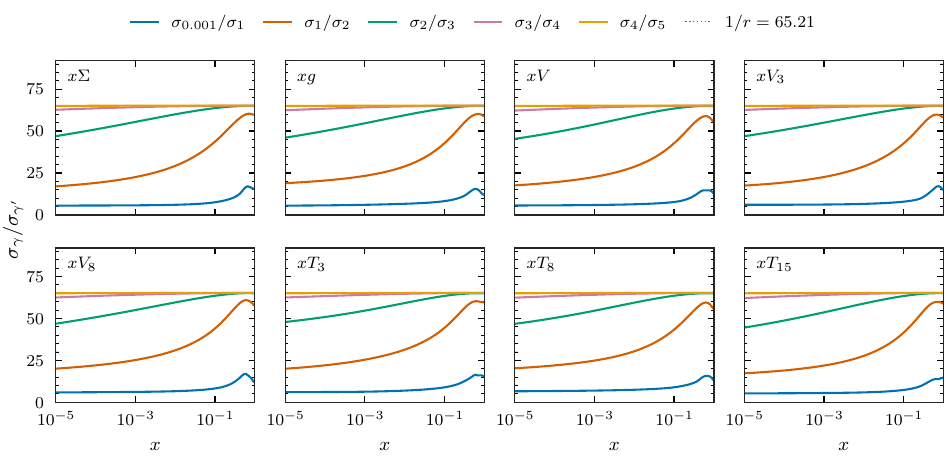}
  \caption{Ratio of the widths of \cref{fig:InitPDFsGamma} for consecutive
  values $\gamma < \gamma'$ of the initialisation exponent, one panel per fitted
  flavour. The horizontal dotted line is the value $1/r = 65.21$ predicted in
  the linear regime by \cref{eq:InitAmplitude}. The ratio is close to this value
  for $\gamma \gtrsim 3$, whereas for smaller $\gamma$ it falls below it, most
  strongly at small $x$, where the $\log x$ input is largest and the hyperbolic
  tangents saturate.}
  \label{fig:InitPDFsGammaRatio}
\end{figure}

\subsection{A linear model at initialisation}
\label{sec:init_gamma}

We begin our discussion by studying the dependence on $\gamma$ of the PDFs at
initialisation. In \cref{fig:InitPDFsGamma} we show the one-sigma widths of the
network at initialisation as functions of $x$ for the six values of $\gamma$
here considered, namely $\gamma=0.001,\,1,\,2,\,3,\,4,\,5$, and in
\cref{fig:InitPDFsGammaRatio} the ratio of the one-sigma curves for consecutive
values of $\gamma$. The central values have been omitted for clarity, as they
are not dissimilar from the baseline of \cref{fig:dis_initialisation}, and the
dependence on $\gamma$ is negligible. Inspected together, these figures reveal
two notable facts.

First, the width of the ensemble at initialisation decreases systematically with
increasing values of $\gamma$ across all flavours. The suppression towards the
large-$x$ region is an artefact of the preprocessing factors, which make the
PDFs vanish at $x=1$. In the small-$x$ region the width spans several orders of
magnitude depending on the choice of $\gamma$. Taking $x\Sigma$ as an example,
the width goes from $\sigma \sim 10^{0}\div10^{1}$ for $\gamma = 0.001$ and
$\gamma=1$, down to $\sigma \sim 10^{-7}\div10^{-6}$ for $\gamma = 5$.

Second, the ratio of the widths for consecutive values of $\gamma$ tends to a
constant as $\gamma$ is increased, independently of the value of $x$ and of the
flavour. This is particularly evident for $\gamma \gtrsim 3$, where the ratio
converges to $\sigma_{\gamma}/\sigma_{\gamma+1} \approx 65.21$ -- shown in
\cref{fig:InitPDFsGammaRatio} with a dotted horizontal line -- whereas smaller
values of $\gamma$ break this scaling with large deviations at small-$x$. In
other words, $\gamma$ does not modify the shape of the curves, but rather
rescales their amplitude by an $x$-independent factor, which on the logarithmic
scale of \cref{fig:InitPDFsGamma} appears as a rigid vertical translation.

Both facts can be understood by noting that, for large values of $\gamma$, the
typical size of the weights is small enough for the hyperbolic tangents to be
well approximated by their argument. Since the biases vanish at initialisation,
the network reduces to a DLN, \ie\ a product of weight matrices
acting on the input vector $X = (x, \log x)$. Writing each weight matrix as
$\wght{\ell} = \sigma_{\ell}\,\uwght{\ell}$, where the entries of $\uwght{\ell}$
are standard normal variables that do not depend on $\gamma$, the dependence on
$\gamma$ factors out of the network at initialisation, which we denote by
$f_{\gamma}$,
\begin{equation}
  f_{\gamma}(x) = r^{\gamma}\, \uwght{L}\cdots\uwght{1} X\,, \qquad
  r \equiv \left[
    \prod_{\ell=1}^{L} \frac{2}{n_{\ell} + n_{\ell-1}}
  \right]^{1/2}\,,
  \label{eq:InitAmplitude}
\end{equation}
where $r \leq 1$ depends only on the architecture. Since the random seeds are
held fixed across fits, the relation $f_{\gamma} = r^{\gamma - \gamma'}
f_{\gamma'}$ holds replica by replica and flavour by flavour, and it is not
spoiled by the preprocessing factors, which are independent of $\gamma$. As a
consequence, the width of the ensemble decreases as the pure power $r^{\gamma}$,
and the ratio between consecutive values of $\gamma$ is $1/r = 65.21$ for the
architecture of \cref{tab:dis_fit_settings}, which is the value marked by the
dotted line in \cref{fig:InitPDFsGammaRatio}. For smaller values of $\gamma$ the
weights are larger and the hyperbolic tangents saturate, first at small $x$
where $|\log x|$ is largest. The linear approximation then fails, and the
spacing between consecutive curves is compressed. The derivation of
\cref{eq:InitAmplitude} and the saturation of the internal layers are discussed
in detail in \cref{app:dln_ntk}.

In summary, as long as the network operates in its linear regime, $\gamma$
controls the amplitude of the prior but not its shape in $x$. Whether this prior
survives the training, and how it affects the final result, is discussed below
and further in \cref{app:init_dynamics}.

\subsection{Training dynamics at different initialisations}
\label{sec:init_speed}
\begin{figure}[t]
  \centering
  \includegraphics[width=0.7\linewidth]{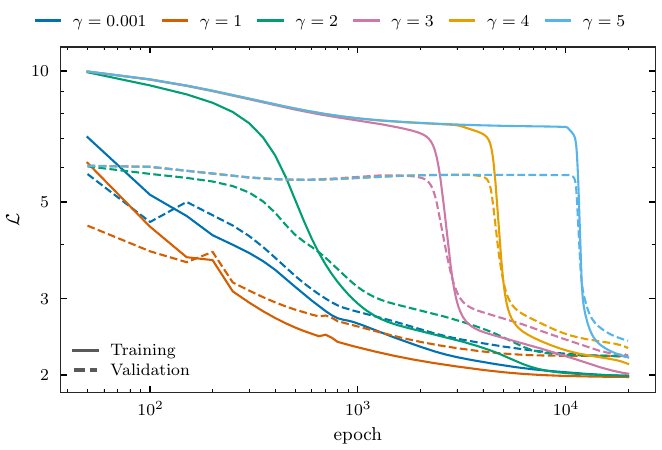}
  \caption{Training (solid) and validation (dashed) losses as functions of the
  epoch, for a single representative replica. Each curve represents one of the
  six values of $\gamma$ considered in this section. Both axes are logarithmic.
  For $\gamma \gtrsim 3$ the trajectory develops a plateau, on which the loss
  stalls close to its initial value, followed by a rapid drop. Increasing
  $\gamma$ delays that drop and slightly lifts the level reached afterwards.}
  \label{fig:LossGamma}
\end{figure}

We now turn to the implications of the initialisation on the training dynamics.
We train six fits with plain GD and the same settings as the baseline fit,
\cref{tab:dis_fit_settings}, which differ only in the value of $\gamma$. The
resulting training (solid) and validation (dashed) losses are shown in
\cref{fig:LossGamma} as functions of the training epoch for a single
representative replica of each of the six values of $\gamma$ considered before.

Two regimes can be clearly distinguished from this comparison. For $\gamma
\lesssim 2$, the loss function decreases monotonically, as originally observed
in \cref{fig:loss-functions}. By contrast, the cases with $\gamma \gtrsim 3$
exhibit a behaviour that is more similar to that already observed in
Ref.~\cite{Chiefa:2025cap}: after an initial plateau where the loss stalls, the
trajectories are characterised by a sudden drop followed by a steady descent
that converges to an asymptotic value. In this regime, increasing $\gamma$
delays the drop of the loss, which we quantify by means of the escape epoch
$\tesc$, defined as the epoch at which the training loss falls below $\loglhd =
3$. Taking the median over the replicas, we find $\tesc \approx 2500$, $5400$
and $\gtrsim 10^4$ for $\gamma = 3$, $4$ and $5$, respectively.\footnote{The
case with $\gamma = 5$ is a lower bound because 22 of the 120 replicas do not
escape within the $2\times10^4$ epochs.} The case $\gamma=2$ sits at the
boundary between these two regimes, as its drop is not as abrupt as for larger
$\gamma$, but still dissimilar from the monotonic behaviour of the fits with
smaller $\gamma$. 

\begin{figure}[p!]
  \centering
  \includegraphics[width=\linewidth]{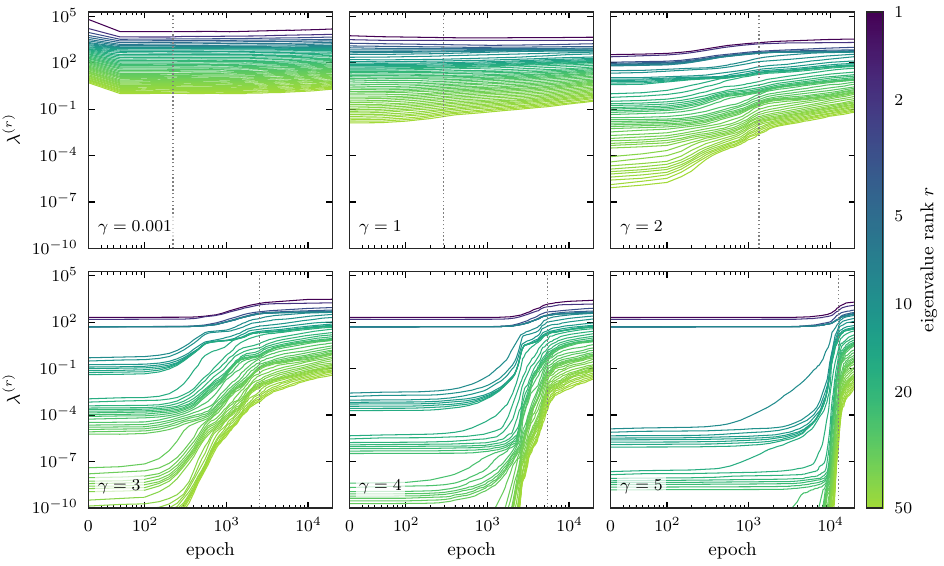}
  \caption{Training trajectories of the leading $50$ eigenvalues $\lambda^{(r)}$
  of the NTK, with preprocessing factors excluded. Each panel corresponds
  to a different initialisation exponent $\gamma$ of \cref{eq:init_variance}.
  The colour encodes the eigenvalue rank $r$, as in \cref{fig:EigvalsNTK}. The
  curves are medians over the $\nreps=120$ replicas of each fit, and bands are
  omitted for clarity; the red dashed line marks the median escape epoch $\tesc$
  of each fit. The abscissa is linear below $50$ epochs and logarithmic above,
  so that the initialisation at $\tau=0$ fits on the axis. The ordinate is cut
  at $10^{-10}$, the smallest eigenvalue that still lies inside $\ntkimg$ at the
  tolerance of \cref{fig:DimNTKSpace} for any of these fits.}
  \label{fig:EigValsGamma}
\end{figure}

The two regimes are mirrored by the spectrum of the NTK, shown in
\cref{fig:EigValsGamma} for the leading 50 eigenvalues. For $\gamma \lesssim 1$
the eigenvalues are almost constant throughout the training, apart from a short
initial transient for $\gamma = 0.001$ and a slow, monotonic growth of the
smallest eigenvalues for $\gamma = 1$. For $\gamma \gtrsim 3$, instead, the
leading eigenvalues remain essentially constant along the plateau, and the drop
of the loss is accompanied by a growth of the bulk of the spectrum by several
orders of magnitude, which is steepest around $\tesc$. Once again, the case
$\gamma = 2$ sits in between, with a gradual growth of the spectrum spread over
the first $10^3$ epochs. In \cref{app:init_dynamics}, and specifically in
\cref{fig:QmodesGamma}, we show that during this first plateau the NN is
operating in its linear regime, and therefore it is unable to reconstruct the
underlying law. The sharp transient observed in both the loss function
(\cref{fig:LossGamma}) and eigenvalues (\cref{fig:EigValsGamma}) signals the
departure from this linear regime.

\begin{figure}[p!]
  \centering
  \includegraphics[width=\linewidth]{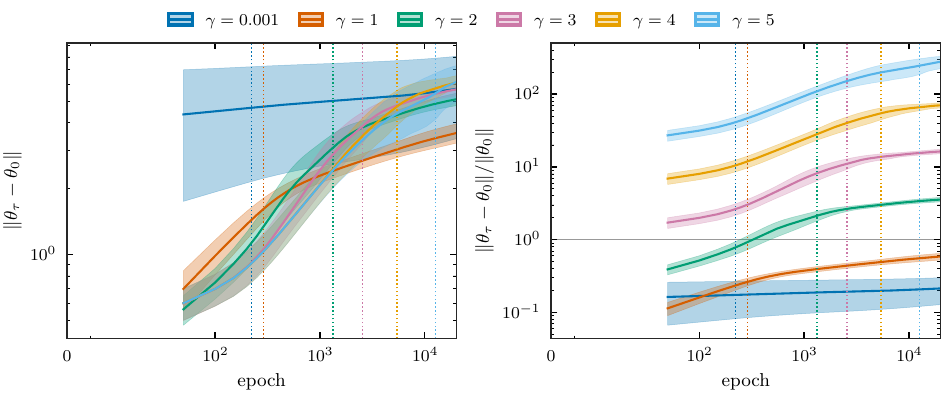}
  \caption{Displacements of the parameters in Euclidean space as a function of
  training time and for the different initialisation widths. (Left) The distance
  $\|\theta_{\tau} - \theta_{0}\|$ separating the parameters from their position
  at initialisation. (Right) The same distance normalised to the size of the
  initialisation, $\|\theta_0\|$. Only the network parameters are shown, the
  preprocessing factors being frozen after initialisation. Solid lines are means
  over the $\nreps = 120$ replicas of each fit and the shaded areas are
  one-sigma bands. The dotted vertical lines mark the median escape epoch
  $\tesc$ of each fit. The abscissa is linear below $50$ epochs and logarithmic
  above, so that the initialisation at $\tau=0$ sits on the axis. The first
  stored epoch is omitted, the displacement vanishing there by construction.}
  \label{fig:ParamaterDisplacementGamma}
\end{figure}

The displacement of the parameters from their initial values, shown in
\cref{fig:ParamaterDisplacementGamma}, completes the picture. By the end of the
training the parameters have moved by an amount of the same order for all values
of $\gamma$ (left panel) but, relative to the size of the initialisation
$\|\theta_0\|$ (right panel), the displacement spans three orders of magnitude.
For $\gamma \lesssim 1$ the parameters keep moving throughout the training, but
in a contained way, remaining within a fraction of $\|\theta_0\|$:
$\|\theta_{\tau} - \theta_0\|/\|\theta_0\| \approx 0.2$ and $0.6$ at the end of
the training for $\gamma = 0.001$ and $\gamma = 1$, respectively. Together with
the slowly varying eigenvalues of the NTK, this is the signature of the lazy (or
NTK) regime of Refs.~\cite{jacot2018neural,jacot2022sts,LazyNTKChizat:2019}, in
which the NTK remains approximately constant and the training is well described
by the NTK equations. This also explains why, in \cref{fig:AnltSolSigma}, the
analytical solution with the NTK frozen at initialisation provides a good
description of the training. For $\gamma \geq 2$, instead, the parameters travel
several times $\|\theta_0\|$, and more than two orders of magnitude for $\gamma
= 5$, so that the NTK at initialisation is not expected to describe the whole
training. We come back to this point in \cref{sec:init_frozen}.

We remark that, for $\gamma \gtrsim 3$, the plateau followed by a sudden drop of
the loss is reminiscent of the saddle-to-saddle dynamics described in
Ref.~\cite{jacot2022sts} for deep linear networks with a small initialisation.
We however clarify that, despite the qualitative similarity, the two phenomena
differ in nature. In Ref.~\cite{jacot2022sts}, the network is linear by
construction: the successive plateaus separate saddles corresponding to linear
maps of increasing rank, and the transitions between them remain linear maps
throughout. Here, instead, we observe a single plateau, along which the network
operates in the linear regime discussed in \cref{app:dln_ntk}, and
the escape coincides with the saturation of the hidden layers (see the right
panel of \cref{fig:DimNTKGamma}). What we observe is therefore the departure
from the deep linear description, rather than a transition between two of its
saddles.

We therefore conclude that changing $\gamma$ has a considerable effect on the
training trajectory of the NN. In particular, starting from a small
initialisation makes the training longer: the network first needs to escape the
regime where it acts as a linear model before gaining enough expressivity to fit
the data. In this respect, the inspection of the NTK paired with the loss curve
proves to be a good diagnostic tool to identify such scenarios.

\subsection{Validity of the frozen-NTK description}
\label{sec:init_frozen}
An immediate question is whether and to what extent the analytical solution
discussed in \cref{sec:analytical_training} can be equally applied to the fits
with different initialisations. We answer this question by investigating two
complementary aspects: the first is whether the analytical solution can
reproduce the \emph{numerical training}, and the second is whether it can
reproduce the underlying law $\fin$. In fact, the two questions can have
different answers depending on the value of $\gamma$. To show this, we freeze
the NTK at the two ends of the training, $\tntk=0$ and $\tntk = \tend$
respectively, and compare the resulting analytical solution with the trained
solution. In both cases, the analytical solution starts from the same initial
condition as the trained network, namely the replica at initialisation $f_0$,
and uses the NTK of that same replica at $\tntk$, so that the two solutions are
compared replica by replica. In the following, the preprocessing factors are
always included in the NTK.

\subsubsection*{Freezing at initialisation}
\begin{figure}[h!]
  \centering
  \includegraphics[width=\linewidth]{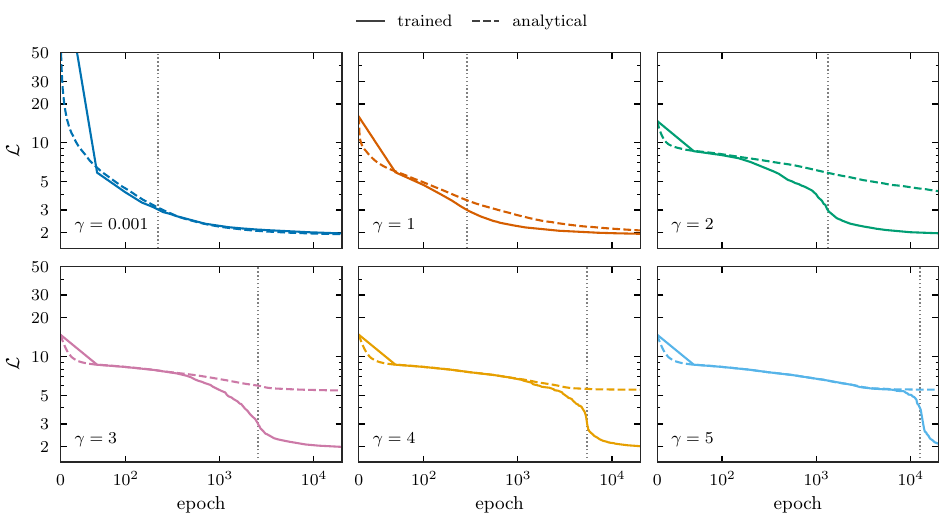}
  \caption{Training loss of the analytical solution of \cref{eq:analytic_sol}
  with the NTK frozen at initialisation (dashed) against the loss of the
  empirically trained network (solid), one panel per initialisation exponent
  $\gamma$. The ordinate is logarithmic while the abscissa is linear below epoch
  $50$ and logarithmic above. The analytical loss is evaluated by feeding the
  evolved solution back through the same FK tables and inverse covariance that
  define the trained loss, so that the two curves are the same observable. Both
  are medians over the $\nreps = 120$ replicas of each fit. The dotted vertical
  line marks the median escape epoch $\tesc$ of the ensemble. The trained curves
  are those of \cref{fig:LossGamma}, up to the replacement of the single
  representative replica by the ensemble median.}
  \label{fig:LossGammaAnalyticalInit}
\end{figure}

We begin with the NTK frozen at initialisation. In
\cref{fig:LossGammaAnalyticalInit} we compare the loss function of the trained
network with that obtained with the analytical solution along the whole training
($\tend=20000$) and for six values of $\gamma$. For $\gamma = 0.001$ and $\gamma
= 1$, the two curves remain close to each other over the whole trajectory, thus
supporting the observation -- already made in \cref{fig:AnltSolSigma} and
supported by \cref{fig:ParamaterDisplacementGamma} -- that the fit is
\emph{lazy} from the beginning with $\gamma\sim1$. On the other hand, from
$\gamma = 2$ onwards, the two curves coincide along the plateau but depart at
the escape epoch: the trained loss drops, while the analytical one continues
along the shallow descent it was already on.

\begin{figure}[t]
  \centering
  \includegraphics[width=0.85\linewidth]{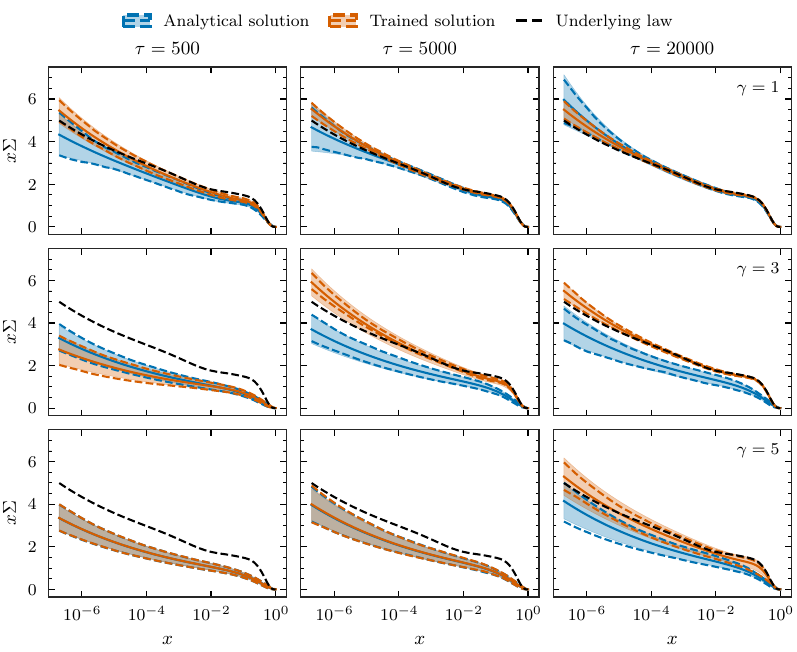}
  \caption{Analytical (blue) versus empirically trained (orange) solution of
  the singlet $\Sigma$, with the NTK frozen at initialisation, $\tntk=0$. The
  rows show the analytical solution obtained with $\gamma=1$ (top), $\gamma=3$
  (middle), $\gamma=5$ (bottom). The columns represent three different evolution
  epochs $\tau$ common to the three rows. Analytical and trained solutions are
  compared at the same epoch $\tau$. The solid line represents the mean over the
  ensemble, the shaded area the one-sigma band and the dashed lines the $68\%$
  \cl~over the $\nreps = 120$ replicas of each fit. The dashed black line is the
  underlying law $\fin$ used to generate the data.}
  \label{fig:AnltSolGammaInit}
\end{figure}

The origin of this behaviour can be understood by inspecting the shapes of the
resulting PDFs. In \cref{fig:AnltSolGammaInit} we compare the analytical and
trained solutions of the singlet for $\gamma = 1,\,3$ and $5$ (rows) at three
common evolution epochs (columns), with the two solutions compared at the same
epoch. The underlying law is shown as a dashed line for reference. In the top
row ($\gamma=1$), both solutions converge onto the underlying law at the same
pace, thus reproducing the behaviour already observed in \cref{fig:AnltSolSigma}
for the baseline fit and explaining the agreement of the two loss functions in
\cref{fig:LossGammaAnalyticalInit}. On the other hand, for $\gamma=5$ (last row)
the analytical solution and the trained one are almost indistinguishable at
$\tau=500$ and $\tau=5000$, well before the escape epoch, but both undershoot
the underlying law across the whole small-$x$ region. At $\tau=20000$, when the
network has already escaped from the linear regime, the two curves depart. The
middle row ($\gamma=3$), whose escape occurs at $\tesc\approx2500$,
interpolates between the two: at $\tau=500$ the two bands still overlap but are
visibly offset, and by $\tau=5000$ the trained solution has escaped and left the
analytical one behind.

The two examples given above show that, even for large values of $\gamma$, the
analytical solution with the NTK frozen at initialisation reproduces the trained
network as long as the latter remains on the plateau, while neither of them
reproduces the underlying law. In fact, the NTK is stationary on the initial
plateau, its eigenvalues and eigenvectors keeping their initial values until the
escape (see \cref{fig:EigValsGamma,fig:QmodesGamma}), and the analytical
solution provides an accurate description of the training dynamics. However,
the dimension of the image of the NTK of the full model at initialisation,
computed with the tolerance of \cref{fig:DimNTKSpace}, 
\begin{equation}
  \begin{array}{lcccccc}
    \gamma        & 0.001 & 1  & 2  & 3  & 4  & 5 \\[2pt]
    \dim\ntkimg   & 134   & 77 & 51 & 33 & 25 & 15\,,
  \end{array}
  \label{eq:DimNTKInitModel}
\end{equation}
decreases rapidly with $\gamma$. The few directions available at initialisation
when $\gamma$ is large are therefore not enough to reproduce the shape of
$\fin$. Moreover, in the linear regime, where the network operates before the
escape, the tangent vectors lie, flavour by flavour, in the three-dimensional
subspace
\begin{equation}
  \vlin \equiv \mathrm{span}\{1,\,x,\,\log x\}\,,
  \label{eq:SpanLinearFeatures}
\end{equation}
which bounds the image of the network alone to $3\nout = 24$ dimensions, reduced
to $15$ once the preprocessing factors are included.\footnote{A detailed
derivation is given in \cref{app:dln_ntk}. Note that the preprocessing factors
multiply the tangent vectors of the network flavour by flavour,
\cref{eq:TangentFeaturesDLNModel}, so that the functions available to the
analytical solution are not straight lines in $x$ and $\log x$, but inherit the
shape of $x^{1-\delta}(1-x)^{\epsilon}$. This is why the analytical solution in
the $\gamma = 5$ row of \cref{fig:AnltSolGammaInit} has a non-trivial shape,
even though the tangent vectors of the network lie in $\vlin$.} The analytical
solution can thus only move along these few directions, which are not expressive
enough to reproduce the underlying law. The trained network, instead, reaches
the underlying law only after the escape, when the saturation of the hidden
layers moves the NN out of $\vlin$ and new directions enter the image of the
NTK. More details on this dynamics are given in \cref{app:init_dynamics}.

\subsubsection*{Freezing at the end of training}
\begin{figure}[p!]
  \centering
  \includegraphics[width=\linewidth]{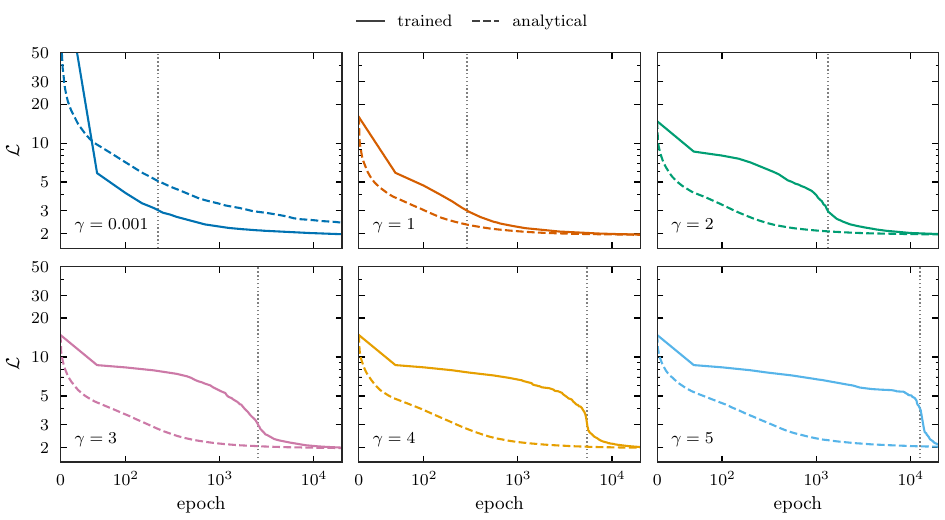}
  \caption{Same as \cref{fig:LossGammaAnalyticalInit}, but with the NTK frozen
  at the last stored epoch, $\tntk = \tend = 2\times10^{4}$.}
  \label{fig:LossGammaAnalyticalEnd}
\end{figure}

At the opposite end of the training trajectory the situation is reversed. Also
in this case we show, in \cref{fig:LossGammaAnalyticalEnd}, the analytical loss
function compared with the trained one. We clearly see that for $\gamma=0.001$
and $\gamma=1$ the two curves remain close along the entire training. The same
behaviour is not observed -- nor expected -- for the larger values of $\gamma$.
In fact, the analytical curve does not reproduce the sudden drop that
characterises the numerical loss. Interestingly, in all cases both curves
converge to the same value.

\begin{figure}[p]
  \centering
  \includegraphics[width=0.85\linewidth]{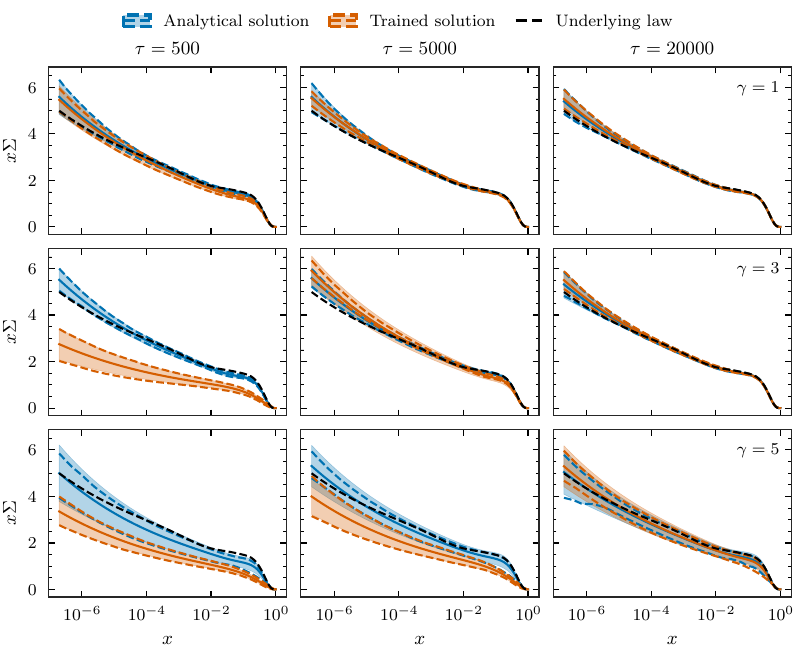}
  \caption{Same as \cref{fig:AnltSolGammaInit}, but with the NTK frozen at the
  last stored epoch, $\tntk = \tend = 2\times10^{4}$.}
  \label{fig:AnltSolGammaEnd}
\end{figure}

The resulting PDFs are shown in \cref{fig:AnltSolGammaEnd} with the same layout
as in \cref{fig:AnltSolGammaInit}, now with the NTK frozen at the last stored
epoch. We observe that in the last column, where the dominant modes are indeed
learned by the analytical solution, the two solutions agree to a good
approximation for all three values of $\gamma$, being also compatible with the
underlying law. This behaviour is indeed expected: once the eigenvectors of the
NTK have departed from the linear regime, the frozen-NTK description is as good
at large $\gamma$ as it is at $\gamma=1$. This is consistent with the late-time
behaviour of the subspace drift, \cref{fig:SubspaceDriftGamma} and $k\geq 16$,
which flattens after $\tesc$ for every $\gamma$: past the escape the
eigenvectors stop turning, and \cref{eq:frozen_ntk} becomes an accurate
approximation again. The analytical solution therefore converges much faster
than the trained network, and the behaviour of Ref.~\cite{Chiefa:2025cap} is
recovered.

Finally, the first two columns of \cref{fig:AnltSolGammaEnd} reveal a
complementary aspect of this story. Evolved for only $500$ epochs, the
analytical solution built on the \emph{late} NTK has already reached the
underlying law at $\gamma=3$ and $\gamma=5$, while the trained network at that
epoch is still on its plateau, a factor of two below. Had the network possessed
the late NTK from the first epoch, it would have learned the underlying law
almost immediately. Once again, this shows that the escape is a non-linear event
driven by the network leaving $\vlin$.


\subsection{Is there an optimal initialisation?}
\label{sec:init_optimal}
\begin{figure}[t]
  \centering
  \includegraphics[width=\linewidth]{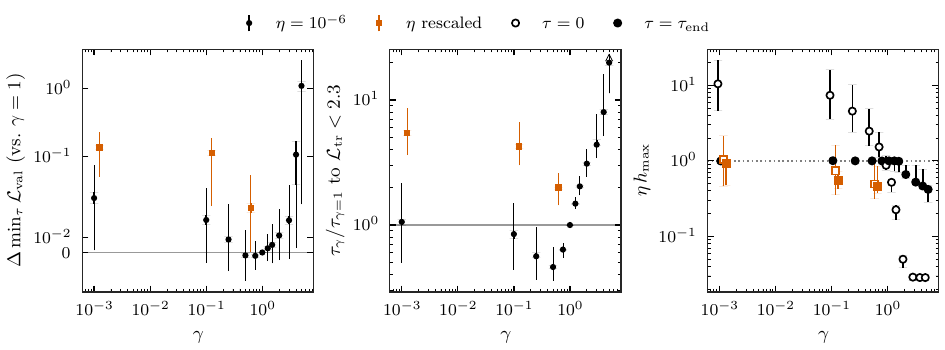}
  \caption{(Left) Minimum over the first $2\times10^{4}$ epochs of the
  validation loss, minus that of $\gamma = 1$ for the same replica: mean over
  $120$ replicas with its bootstrap error (thick bar), and $68\%$ interval of
  the replicas (thin line). The ordinate is linear within $\pm 0.05$ and
  logarithmic outside. (Middle) Number of epochs for the training loss
  to fall below $2.3$, divided by that of $\gamma = 1$ for the same replica:
  median with its bootstrap $68\%$ interval (thick bar), and $68\%$ interval of
  the replicas (thin line); the arrow marks $\gamma = 5$, where $31\%$ of the
  replicas never reach $2.3$ within $2\times10^{4}$ epochs. (Right)
  $\eta\,h_{\max}$, with $h_{\max}$ the largest eigenvalue of $H^{\perp}$, at
  initialisation (open) and at the end of the training (filled), median and
  $68\%$ interval over five replicas. The dotted line represents the stability
  condition of GD, \cref{eq:gd_stability}. We use black circles for the fixed
  learning rate, $\eta = 10^{-6}$, and orange squares for the rescaled
  learning rates so that $\eta\,h_{\max}$ at initialisation matches that of
  $\gamma = 1$ ($\eta = 10^{-7}$ for $\gamma = 0.001, 0.1$ and $2\times10^{-7}$
  for $\gamma = 0.5$), drawn slightly to the right.}
  \label{fig:OptimalGamma}
\end{figure}

\begin{figure}[p!]
  \centering
  \includegraphics[width=\linewidth]{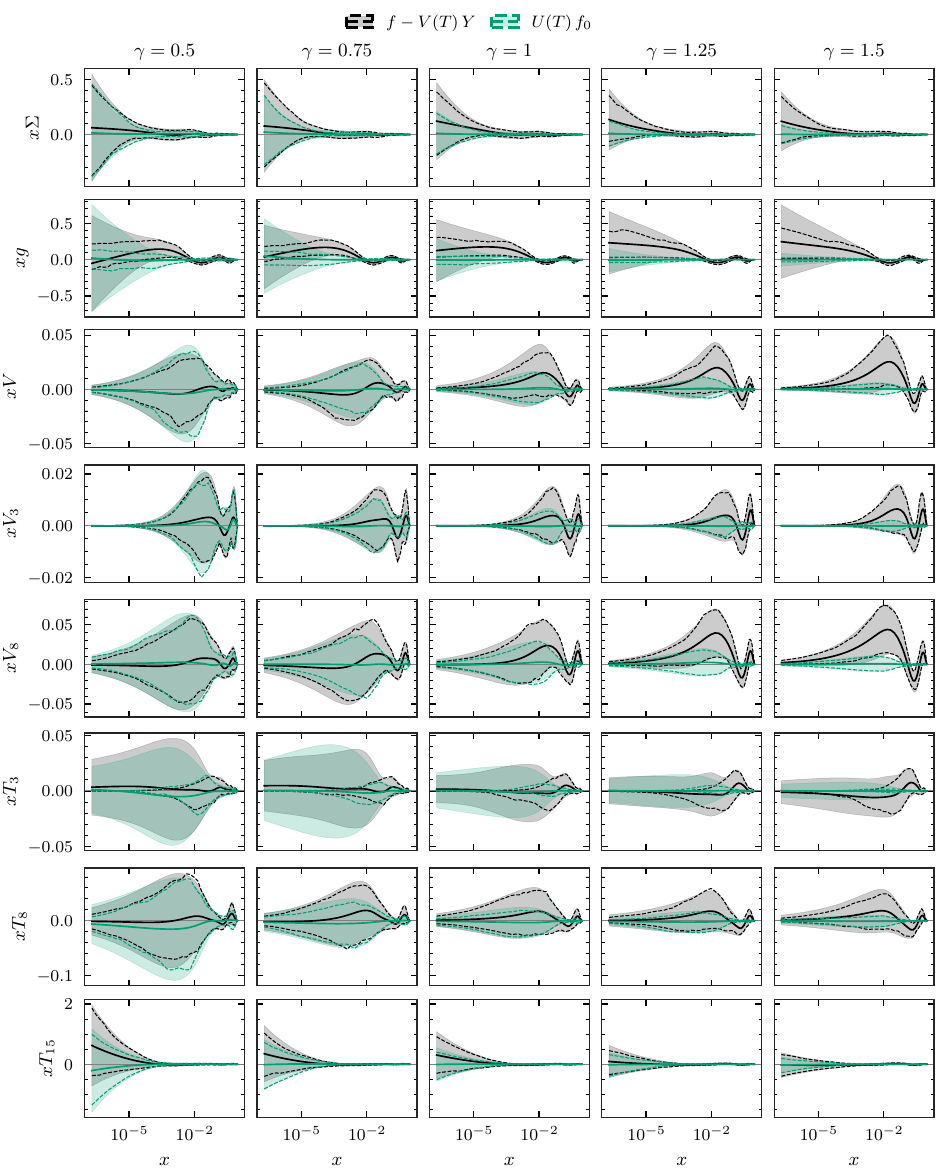}
  \caption{Comparison between the trained solution with the contribution of the
  data removed, $f - V(T)\obs$ (black), and the contribution of the
  initial condition to the analytical solution, $U(T)f_0$ (green), for the eight
  fitted flavours (rows) and for $\gamma = 0.5$, $0.75$, $1$, $1.25$ and $1.5$
  (columns). For each fit the NTK is frozen at the last epoch, $\tntk = \tend =
  2\times10^{4}$, and both quantities are evaluated at the end of the training,
  with $T$ corresponding to $\tau = \tend$. The two coincide only if the
  analytical solution reproduces the trained network. The solid lines represent
  the mean over the ensemble, the shaded areas the one-sigma band and the dashed
  lines the $68\%$ \cl~over the $\nreps = 120$ replicas of each fit.}
  \label{fig:InitComponentGammaSmall}
\end{figure}

We provided evidence that the choice of $\gamma$ -- and hence of the initial
condition -- affects both the efficiency of the training process
(\cref{fig:LossGamma}) and the expressiveness of the network in its early
learning steps (\cref{fig:AnltSolGammaInit,app:frozen_ntk_init}). We also showed
that these two aspects are related, so that changing one inevitably affects the
other. We therefore ask whether an optimal value of $\gamma$ exists, with
optimality measured here by the lowest validation loss attained, the efficiency
of training, and the analytical tractability of the training dynamics. We
illustrate these properties through
\cref{fig:OptimalGamma,fig:InitComponentGammaSmall}, which we discuss in turn.

We start from the two left panels of \cref{fig:OptimalGamma}, which show, for
different values of $\gamma$, the minimum validation loss -- with respect to the
baseline $\gamma=1$ -- and the number of epochs required by the training loss to
drop below $\loglhd_{\rm tr}=2.3$, respectively.\footnote{The threshold $2.3$
lies just above the asymptotic loss, $\loglhd_{\tend} = 1.96$--$2.11$, so the
count measures how quickly a fit approaches its end point.} Inspecting the black
points of both panels, we observe that choosing $\gamma_{\rm min}\sim0.6$
results in fits that have the lowest minimum of the validation loss and that
take the least time to reach it, differing by orders of magnitude from those
with larger values of $\gamma$. Note that the comparison with the baseline
$\gamma=1$ is made replica by replica (same pseudodata and same draw of the
weights once rescaled), so that the error bars measure how the effect of
changing $\gamma$ varies across replicas relative to the baseline case, and
vanish at $\gamma=1$ by construction. We can thus conclude that the fits depend
smoothly on $\gamma$, as far as the points shown in the panel are concerned.

Comparing different initialisations at a fixed learning rate requires, however,
some care. In fact, the analytical solution shows that the size of a GD step
along a given mode $q^{(i)}$ is set by the timescale of \cref{eq:tau}, with $t =
\eta\tau$. Schematically, we can denote by $e^{(i)}$ the distance of the model
from its asymptotic value along the $i$-th eigenvector. A single GD step along
this direction then is
\begin{equation}
  e^{(i)}_{\tau+1} = \left(1 - 2\eta h^{(i)}\right) e^{(i)}_{\tau}\,,
  \label{eq:gd_step}
\end{equation}
so that every mode converges only if $|1 - 2\eta h^{(i)}| < 1$, that is
\begin{equation}
  \eta\, h_{\max} < 1\,,
  \label{eq:gd_stability}
\end{equation}
where $h_{\max}$ is the largest eigenvalue of $H^{\perp}$. A mode with $1/2 <
\eta h < 1$ converges while oscillating, whereas for $\eta h > 1$ it grows.

Being derived for a frozen NTK, \cref{eq:gd_stability} does not imply that a
network starting above the bound diverges: the large steps also modify the NTK,
and can lower $h_{\max}$ until $\eta\, h_{\max} \simeq 1$, a regime known as the
\emph{edge of stability}~\cite{cohen2021eos,damian2023ssi,lewkowycz2020llr}. In
the right-most panel of \cref{fig:OptimalGamma} we show $\eta\, h_{\max}$ at
initialisation (open circles) and at the end of the training (filled circles)
for the different values of $\gamma$ considered here. We observe that the fit
with $\gamma=1$ sits right at the stability bound already at initialisation
($\eta\, h_{\max} = 0.95$), and moves towards unity by the end of the training.
For $\gamma<1$, the value of $\eta\, h_{\max}$ at initialisation is above unity,
and collapses to one by the end of the training. For $\gamma>1$ it is instead
below unity at initialisation and grows during training, but it reaches one only
for the values of $\gamma$ closest to unity: for $\gamma\geq2$ it is still below
the bound after $2\times10^{4}$ epochs, the more so the larger $\gamma$. This
indicates that GD tends to attain the stability condition during
training~\cite{cohen2021eos,damian2023ssi}, and that this condition is reached
within the first few steps in those fits for which the NTK is almost
constant.\footnote{This would also explain the short decreasing transient of the
eigenvalues for $\gamma=0.001$ observed in \cref{fig:EigValsGamma}: given the
instability caused by the large eigenvalues at initialisation, GD uses the first
steps to lower the eigenvalues until the stability condition is achieved.} Note
that if the learning rate is rescaled so that the stability condition is
satisfied already at initialisation (open orange squares), the value of $\eta\,
h_{\max}$ at the end of the training (filled orange squares) remains much closer
to its initial value.

For a frozen NTK, the stability condition of \cref{eq:gd_stability} thus guarantees
that gradient descent converges to the same solution as the gradient flow of
\cref{eq:analytic_sol}, the two differing only by a short transient in the
directions of the larger modes. Any discrepancy between the analytical and the
trained solutions must therefore be ascribed to the evolution of the NTK: if the
latter is constant -- or slowly varying -- during training, then the analytical
solution is guaranteed to follow the trained network. To check for which values
of $\gamma$ we gain the most analytical control, we subtract from the trained
network at the end of the training the component of the analytical solution that
carries the information of the data, namely $f - V(t)\obs$, and compare the
result with the component that propagates the initial condition, $U(t)f_0$. The
two quantities coincide only if the analytical solution reproduces the trained
network. This comparison is shown in \cref{fig:InitComponentGammaSmall}. We
observe that the residual obtained by removing the contribution of the data from
the trained network is closest to the analytical prediction for $\gamma=0.5$ and
$\gamma=0.75$, and that the agreement deteriorates as $\gamma$ increases. It is
worth emphasising that these are the same two values of $\gamma$ that minimise
both the validation loss and the training time in the two left panels of
\cref{fig:OptimalGamma}. The initialisations that yield the best and the fastest
fits are therefore also those for which we gain the most analytical control over
the training, and for which the contribution of the initial condition to the
final solution can be predicted. The only exception is $T_3$, for which
discrepancies are observed also in the baseline case, for reasons that we do not
understand yet and that we leave for future investigation.

We finally note that the observations made here hold insofar as vanilla GD is
adopted. Whether the same behaviour is expected once the optimisation algorithm
is replaced by a more efficient one (\eg\ Adam or Nadam as in the NNPDF
case) is unclear from the present analysis. This aspect will also be
investigated in future work.

\subsection{A wide NN is not a substitute for a small initialisation}
\label{sec:init_width}
We devote the last part of this section to laying out the connection between the
parameter $\gamma$ introduced in \cref{eq:init_variance} and the width of the
NN, that is, the number of nodes in its hidden layers. In fact, one may be led to
believe that increasing the number of internal nodes in \cref{eq:init_variance}
is equivalent to increasing the value of $\gamma$, as far as the Glorot normal
initialisation or other equivalent prescriptions are concerned. We now show
that, while these two approaches can lead to identical distributions of the
weights at initialisation, they differ in the scaling of the NTK and in the
resulting training dynamics.

\begin{figure}[t!]
  \centering
  \includegraphics[width=0.7\linewidth]{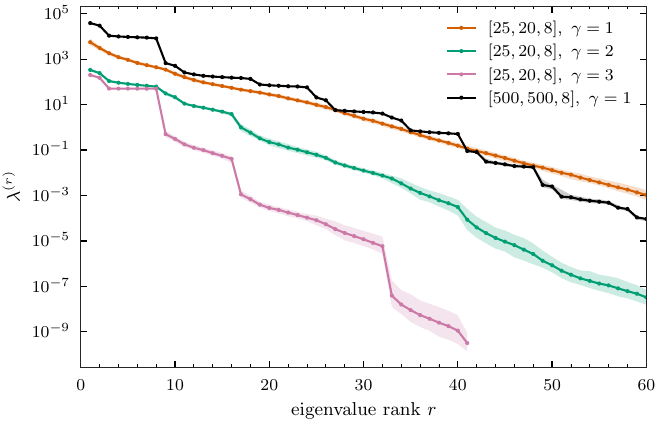}
  \caption{Eigenvalues of the NTK of the neural network (preprocessing factors
  excluded) at initialisation, against their rank. Coloured curves are the
  baseline architecture $[25, 20, 8]$ at $\gamma = 1, 2, 3$. The black curve is
  the wide architecture $[500, 500, 8]$ at $\gamma = 1$. Lines are medians over
  the replicas of each fit ($\nreps = 120$ for the baseline architecture, $5$
  for the wide one), bands are $68\%$ intervals. Each curve stops at the
  dimension of the image of the NTK, with the tolerance of
  \cref{fig:DimNTKSpace}.}
  \label{fig:EigValsWidthInit}
\end{figure}

In \cref{fig:EigValsWidthInit} we compare the spectrum of the NTK at
initialisation for $\gamma=1,\,2$ and $3$ with that of a network initialised
with standard Glorot normal ($\gamma=1$) but with the width of the two hidden
layers increased to 500 nodes each -- everything else remains unchanged. Two
important aspects emerge from inspecting the figure. First, the size of the NTK
eigenvalues of the wide network (black points) is compatible with that of the
baseline network (orange points), and much larger than that of the networks with
the baseline architecture but larger $\gamma$. This contradicts the naive
expectation. Indeed, based on the definition of \cref{eq:init_variance}, the
variance of the parameter distribution of the wide network is $\sigma^2_{\rm
wide} = 2\times10^{-3}$, the same variance being obtained by setting $\gamma\sim
2$ with the baseline architecture. Yet, the figure clearly shows that the two
spectra are not compatible as far as their size is concerned. Second, the
spectrum of the wide network shows a tier structure which is very similar -- if
not identical -- to that observed for larger values of $\gamma$, in particular
for $\gamma=2$. This tier structure is extensively discussed in
\cref{app:dln_ntk} in the limit of large $\gamma$, although at present we cannot
explain why a similar structure characterises the spectrum of the NTK for a wide
NN. Therefore, we give an intuitive explanation of the first aspect, deferring
the second to future work. 

The size of the spectrum of the wide network can be understood by considering
the kernel of the inner layers, defined as~\cite{Chiefa:2025cap}
\begin{equation}
  K_{i\alpha,i'\alpha'}^{(\ell)} = \expct\left[\preact{\ell}_{i\alpha} \preact{\ell}_{i'\alpha'}\right]\,,
  \qquad \mathrm{with} \qquad
  \preact{\ell}_{i\alpha} = \sum_{j=1}^{n_{\ell-1}}\wght{\ell}_{ij} \act{\ell-1}_{j\alpha}
      + \bias{\ell}_i\,,
\end{equation}
where $\act{\ell}_{i\alpha} \equiv \rho\left(\preact{\ell}_{i\alpha}\right)$ is
the activation function $\rho$ applied to the pre-activation
$\preact{\ell}_{i\alpha}$ of layer $\ell$. A simple calculation shows that
\begin{equation}
  \begin{split}
   K_{i\alpha,i'\alpha'}^{(\ell)}
    & = \sum_{j=1}^{n_{\ell-1}} \sum_{j'=1}^{n_{\ell-1}}
        \expct \left[\wght{\ell}_{ij}\wght{\ell}_{i'j'}\right]
        \expct \left[\act{\ell-1}_{j\alpha}\act{\ell-1}_{j'\alpha'}\right] \\
    & = \delta_{ii'} \sigma^2_{\ell} \sum_{j=1}^{n_{\ell-1}}
        \expct \left[\act{\ell-1}_{j\alpha}\act{\ell-1}_{j\alpha'}\right] \\
    & = \delta_{ii'} \sigma^2_{\ell}\, n_{\ell-1}\,
        \expct \left[\act{\ell-1}_{1\alpha}\act{\ell-1}_{1\alpha'}\right]\,.
  \end{split}
  \label{eq:kernel}
\end{equation}
The first equality holds because the weights are \emph{i.i.d.} within and across
layers and the biases are initialised to zero. The second one follows from the
definition of $\sigma_{\ell}$ and the application of the resulting Kronecker
delta. Finally, since the weights of layer $\ell-1$ are \emph{i.i.d.} across
neurons, all neurons in that layer are statistically equivalent, so the
expectation value $\expct
\left[\act{\ell-1}_{j\alpha}\act{\ell-1}_{j\alpha'}\right]$ is the same for
every $j$.\footnote{Note that in the limit of large width this expectation value
can be computed exactly. Such a derivation is beyond the scope of this work, and
we refer to Refs.~\cite{jacot2018neural,lee:2020,Roberts:2021fes} for more
details on the theory of wide networks and their relation with Gaussian
processes.} Therefore, the sum over $j$ reduces to the expectation for a single
neuron multiplied by $n_{\ell-1}$, which explains the third and last equality.

We also mention that the NTK can be computed by iterating expressions similar to
\cref{eq:kernel} layer by layer, a recursion also known as the \emph{forward
equation} of the NTK (see \eg\ Ref.~\cite{Roberts:2021fes}). Being recursive,
this construction is such that the scaling of each layer accumulates through the
layers. We therefore conclude that the reduction of the variance induced by
$\gamma$ and that induced by the width of the network have different scaling
properties in \cref{eq:kernel}. Specifically, we can show that
\begin{equation}
  K_{\gamma}^{(\ell)} \sim O\left(n^{1-\gamma} \right)\,,
  \qquad \mathrm{and} \qquad
  K_{\rm wide}^{(\ell)} \sim O\left( 1 \right)\,,
  \label{eq:kernel_scaling}
\end{equation}
with $n$ the typical width of the layers, so that the NTK of the wide network
stays $O(1)$ with depth, while for $\gamma>1$ it shrinks geometrically. This
explains why the spectrum remains almost unchanged in size when the width is
increased, and also why the typical size of the eigenvalues for larger $\gamma$
is much smaller.

\begin{figure}[t]
  \centering
  \includegraphics[width=\linewidth]{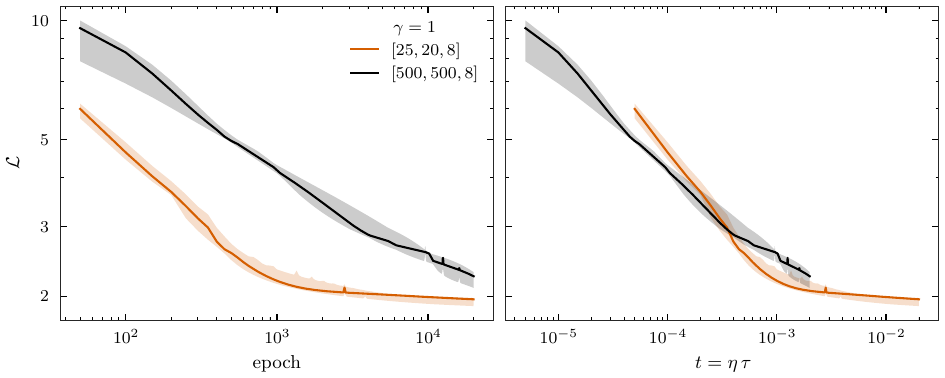}
  \caption{Training loss at $\gamma = 1$ for the baseline architecture $[25, 20,
  8]$ (learning rate $\eta = 10^{-6}$) and the wide architecture $[500, 500, 8]$
  ($\eta = 10^{-7}$), as a function of the epoch (left) and of the
  gradient-flow time $t = \eta\,\tau$ (right). Lines are medians and
  bands $68\%$ intervals over the first five replicas of each fit. Both axes are
  logarithmic.}
  \label{fig:LossWidth}
\end{figure}

The different scaling properties affect the training dynamics as well. In
\cref{fig:LossWidth} we show the training losses of the baseline fit and of the
wide network just discussed ($n_1=n_2=500$), both with $\gamma=1$, as functions
of the epoch (left) and of the gradient-flow time $t=\eta\tau$ (right). The
learning rate of the wide network has been reduced to $\eta=10^{-7}$ in order to
account for the increase of the leading eigenvalues of the NTK in
\cref{fig:EigValsWidthInit}, in accordance with the stability condition of
\cref{eq:gd_stability}. We clearly see that the wide network has not reached
convergence within the $2\times10^4$ epochs, and also that its monotonic
decrease has a different functional form compared to the baseline case.
Nonetheless, the wide network does not show the plateau followed by a sudden
drop observed for large $\gamma$ in \cref{fig:LossGamma}, confirming that the
two prescriptions lead to different training dynamics. We also note that the
fair like-for-like comparison is actually given by the right panel, where the
difference in the learning rates is accounted for.

\FloatBarrier

%% file: sections/conclusions.tex
\section{Conclusions and outlook}
\label{sec:conclusions}

In this paper we have investigated the training dynamics of a NN through a
quantitative analysis of the NTK in a realistic PDF determination, extending a
previous study based on a single-flavour toy model. We have generalised the
analytical solution of the gradient flow of Ref.~\cite{Chiefa:2025cap} to the
multi-flavour case and validated it against the numerically trained network.
Within the NNPDF4.0 methodology~\cite{NNPDF:2021njg,NNPDF:2021uiq}, we have
exposed the implicit regularisation induced by both gradient descent and
parametrisation, and characterised the behaviour of the NN during the training
process in terms of the NTK and a second matrix $M$. In addition to our
quantitative analysis, we have shown that the matrix $M$ is the Fisher
information matrix of the PDFs, derived its connection with the Shapley value of
Ref.~\cite{Bonnet-Guerrini:2026thq}, and outlined a geometric interpretation of
training. Here we summarise our findings.

First, the analysis of Ref.~\cite{Chiefa:2025cap} can be extended to a more
realistic case, proving the NTK a valuable diagnostic tool to characterise the
features learned by the NN during training. Consistent with the previous study,
the spectrum of the NTK exhibits a hierarchical structure
(\cref{fig:EigvalsNTK}). Contrary to it, however, the \emph{rich} phase is
confined to the first few epochs, where sub-leading modes populate the image of
the NTK incrementally; after this short initial transient, the NTK becomes
\emph{lazy} and the training flow can be integrated analytically
(\cref{eq:analytic_sol_compact}). A similar behaviour is also reflected in the
modes of the training Hamiltonian (\cref{eq:hamiltonian_eigvec_pdf}), which are
able to capture much of the shape of the underlying law already at
initialisation (\cref{fig:QmodesEpch1}).

Second, the extension of the analytical solution to the multi-flavour case
reproduces the numerically trained network to a good approximation for all
flavours, except for the uncertainty of $T_3$ at small-$x$
(\cref{fig:AnltSolInit}). Within the lazy training regime, we have used the
analytical solution to quantify the impact of the initial condition and of the
data on the final solution (\cref{fig:UVContributions}). We have found that the
central value of the PDFs is essentially shaped by the contribution of the data,
while the bias induced by the initialisation prescription, as defined in
\cref{sec:framework}, is negligible. For the uncertainties, the contribution of
the initial condition is suppressed in the region covered by the data --
although not for $T_3$ -- and remains comparable to that of the data in the
extrapolation regions, with a relative weight that depends on the flavour and on
the training time (\cref{fig:ErrorBudget}).

Third, the implicit regularisation of gradient descent amounts to the
incremental inclusion of NTK modes, each learned at a timescale set by the size
of the corresponding eigenvalue (\cref{eq:analytic_sol_compact}). We have
demonstrated that only about 30 of the leading modes are needed to reconstruct
the underlying law (\cref{fig:AnltSolInit}), which is in stark contrast with the
$\parameters$ parameters of the network. This suggests that the total number of
parameters largely overestimates the effective degrees of freedom used to
reconstruct the underlying law, and provides an empirical evidence for the
information-theoretic arguments of Ref.~\cite{sun_geometric_2025}. Beyond the
leading modes, we have shown that the sub-leading ones are poorly constrained by
the data (\cref{eq:analytic_sol_compact_l2}) and their inclusion carries noise
into the solution (\cref{fig:AnltSolInit}). This shows that early stopping sets
the trade-off between the signal and the noise of the data.

Finally, we have shown that the width of the distribution of the parameters at
initialisation has a considerable impact on the dynamics and efficiency of
training. A distribution that is too narrow forces the network to operate in a
regime linear in its input variables (\cref{eq:InitAmplitude}). The loss
function then stalls on a plateau until the inner layers saturate
(\cref{fig:DimNTKGamma}), the loss function drops (\cref{fig:LossGamma}), and
the NN departs from the linear regime. This is reflected in the spectrum of the
NTK: the eigenvalues closely follow the behaviour of the loss function
(\cref{fig:EigValsGamma}), while the corresponding eigenvectors are confined to
the span $\{1,\,x,\,\log x\}$ (\cref{fig:QmodesGammaLinearity}). Conversely, an
initialisation slightly wider than the standard one yields the best and fastest
fits, which are also those over which we gain the most analytical control. We
have noted that the sharp transition in both loss function and eigenvalues was
already observed in Ref.~\cite{Chiefa:2025cap}, and that, despite the
qualitative similarity, it is different in nature from the saddle-to-saddle
dynamics of Refs.~\cite{jacot2022sts,FUKUMIZU2000317,zhang2026sts}. We have then
clarified that a small initialisation is not equivalent to a large width, as the
NTK scales differently in the two cases. 

Our analysis can be extended in many respects. The analytical treatment can be
improved by solving the gradient flow with more sophisticated optimisers or
stochastic variants of gradient descent, and by restoring the sum rules and
physical constraints removed here. Concerning possible applications, the
operator $M$ and the NTK can be used to revisit the closure tests with
inconsistent data of Ref.~\cite{Barontini:2025lnl}, identifying the directions
in PDF space affected by the inconsistencies. Moreover, the NTK can be computed
for any parametrisation, such as the fixed functional forms of
FPPDF~\cite{Cruz-Martinez:2026rct}, as well as for NNs used in other tasks in
particle physics.

We view these results as part of a broader change of perspective aiming at a
quantitative interpretation of ML~\cite{Kramer:2026eoq}, where the relationship
with physics is reciprocal. In fact, physics can help ML with formal tools
inherited from
it~\cite{Winterhalder:2026jcu,Halverson:2020trp,Demirtas:2023fir,Roberts:2021fes,Kuntz:2026kuv},
and offers in turn realistic case studies -- such as the PDF fit analysed here
-- against which these ideas can be tested, contributing to an emerging theory
of deep learning~\cite{simon2026stdl}.

%% file: sections/dln.tex
\section{The deep linear network approximation}
\label{app:dln_ntk}

Throughout \cref{sec:init_dependence} we relied on the DLN approximation, in
which the network is assumed to operate in the linear regime of its activations.
Here we collect the derivations behind the statements made there. We first
derive the scaling with $\gamma$ of the network at initialisation, and quantify
the saturation of the hidden layers that limits the validity of the
approximation. We then derive the tangent subspace $\vlin$ of
\cref{eq:SpanLinearFeatures}, compute the leading tier of the NTK spectrum
exactly, and finally derive a compact form for the NTK of a DLN that reduces its
eigenvalue problem to $3\nout$ dimensions.

\subsection{Deep linear networks}

In the linear regime, the pre-activations can be written as
\begin{equation}
  \begin{split}
    \preact{1}_{i\alpha} &= \sum_{j=1}^{2}\wght{1}_{ij} X_{j\alpha}
      + \bias{1}_i\,,\\
    \preact{\ell}_{i\alpha} &= \sum_{j=1}^{n_{\ell-1}}\wght{\ell}_{ij} \preact{\ell-1}_{j\alpha}
      + \bias{\ell}_i\,,
    \quad \text{with} \quad
    \preact{L}_{i\alpha} \equiv \NN_{i}(x_{\alpha})\,,
  \end{split}
  \label{eq:LinearPreactivations}
\end{equation}
where $X_\alpha = (x_\alpha, \,\log x_\alpha)$ is the vector of input variables,
while $\wght{\ell}\in\real^{n_{\ell}\times n_{\ell-1}}$ and
$\bias{\ell}\in\real^{n_{\ell}}$ are the weights and biases of layer $\ell$,
respectively. We have restored the bias terms because, although their values
vanish at initialisation, their gradients do not: they generate the constant
sector of the NTK computed below. Note that in the following we suppress the
indices for clarity in favour of the vector notation, so that
\cref{eq:LinearPreactivations} becomes matrix-vector multiplications and
additions. With this convention, one can
verify that the output of the network can be written as
\begin{equation}
  \begin{split}
    \preact{L} 
    &= \wght{L}\preact{L-1} + \bias{L} \\
    &= \wght{L}\left(\wght{L-1}\preact{L-2} + \bias{L-1}\right) + \bias{L} \\
    &= \wght{L}\left(\wght{L-1}\left(\wght{L-2}(\dots) + \bias{L-2}\right) + \bias{L-1}\right) + \bias{L} \\
    &\;\;\vdots \\
    &= 
      \bias{L} 
      + \sum_{\ell=1}^{L-1} \left(\prod_{m=L}^{\ell+1} \wght{m}\right) \bias{\ell}
      + \left(\prod_{\ell=0}^{L-1} \wght{L-\ell}\right) X\,,
  \end{split}
  \label{eq:GenLinearNN}
\end{equation}
which, once the biases vanish as they do at initialisation, reduces to
\begin{equation}
  \NN(x) = \wght{L}\cdots\wght{1} X = A\, x + B \log x\,,
  \label{eq:DLNAtInit}
\end{equation}
with $A, B \in \real^{n_L}$. The network is linear in its inputs, though not in
its parameters.

The output of the network in \cref{eq:GenLinearNN} decomposes into three terms. The first one is
$x$-independent and collects the $\nout$ biases of the last layer, one for each
fitted flavour. The monomials in the second term, still $x$-independent,
multiply the bias at inner layer $\ell$ with the chain of weights from the last
layer down to layer $\ell + 1$. Finally, the last term is exactly the linear
approximation of \cref{eq:DLNAtInit}, which carries the $x$-dependence. Note
that of the three terms in \cref{eq:GenLinearNN}, only the first one is not
affected by the scaling with $\gamma$.

The scaling with $\gamma$ of the network at initialisation follows from
\cref{eq:DLNAtInit}. The weights are independent Gaussian variables with
zero mean and variance $\sigma^2_{\ell}$ given by \cref{eq:init_variance}, so
that we can write $\wght{\ell} = \sigma_{\ell}\,\uwght{\ell}$, where the entries
of $\uwght{\ell}$ are standard normal variables that do not depend on $\gamma$.
Denoting by $f_{\gamma}$ the network at initialisation for a given $\gamma$, the
dependence on $\gamma$ factors out of the chain of weights,
\begin{equation}
  f_{\gamma}(x) = \left(\prod_{\ell=1}^{L} \sigma_{\ell}\right)
  \uwght{L}\cdots\uwght{1} X
  = r^{\gamma}\, \uwght{L}\cdots\uwght{1} X\,,
\end{equation}
with $r$ defined in \cref{eq:InitAmplitude}. The product of the matrices
$\uwght{\ell}$ is a random matrix that is no longer Gaussian, but whose
distribution is independent of $\gamma$.\footnote{The variance of the entries
of the product $P = \uwght{L}\cdots\uwght{1}$ can be computed explicitly. Each
entry $P_{ij} = \sum_k\uwght{L}_{ik_L} \uwght{L-1}_{k_Lk_{L-1}}\dots
\uwght1_{k_1 j}$ is a sum over all the paths $k$ that the indices can take. Each
path is the product of independent, centred random variables with unit
variance, and therefore has unit variance itself, since for independent
variables $\mathrm{Var}(X_1 \cdots X_n) = \prod_{i=1}^{n} \left(
\mathrm{Var}(X_i) + \mathbb{E}[X_i]^2 \right) - \prod_{i=1}^{n}
\mathbb{E}[X_i]^2$~\cite{GVK023117737}. The variance is additive across the sum
over uncorrelated paths, so that $\mathrm{Var}(P_{ij}) = (\# \text{
paths})(\text{path variance})= \prod_{\ell=1}^{L-1} n_{\ell}$, the summed
indices running over the internal layers.} Moreover, since the random seeds are
held fixed across fits, the matrices $\uwght{\ell}$ are the same draw for every
$\gamma$, and the relation $f_{\gamma} = r^{\gamma - \gamma'} f_{\gamma'}$ holds
replica by replica and flavour by flavour. The preprocessing factors do not
spoil this relation, as they multiply the output of the network pointwise and
are themselves independent of $\gamma$. Since $r \leq 1$ for any architecture,
the network at initialisation is suppressed as the pure power $r^{\gamma}$, and
the ratio between consecutive integer values of $\gamma$ is $1/r$. For the
architecture of \cref{tab:dis_fit_settings} this evaluates to
\begin{equation}
  r = 1.533\times 10^{-2}\,, \qquad
  \frac{1}{r} = 65.21\,.
  \label{eq:ScalingRatio}
\end{equation}
This is the asymptotic value marked by the dotted line in
\cref{fig:InitPDFsGammaRatio}, which the ratio between the $\gamma = 4$ and
$\gamma = 5$ curves reproduces, thus providing a numerical validation of the
linear approximation. It also explains why the curves of \cref{fig:InitPDFsGamma}
are parallel on a logarithmic scale for $\gamma \gtrsim 3$: since the widths
obey $\sigma_{\gamma} = r^{\gamma-\gamma'}\sigma_{\gamma'}$, their logarithms
differ by $(\gamma-\gamma')\log_{10} r$, which is independent of $x$.

\subsection{Saturation of the hidden layers}
\label{app:saturation}
The DLN approximation holds as long as the hyperbolic tangents operate in their
linear regime. Decreasing $\gamma$ increases the typical size of the parameters
and therefore saturates the activation functions, so that the DLN approximation
breaks down: saturation effects become predominant and the simple scaling law
predicted by \cref{eq:InitAmplitude} is no longer valid.

\begin{figure}
  \centering
  \includegraphics[width=\linewidth]{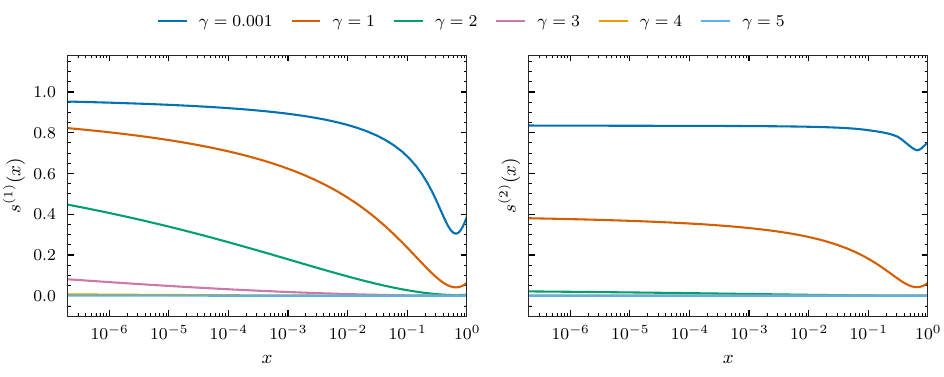}
  \caption{Saturation of the first (left) and second (right) hidden layers as a
    function of $x$ as defined in \cref{eq:SaturationMetric}. The six values of
    $\gamma$ considered in this analysis are represented by the different
    coloured curves. The curves are means over the $\nreps = 120$ replicas of
    each fit. For $\gamma \gtrsim 3$ the saturation is negligible across the
    entire $x$ range, whereas for smaller values of $\gamma$ the saturation
    becomes significant, particularly at small $x$ where the $\log x$ input is
    largest.}
  \label{fig:InitSaturation}
\end{figure}
The degree of saturation can be quantified by measuring in which regime the
activation functions operate at different $\gamma$. We define the saturation
$s^{(\ell)}(x_{\alpha}) \equiv s^{(\ell)}_{\alpha}$ of layer $\ell$ evaluated at
$x_{\alpha}$ as
\begin{equation}
  s^{(\ell)}_{\alpha} 
  \equiv \left\langle 1 - \tanh' \phi^{(\ell)}_{\alpha} \right\rangle
  = \left\langle \tanh^{2} \phi^{(\ell)}_{\alpha} \right\rangle
  \equiv \frac{1}{n_{\ell}}\sum_{i=1}^{n_{\ell}}
    \tanh^2\phi_{i\alpha}^{(\ell)}\,,
  \label{eq:SaturationMetric}
\end{equation}
that is, one minus the average of the derivative of the hyperbolic tangent over
the units of the layer. Here, $\phi^{(\ell)}_i(x_{\alpha})\equiv\phi^{(\ell)}_{i\alpha}$
is the $i$-th pre-activation node of hidden layer $\ell$ evaluated at
$x_{\alpha}$. This average equals zero when the layer operates in the linear
regime (not saturated) and tends to one as its units saturate. In
\cref{fig:InitSaturation} we show the saturation of the first (left panel) and
second (right panel) hidden layers as functions of $x$ for the six values of
$\gamma$ considered in this analysis. Below we tabulate the values of
\cref{eq:SaturationMetric} averaged over $x$ and replicas for the first two layers,
\begin{equation}
  \begin{array}{lcccccc}
    \gamma  & 0.001 & 1 & 2 & 3 & 4 & 5 \\[2pt]
    s^{(1)} & 0.732 & 0.453 & 0.162 & 0.022 & 0.002 & 0.000 \\[2pt]
    s^{(2)} & 0.806 & 0.245 & 0.008 & 0.000 & 0.000 & 0.000
  \end{array}\, ,
  \label{eq:SaturationInit}
\end{equation}
which provide a more compact summary of the saturation at initialisation. Taken
together, \cref{fig:InitSaturation} and \cref{eq:SaturationInit} support our
previous observations: for $\gamma \gtrsim 3$ the saturation is negligible
across the entire $x$ range, whereas for smaller values of $\gamma$ the
saturation is significant, particularly at small $x$ where the $\log x$ input is
largest. Furthermore, the layer-by-layer account shows that the saturation of
the second layer is larger than that of the first layer for $\gamma \geq 1$,
whereas the opposite is true for $\gamma = 0.001$. 

The saturation of the internal layers is thus closely related to the regime in
which the network operates. In fact, we can identify a non-linear regime for
$\gamma \lesssim 2$, where the saturation is significant, and a linear regime
for $\gamma \gtrsim 3$, where the saturation is negligible. However, we believe
that between these two extreme cases there might be a family of intermediate
regimes, where small but non-negligible saturations allow the activation
function to be expanded in a Taylor series up to some fixed order. Here we do
not explore this possibility further, but leave it for future work.

\subsection{The span of the tangent plane}
It can be easily verified from \cref{eq:GenLinearNN} that the derivatives of the
network with respect to the parameters (the \emph{tangent vectors} of
\cref{sec:geometric_interpretation}) can be written as
\begin{equation}
  J^{(\rm net)}_{i'\alpha,\mu} \equiv \partial_{\mu} \preact{L}_{i'\alpha}
  = u_{\mu i'} + v_{\mu i'}\,x_{\alpha} + w_{\mu i'} \log x_{\alpha}\,,
  \quad
  \mu=1,\dots, \nparams\,,
  \label{eq:TangentFeaturesDLN}
\end{equation}
where we have restored the indices for the sake of clarity. Here we use primed
indices to label the $\nout=8$ outputs of the network. The same form holds for
the full model in \cref{eq:nnpdf_model}, whose $\nf$ flavours are obtained from
the outputs through the rotation $\rfk\in\real^{\nout\times\nf}$ from the
fitting basis to the FK basis and by multiplying by the large- and small-$x$
preprocessing factors. We can then write
\begin{equation}
  J_{i\alpha,\mu} =
  \sum_{i'} \left(\rfk\right)_{i'i} \, x_{\alpha}^{1-\delta_{i'}} (1-x_{\alpha})^{\epsilon_{i'}} \, J^{(\rm net)}_{i'\alpha,\mu}\,,
  \label{eq:TangentFeaturesDLNModel}
\end{equation}
where the un-primed indices label the $\nf=14$ outputs of the full model. It is
easily verified that when the preprocessing factors are set to one,
\ie\ $\delta_{i'}=1$ and $\epsilon_{i'}=0$,
\cref{eq:TangentFeaturesDLNModel} is again a combination of constant and
non-constant terms whose coefficients are linear in $u$, $v$ and $w$. Note that
the rows of $\rfk$ are the $\nout$ independent flavours and therefore are all
disjoint, and the Gram matrix is simply diagonal
\begin{equation}
  \rfk \left(\rfk\right)^T = \mathrm{diag}(m_{i'})\,,
  \quad i'=1,\dots,\nout\,.
  \label{eq:FKBasisGram}
\end{equation}
Here $m_{i'}$ counts the degeneracy of the fitted flavours
once mapped onto the full evolution basis of the FK tables, which sets
\begin{equation}
  \begin{split}
    &V = V_{15} = V_{24} = V_{35}\,,\\
    &\Sigma = T_{24} = T_{35}\,,\\
    &\mathrm{photon} = 0\,,
  \end{split}
  \label{eq:FlavReps}
\end{equation}
and all the other flavours unchanged. We then conclude that $m_{V}=4$,
$m_{\Sigma}=3$, and one for the other fitted flavours. In the following, we work
with the tangent vectors of the network for the sake of simplicity, and apply
\cref{eq:TangentFeaturesDLNModel} when needed.

The tangent vectors in \cref{eq:TangentFeaturesDLN} are generated by derivatives
with respect to three groups of parameters:
\begin{enumerate}
  \item[\textbf{G1})] The $\nout$ biases of the last layer, which only contribute to the constant
  terms $u_{\mu i}$. The tangent vectors in this class can be written as
  \begin{equation}
    \frac{\partial \preact{L}_{i'\alpha}}{\partial \bias{L}_{j'}} = \delta_{i'j'}\,,
    \label{eq:TanVecBiasL}
  \end{equation}
  that is, the identity $\id_{\nout}$. Note that these vectors are independent of
  $\gamma$ and are non-zero even for vanishing biases.
  \item[\textbf{G2})] The biases of the hidden layers $\bias{\ell}$, which contribute to the constant term only.
  Their tangent vectors are
  \begin{equation}
    \frac{\partial \preact{L}_{i'\alpha}}{\partial \bias{\ell}_{j}} 
    = \left[\wght{L}\dots\wght{\ell+1}\right]_{i'j}
    = r^{\gamma}_{\ell+1} \left[\uwght{L}\cdots\uwght{\ell+1}\right]_{i'j}\,,
    \quad
    \ell \neq L\,,
    \label{eq:TanVecBiasLHidden}
  \end{equation}
  where we made explicit the scaling with $\gamma$ by using $\wght{\ell} =
  \sigma_{\ell}\uwght{\ell}$ and defining $r_{\ell}^2 = \prod_{m=\ell}^{L}
  2/(n_{m}+n_{m-1})$ such that $r_{L-1} > r_{L-2} > \dots > r_1 \equiv r$,
  \cref{eq:InitAmplitude}. These vectors are also non-zero for vanishing biases,
  but the chain of weights introduces a dependence on $\gamma$.
  \item[\textbf{G3})] The weights of the model for all layers, $\wght{\ell}$. In this case the
  derivatives generate contributions for both constant and non-constant terms of
  \cref{eq:TangentFeaturesDLN}. The tangent vectors are
  \begin{equation}
    \begin{split}
      \frac{\partial \preact{L}_{i'\alpha}}{\partial \wght{\ell}_{jk}}
      &= \sum_{\ell'=1}^{\ell-1} \sum_{p=1}^{n_{\ell'}}
        \left[\wght{L}\dots \wght{\ell+1} \right]_{i'j}
        \left[\wght{\ell-1}\dots \wght{\ell'+1} \right]_{kp}
        \bias{\ell'}_{p} \\
      & \hspace{50mm} + \sum_{p=1}^{n_{\rm in}}\left[\wght{L}\dots \wght{\ell+1} \right]_{i'j}
        \left[\wght{\ell-1}\dots \wght{1} \right]_{kp} X_{p\alpha} \\
      &= \sum_{\ell'=1}^{\ell-1} \frac{r^{\gamma}_{\ell'+1}}{\sigma_{\ell}} \sum_{p=1}^{n_{\ell'}}
        \left[\uwght{L}\dots \uwght{\ell+1} \right]_{i'j}
        \left[\uwght{\ell-1}\dots \uwght{\ell'+1} \right]_{kp}
        \bias{\ell'}_{p} \\
      & \hspace{50mm} + \frac{r^{\gamma}}{\sigma_{\ell}}
        \sum_{p=1}^{n_{\rm in}}\left[\uwght{L}\dots \uwght{\ell+1} \right]_{i'j}
        \left[\uwght{\ell-1}\dots \uwght{1} \right]_{kp} X_{p\alpha}\,,
    \end{split}
    \label{eq:TanVecWght}
  \end{equation}
  where we made the dependence on $\gamma$ explicit in the second equality. Note
  that when biases vanish, the constant term in \cref{eq:TanVecWght} cancels
  out.
\end{enumerate}
We conclude by observing that \cref{eq:TangentFeaturesDLN} serves as an example
to demonstrate the difference between the span of the tangent plane and the
frame induced by the parametrisation. Here, the frame is made of the
$\nparams=\parameters$ vectors of the form above, but they all lie in the
three-dimensional subspace $\vlin$ of \cref{eq:SpanLinearFeatures}, which is the
span. Since the eigenvectors of the NTK are linear combinations of the tangent
vectors, they are an aggregation of these directions and are therefore contained
in $\vlin$ as well. As we now show in more detail, this is what constrains the
dimension of $\ntkimg$ and fixes the functional form of the modes $q^{(r)}$ of
\cref{eq:hamiltonian_eigvec_pdf} for the DLN.

\subsection{The NTK of a deep linear network}
In the DLN case the computation of the NTK is analytically tractable. For the
sake of simplicity, in the derivation we neglect preprocessing factors. They can
be added back in using \cref{eq:TangentFeaturesDLNModel}.

We start by separating the $x$-dependence of \cref{eq:TangentFeaturesDLN} from
its coefficients $u$, $v$ and $w$, which can be collected into a single matrix
$\coeff\in\real^{\nparams \times 3\nout}$. Here the rows of $\mathcal{C}$ run
over the parameter index $\mu=1,\dots,\nparams$ and the columns are labelled by
the composite index $(r,i')$ in \emph{input-major} ordering, the input index $r$
being the slowest,
\begin{equation}
  (r,\, i') \;\longmapsto\; \nout(r-1) + i'\,,
  \qquad r = 1,\, 2,\, 3\,,
  \quad i' = 1,\dots,\nout\,,
  \label{eq:FeatureMajor}
\end{equation}
where primed indices label the outputs of the network as in
\cref{eq:TangentFeaturesDLN}. Note that this is not the flavour-major convention
defined in \cref{sec:dataset_method} where the pair $(i,\alpha)$ labels the
vectorised PDF space. This choice makes the structure of the NTK more
transparent, as we now show. With \cref{eq:FeatureMajor} the matrix $\coeff$ can
be defined by stacking side-by-side the three matrices,
\begin{equation}
  \coeff = \left[\, u \;\big|\; v \;\big|\; w \,\right]\,,
  \qquad
  u,\,v,\,w \in \real^{\nparams\times\nout}\,,
  \label{eq:CoeffTensor}
\end{equation}
where each block collects the coefficients for all $\nout$ flavours of one of
the functions spanning $\vlin$, \cref{eq:SpanLinearFeatures}.

In order to introduce the $x$-dependence of \cref{eq:TangentFeaturesDLN}, we
define the matrix of \emph{input variables} $\Phi\in\real^{\ngrid\times 3}$ as
the juxtaposition of the three functions spanning $\vlin$
\begin{equation}
  \Phi = \left[\, \phi_1 \;\big|\; \phi_2 \;\big|\; \phi_3 \,\right]\,,
  \qquad
  \phi_1,\,\phi_2,\,\phi_3 \in \real^{\ngrid}\,,
  \label{eq:InputFeatures}
\end{equation}
where each $\phi_r$ is a column vector containing the $r$-th input function of
$\vlin$ evaluated on the $x$-grid. The basis $\Phi$ is the same for each output
of the network, so that we define its extension to the $\nout$ outputs as
\begin{equation}
    \Psi^{(\rm net)}
    = \left[\, \id_{\nout}\!\otimes\phi_{1} \;\big|\;
               \id_{\nout}\!\otimes\phi_{2} \;\big|\;
               \id_{\nout}\!\otimes\phi_{3} \,\right]
      \in\real^{\nout\ngrid\times 3\nout}\,,
  \label{eq:PsiNet}
\end{equation}
where each block of \cref{eq:PsiNet} is a block-diagonal matrix with elements
\begin{equation}
   \id_{\nout}\!\otimes\phi_{r} = \diag\left(\phi_r,\, \cdots,\, \phi_r\right)_{\nout} \,,
   \qquad
   r= 1,\, 2,\, 3 \,.
   \label{eq:PsiNetBlock}
\end{equation}
Equivalently, adopting the input-major notation of \cref{eq:FeatureMajor},
\cref{eq:PsiNet} can be expressed as
\begin{equation}
  \Psi^{(\rm net)}_{i'\alpha,rj'} = \delta_{i'j'}\Phi_{\alpha r}\,.
  \label{eq:PsiNetIndex}
\end{equation}
The rotation to the $\nf$-dimensional FK basis is achieved by first expanding the rotation
$\rfk$ of \cref{eq:TangentFeaturesDLNModel} to all points in the grid
\begin{equation}
  \emb = \left(\rfk\right)^T \otimes \id_{\ngrid}
  \in\real^{\nvec\times\nout\ngrid}\,,
  \label{eq:FlavEmbedding}
\end{equation}
and then by applying the resulting matrix to \cref{eq:PsiNetIndex}, which yields
\begin{equation}
  \Psi
  = \emb\,\Psi^{(\rm net)}
  = \left[\, \left(\rfk\right)^T\!\otimes\phi_{1} \;\big|\;
             \left(\rfk\right)^T\!\otimes\phi_{2} \;\big|\;
             \left(\rfk\right)^T\!\otimes\phi_{3} \,\right]
  \in\real^{\nvec\times 3\nout}\,,
  \label{eq:Psi}
\end{equation}
where each block follows from the mixed-product rule
\begin{equation}
  \left( A \otimes B \right) \left( C \otimes D \right) 
  = \left(AC \right) \otimes \left(BD\right)\,,
  \label{eq:KroneckerRule}
\end{equation}
and can be seen as the application of the rotation matrix $\rfk$ to the block
\cref{eq:PsiNetBlock}.

With this notation, it can be easily verified that the tangent vectors factorise into the following
product
\begin{equation}
  J = \Psi \coeff^T 
    = \emb\,\Psi^{(\rm net)} \coeff^T \,,
  \label{eq:JacDLNFact}
\end{equation}
and the NTK is then expressed in the convenient form
\begin{equation}
  \ntk = JJ^T = \Psi \coeff^T \coeff \Psi^T = \Psi \mathcal{G} \Psi^T\,.
  \label{eq:NTKDLN}
\end{equation}
Here $\mathcal{G} = \coeff^T \coeff\in\real^{3\nout\times3\nout}$ is the matrix
of the coefficients in \cref{eq:CoeffTensor} contracted over the parameter
index. With the vectorisation convention adopted in \cref{eq:CoeffTensor}, we
can express this matrix as
\begin{equation}
  \mathcal{G} \equiv
  \left[\begin{matrix}
      \,\mathcal{G}_{11} & \mathcal{G}_{12} & \mathcal{G}_{13} \,\\
      \,\mathcal{G}_{21} & \mathcal{G}_{22} & \mathcal{G}_{23} \,\\
      \,\mathcal{G}_{31} & \mathcal{G}_{32} & \mathcal{G}_{33} \,\\
  \end{matrix}\right]
  =
  \left[\begin{array}{c}
    u^T \\
    v^T \\
    w^T
  \end{array}\right]
  \;
  \left[\, u \;\big|\; v \;\big|\; w \,\right]
  = \left[\begin{matrix}
      \,u^T u & u^T v & u^T w \,\\
      \,v^T u & v^T v & v^T w \,\\
      \,w^T u & w^T v & w^T w \,\\
  \end{matrix}\right]\,,
  \qquad
  \mathcal{G}_{rs} \in \real^{\nout\times \nout}\,,
  \label{eq:MatG}
\end{equation}
which ultimately motivates our choice of the input-major ordering: taking the
input index as the slowest is what turns $\mathcal{G}$ into a $3\times3$ array
of blocks of size $\nout$, labelled by the functions spanning $\vlin$.

By introducing an auxiliary vector of coefficients
$\zeta\in\real^{3\nout}$, whose components are labelled by the same composite
index of \cref{eq:FeatureMajor},
\begin{equation}
  \zeta = \left( \zeta_{ri'} \right)\,,
  \qquad r = 1,\, 2,\, 3\,,
  \quad i' = 1,\dots,\nout\,,
  \label{eq:ZetaCoeffs}
\end{equation}
and writing the NTK eigenvectors as $z = \Psi \zeta$, we can recast the
eigenvalue problem $\ntk z = \lambda z$ in the following form
\begin{equation}
  \mathcal{G}S\zeta = \lambda \zeta\,,
  \label{eq:NTKEigProbl}
\end{equation}
where we used \cref{eq:KroneckerRule} and \cref{eq:FKBasisGram} to define
\begin{equation}
  S = \Psi^T\Psi
    = \Phi^T\Phi \otimes \rfk\left(\rfk\right)^T
    = S_3 \otimes \mathrm{diag}\left(m_{i'}\right)_{\nout}
  \in\real^{3\nout\times 3\nout}\,,
  \label{eq:SMatrix}
\end{equation}
where $m_{i'}$ are the multiplicities of \cref{eq:FKBasisGram} and $S_3$ is the
matrix of grid moments
\begin{equation}
  S_3
  = \Phi^T \Phi 
  = \ngrid \begin{pmatrix}
    1 & \langle x \rangle & \langle \log x \rangle \\
    \langle x \rangle & \langle x^2 \rangle & \langle x\log x \rangle \\
    \langle \log x \rangle & \langle x\log x \rangle & \langle \log^2 x \rangle
  \end{pmatrix}\,,
  \quad
  \langle g(x) \rangle \equiv \frac{1}{\ngrid}\sum_{\alpha=1}^{\ngrid} g(x_{\alpha}) \,.
  \label{eq:GridMoments}
\end{equation}
Note that, similarly to $\mathcal{G}$, $S$ is a $3\times3$ block of
$\nout\times\nout$ blocks, the $(r,s)$ block being
\begin{equation}
  S_{rs} = \left(S_3\right)_{rs}\mathrm{diag}(m_{i'})_{\nout}\,.
  \label{eq:SMatrixBlocks}
\end{equation}
We pause here to make a few comments.

\paragraph{Factorisation.}
The NTK in \cref{eq:NTKDLN} factorises into a quadratic form where $\Psi$
depends only on the grid, on the input functions in $\Phi$, and on the flavour
map $\rfk$, whereas the dependence on the parameters is carried entirely by the
matrix $\mathcal{G}$. In a DLN, the NTK thus changes only through the latter,
while the span of the tangent vectors is fixed by $\Psi$: the orientation of the
tangent plane remains constant. It also follows that the NTK is invariant along
any direction in parameter space that leaves $\mathcal{G}$ unchanged, and that
such directions are many: being symmetric of size $3\nout$, $\mathcal{G}$ has at
most $3\nout(3\nout+1)/2 = 300$ independent entries, against the
$\nparams=\parameters$ parameters of the network. At least $463$ directions
therefore leave the NTK unchanged, and are inner symmetries of the
parametrisation.

\paragraph{Reduced NTK.}
The $\nvec$-dimensional NTK eigenvalue problem reduces to the lower-dimensional
problem defined by \cref{eq:NTKEigProbl}. The non-zero eigenvalues -- and the
corresponding eigenvectors -- are then completely specified by the $3\times\nout
= 24$ equations encoded in \cref{eq:NTKEigProbl}. Note that although
$\mathcal{G}S$ is not symmetric, it is similar to $S^{1/2}\mathcal{G}S^{1/2}$,
so that its eigenvalues are real and non-negative, and the $\zeta$ are
orthogonal with respect to $S$ -- the NTK eigenvectors $z=\Psi\zeta$ are
orthogonal. Therefore, the whole non-zero spectrum of the NTK is obtained by
diagonalising a $24\times24$ matrix rather than the $\nvec\times\nvec$ one.

\paragraph{Different set of input functions.}
The expressiveness of a DLN is uniquely determined by the input functions that
make up $\Phi$. In our case, these are simple functions and therefore the linear
regime can only approximate curves that are proportional to $x$ and $\log
x$.\footnote{Note that, when preprocessing factors are included, the
expressiveness changes accordingly as in \cref{eq:TangentFeaturesDLNModel}.}
However, nothing prevents us from using more complicated input functions. In
this case, the generalisation of \cref{eq:InputFeatures} to arbitrary $n_1$
input functions becomes
\begin{equation}
  \Phi = \left[\, \phi_1 \;\big|\; \phi_2 \;\big|\; \cdots \;\big|\; \phi_{n_1} \,\right]\,,
  \qquad
  \phi_r \in \real^{\ngrid}\,,
  \qquad
  r=1,\,\cdots,\,n_1\,,
\end{equation}
where $\phi_r$ is the $r$-th basis function in $x$ (\eg, polynomials).
The DLN -- and the corresponding tangent vectors -- then become equivalent to
any fixed functional form that is a linear combination of basis functions
weighted by free parameters, that is
\begin{equation}
  f(x;\, \theta)  = \sum_{r=1}^{n_1} \theta_r \Phi^{(r)}(x)\,,
  \label{eq:FixedForm}
\end{equation}
which is very similar to the expansion used in Ref.~\cite{Costantini:2025wxp}.
The only difference between the DLN case and \cref{eq:FixedForm} is that, in the
former, there are many equivalent combinations of weights and biases that
realise the same weight $\theta_r$. Therefore, even though $\vlin$ is the same,
the set of tangent vectors that two almost-equivalent models realise is
intrinsically different, affecting the learning dynamics and the way the model
explores neighbouring regions.

\subsection*{The reduced matrix $\mathcal{G}$}
Turning $\mathcal{G}$ and $S$ into block matrices arranged by input functions
allows us to decompose the eigenvalue problem of \cref{eq:NTKEigProbl} in the following
form
\begin{equation}
  \begin{bmatrix}
    \mathcal{G}_{cc} & \mathcal{G}_{cn} \\
    \mathcal{G}_{nc} & \mathcal{G}_{nn}
  \end{bmatrix}
  \begin{bmatrix}
    S_{cc} & S_{cn} \\
    S_{nc} & S_{nn}
  \end{bmatrix}
  \begin{bmatrix}
    \zeta_{c} \\
    \zeta_{n}
  \end{bmatrix}
  = \lambda
  \begin{bmatrix}
    \zeta_{c} \\
    \zeta_{n}
  \end{bmatrix}\,,
  \label{eq:NTKEigProblBlocks}
\end{equation}
where the blocks of $\zeta$, $S$, and $\mathcal{G}$ are
\begin{gather}
  \zeta_{c} = \zeta_1\,,
  \qquad
  \zeta_{n} = \begin{bmatrix} \zeta_2 \\ \zeta_3\end{bmatrix}\,,
  \label{eq:ZBlockDefs}
  \\[1em]
  S_{cc} = S_{11}\,,
  \qquad
  S_{nn} = \begin{bmatrix}
    S_{22} & S_{23} \\
    S_{32} & S_{33} \\
  \end{bmatrix}\,,
  \qquad
  S_{cn} = \begin{bmatrix} S_{12} & S_{13} \end{bmatrix} \equiv S_{nc}^T\,,
  \label{eq:SBlockDefs}
  \\[1em]
  \mathcal{G}_{cc} = \mathcal{G}_{11}\,,
  \qquad
  \mathcal{G}_{nn} = \begin{bmatrix}
    \mathcal{G}_{22} & \mathcal{G}_{23} \\
    \mathcal{G}_{32} & \mathcal{G}_{33} \\
  \end{bmatrix}\,,
  \qquad
  \mathcal{G}_{cn} = \begin{bmatrix} \mathcal{G}_{12} & \mathcal{G}_{13} \end{bmatrix} \equiv \mathcal{G}_{nc}^T \,.
  \label{eq:GBlockDefs}
\end{gather}
We now compute the entries of $\mathcal{G}$.

\paragraph{Off-diagonal terms.}
The parameter index runs over the three classes of parameters defined above --
last- and hidden-layer biases and the weights -- so that each of the three
matrices in \cref{eq:CoeffTensor} splits by rows as
\begin{equation}
  u = \begin{bmatrix} u^{(\rm b_{L})} \\ u^{(\rm b_{h})} \\ u^{(\rm w)} \end{bmatrix}\,,
  \qquad\text{and likewise for } v \text{ and } w\,,
  \label{eq:CoeffByClass}
\end{equation}
the blocks having $\nout$, $\sum_{\ell<L} n_{\ell}$ and $\sum_{\ell} n_{\ell}n_{\ell-1}$
rows respectively. Since the parameter index is the one contracted in
$\mathcal{G} = \coeff^T\coeff$, every block of $\mathcal{G}$ is a sum of three
terms, one per class. We can then write
\begin{equation}
  \mathcal{G}_{21} = v^{T}u
  = \left(v^{(\rm b_{L})}\right)^{T}\! u^{(\rm b_{L})}
  + \left(v^{(\rm b_{h})}\right)^{T}\! u^{(\rm b_{h})}
  + \left(v^{(\rm w)}\right)^{T}\! u^{(\rm w)}\,.
  \label{eq:Gcn}
\end{equation}
From \textbf{G1} and \textbf{G2}, the last- and hidden-layer biases can only
generate constant terms, implying $v^{(\rm b_{L})}=v^{(\rm b_{h})}=0$.
Therefore, only the last term in \cref{eq:Gcn} survives. However, since $u^{(\rm
w)}$ corresponds to the first term in \cref{eq:TanVecWght}, when biases are set
to zero at initialisation this term also disappears. We can then conclude that
\begin{equation}
  \mathcal{G}_{12} = \mathcal{G}_{21} = 0\,,
  \qquad \mathrm{if}
  \quad \bias{\ell} = 0 \; \forall \,\ell\,.
\end{equation}
Exactly the same argument applies to $\mathcal{G}_{13} = w^{T}u$, since $w$ too
is generated by the weights alone, so that at initialisation the whole
off-diagonal block vanishes,
\begin{equation}
  \mathcal{G}_{cn} = \mathcal{G}_{nc}^{T} = 0\,.
  \label{eq:GcnVanishes}
\end{equation}

\paragraph{Constant term.}
The same decomposition applied to the diagonal block of the constant sector
gives
\begin{equation}
  \mathcal{G}_{cc} = \mathcal{G}_{11} = u^{T}u
  = \left(u^{(\rm b_{L})}\right)^{T}\! u^{(\rm b_{L})}
  + \left(u^{(\rm b_{h})}\right)^{T}\! u^{(\rm b_{h})}
  + \left(u^{(\rm w)}\right)^{T}\! u^{(\rm w)}\,,
  \label{eq:GccByClass}
\end{equation}
where the last term vanishes at initialisation for the reason given above,
provided that biases are set to zero. The first one is fixed by \textbf{G1}:
\cref{eq:TanVecBiasL} gives $u^{(\rm b_{L})} = \id_{\nout}$, so that the $\nout$
biases of the last layer contribute the identity, with no dependence on $\gamma$
-- no chain of weights stands between them and the output. The second one
follows from \textbf{G2} and \cref{eq:TanVecBiasLHidden}, each hidden layer
contributing the Gram matrix of its own chain. Altogether
\begin{equation}
  \mathcal{G}_{cc}
  = \id_{\nout}
  + \sum_{\ell=1}^{L-1} r^{2\gamma}_{\ell+1}
    \left[\uwght{L}\cdots\uwght{\ell+1}\right]
    \left[\uwght{L}\cdots\uwght{\ell+1}\right]^{T}\,.
  \label{eq:GccBlock}
\end{equation}
The correction is exponentially small in $\gamma$, and is dominated by the
shortest chain $\ell = L-1$, of order $n_{L-1}r^{2\gamma}_{L}$. Therefore, the
constant block approaches the identity as the initialisation width shrinks.

\paragraph{Non-constant terms.}
The blocks with $r,\,s \neq 1$ receive contributions from the weights alone,
since $v$ and $w$ vanish on the biases. At initialisation -- when biases are set
to zero -- only the second line of \cref{eq:TanVecWght} survives. Denoting by
\begin{equation}
  \uchain{>}{\ell} = \uwght{L}\cdots\uwght{\ell+1}\,,
  \qquad
  \uchain{<}{\ell} = \uwght{\ell-1}\cdots\uwght{1}\,,
  \label{eq:ChainsOmXi}
\end{equation}
the two chains of rescaled weights that the derivative leaves above and below
the differentiated layer $\ell$, both reducing to the identity when empty, one
finds
\begin{equation}
  \mathcal{G}_{rs}
  = \sum_{\ell=1}^{L} \frac{r^{2\gamma}}{\sigma^{2}_{\ell}}\,
    \left(\uchain{<}{\ell}^{T}\,\uchain{<}{\ell}\right)_{r-1,\,s-1}\;
    \uchain{>}{\ell}\,\uchain{>}{\ell}^{T}\,,
  \qquad r,\,s = 2,\, 3\,,
  \label{eq:GnnBlock}
\end{equation}
the shift by one accounting for the constant direction $r=1$, which has no
counterpart among the input variables $X_{p\alpha}$. The prefactor is
\begin{equation}
  \frac{r^{2\gamma}}{\sigma^{2}_{\ell}} = \prod_{m\neq\ell} \rho^{\gamma}_{m}\,,
  \qquad
  \rho_{m} = \frac{2}{n_{m}+n_{m-1}}\,,
  \label{eq:GnnPrefactor}
\end{equation}
that is, the product of all the layer variances but the one being
differentiated. Every entry of $\mathcal{G}_{nn}$ therefore carries $L-1$ chains
of weights, and is suppressed by a further power of $\rho^{\gamma}$ with respect
to the correction in \cref{eq:GccBlock}.

\paragraph{The matrix $\mathcal{G}$ at initialisation.}
Collecting the three results, at initialisation $\mathcal{G}$ is block diagonal
in the constant and non-constant sectors, with
\begin{equation}
  \mathcal{G}_{cc} = \id_{\nout} + \mathcal{O}\!\left(r^{2\gamma}_{L}\right)\,,
  \qquad
  \mathcal{G}_{cn} = 0 \,,
  \qquad
  \mathcal{G}_{nn} = \mathcal{O}\!\left(\prod_{m\neq\ell}\rho^{\gamma}_{m}\right)\,.
  \label{eq:GOrders}
\end{equation}
The two sectors are thus separated by the initialisation width alone: as
$\gamma$ grows, $\mathcal{G}_{cc}$ tends to the identity while
$\mathcal{G}_{nn}$ is driven to zero exponentially. The matrix governing
\cref{eq:NTKEigProblBlocks} then factorises as
\begin{equation}
  \mathcal{G}S =
  \begin{bmatrix} \mathcal{G}_{cc} & 0 \\ 0 & \mathcal{G}_{nn} \end{bmatrix}
  \begin{bmatrix} S_{cc} & S_{cn} \\ S_{nc} & S_{nn} \end{bmatrix}
  =
  \begin{bmatrix}
    \mathcal{G}_{cc}S_{cc} & \mathcal{G}_{cc}S_{cn} \\
    \mathcal{G}_{nn}S_{nc} & \mathcal{G}_{nn}S_{nn}
  \end{bmatrix}\,.
  \label{eq:GSBlocks}
\end{equation}
The two diagonal blocks are of order unity and of order
$\prod_{m\neq\ell}\rho^{\gamma}_{m}$ respectively. The upper-right block does
not vanish at any $\gamma$, since $S_{cn}\neq0$ in general. However, for the
lower-left block we have
\begin{equation}
  \left(\mathcal{G}S\right)_{nc} = \mathcal{G}_{nn}S_{nc}
  = \mathcal{O}\!\left(\prod_{m\neq\ell}\rho^{\gamma}_{m}\right)\,,
  \label{eq:GSlower}
\end{equation}
which is negligible against the block of order unity sitting above it. To that
accuracy $\mathcal{G}S$ is upper triangular, and its leading $\nout$ eigenvalues
are those of the first diagonal block,
\begin{equation}
  \spec{\mathcal{G}S}\Big|_{\mathcal{O}(1)} = \spec{\mathcal{G}_{cc}S_{cc}}\,.
  \label{eq:GSpecFact}
\end{equation}
The same argument does \emph{not} apply to the remaining $2\nout$ eigenvalues:
the lower-left block of \cref{eq:GSlower} carries one factor of
$\mathcal{G}_{nn}$, exactly as the lower-right one does. Those eigenvalues
require the elimination of $\zeta_{c}$. We now compute the two sectors in turn.

\subsubsection*{The constant sector}
By \cref{eq:GSpecFact} the $\nout$ eigenvalues of order unity are those of
$\mathcal{G}_{cc}S_{cc}$ alone. Using \cref{eq:SMatrixBlocks} for $S_{cc} =
\ngrid\,\mathrm{diag}(m_{i'})$ and \cref{eq:GccBlock} for $\mathcal{G}_{cc}$,
they are read off immediately,
\begin{equation}
  \lambda^{(c)}_{i'} = m_{i'}\ngrid
    \left[1 + \mathcal{O}\!\left(r^{2\gamma}_{L}\right)\right]\,,
  \qquad i' = 1,\dots,\nout\,,
\end{equation}
that is, in the limit of vanishing initialisation width,
\begin{equation}
  \left\{\lambda^{(c)}_{i'}\right\} = \left\{m_{i'}\ngrid\right\}_{i'=1,\dots,\nout}
  = \left\{200,\, 150,\, 50,\,50,\,50,\,50,\,50,\,50\right\}\,,
  \label{eq:NTKBiasSpecEnd}
\end{equation}
which reproduces, once normalised to the largest eigenvalue, the first tier of
the spectrum observed in \cref{fig:EigValsGammaRank}. The multiplicities
$m_{i'}$ of \cref{eq:FKBasisGram} are therefore the only structure in this
tier: they come from the flavour map $\rfk$, not from the network.

The eigenvectors follow at once. To leading order $\mathcal{G}_{cc}=\id_{\nout}$
and $S_{cc}$ is diagonal, so that $\zeta_{c}$ is the $i'$-th unit vector and
$\zeta_{n}=0$. Projecting back with $z = \Psi\zeta$ and using the first column
block of \cref{eq:Psi}, the normalised modes are
\begin{equation}
  z^{(i')}_{i\alpha} = \frac{\left(\rfk\right)_{i'i}}{\sqrt{m_{i'}\ngrid}}\,,
  \label{eq:NTKBiasEigvec}
\end{equation}
independent of the grid point $\alpha$: the modes of the leading tier are
constant in $x$. Their normalisation and orthogonality follow from
$\left(z^{(i')},z^{(j')}\right) = \zeta^{(i')T}S\,\zeta^{(j')}$, that is
\begin{equation}
  \left(z^{(i')},\,z^{(j')}\right)
  = \frac{\left(S_{cc}\right)_{i'j'}}{m_{i'}\ngrid} = \delta_{i'j'}\,.
  \label{eq:ScalarBiasFeatures}
\end{equation}

\subsubsection*{The non-constant sector}
\label{sec:nonconstant}
The remaining $2\nout$ eigenvalues are those whose eigenvectors have a
non-constant component of order unity. For these the second block row of
\cref{eq:NTKEigProblBlocks}, $\mathcal{G}_{nn}\left(S_{nc}\zeta_{c} +
S_{nn}\zeta_{n}\right) = \lambda\zeta_{n}$, carries $\mathcal{G}_{nn}$ as an
overall prefactor, so that $\lambda = \mathcal{O}(\mathcal{G}_{nn})$ and the
eigenvalues are suppressed with respect to those of \cref{eq:NTKBiasSpecEnd}.
The first block row can then be solved for $\zeta_{c}$: it reads
$\mathcal{G}_{cc}\left(S_{cc}\zeta_{c} + S_{cn}\zeta_{n}\right) =
\lambda\zeta_{c}$, whose right-hand side is negligible against
$\mathcal{G}_{cc}S_{cc} = \mathcal{O}(1)$, and cancelling the invertible
$\mathcal{G}_{cc}$ leaves the constant component slaved to the non-constant one,
\begin{equation}
  \zeta_{c} = - S_{cc}^{-1}S_{cn}\,\zeta_{n}\,.
  \label{eq:ZetaCSlaved}
\end{equation}
The two components are therefore of the same order, as anticipated. Substituting
\cref{eq:ZetaCSlaved} into the second block row leaves a problem of dimension
$2\nout$,
\begin{equation}
  \mathcal{G}_{nn}\,\widehat{S}\,\zeta_{n} = \lambda\,\zeta_{n}\,,
  \label{eq:NonConstEigProbl}
\end{equation}
where the elimination has replaced $S_{nn}$ by its Schur complement, which
factorises as in \cref{eq:SMatrixBlocks},
\begin{equation}
  \widehat{S} = S_{nn} - S_{nc}S_{cc}^{-1}S_{cn}
              = \widehat{S}_{2}\otimes\mathrm{diag}(m_{i'})_{\nout}\,,
  \qquad
  \widehat{S}_{2}
  = \ngrid \begin{pmatrix}
    \langle x^{2}\rangle - \langle x\rangle^{2}
      & \langle x\log x\rangle - \langle x\rangle\langle\log x\rangle \\
    \langle x\log x\rangle - \langle x\rangle\langle\log x\rangle
      & \langle \log^{2}x\rangle - \langle\log x\rangle^{2}
  \end{pmatrix}\,.
  \label{eq:CentredMoments}
\end{equation}
The grid moments of \cref{eq:GridMoments} are thus replaced by \emph{centred}
moments.

Recall that the blocks of $\mathcal{G}_{nn}$ are given by \cref{eq:GnnBlock}.
Collected into a single matrix, each layer contributes one Kronecker product of
a $2\times2$ input-space factor and an $\nout\times\nout$ flavour-space factor,
\begin{equation}
  \mathcal{G}_{nn}
  = r^{2\gamma} \sum_{\ell=1}^{L} \frac{1}{\sigma^{2}_{\ell}}\,
    A_{\ell} \otimes B_{\ell}\,,
  \qquad
  A_{\ell} \equiv \Omega_{<\ell}^T\Omega_{<\ell}\in\real^{2\times2}\,,
  \qquad
  B_{\ell} \equiv \Omega_{>\ell}\Omega_{>\ell}^T\in\real^{\nout\times \nout}\,.
  \label{eq:GnnBlock2}
\end{equation}
Note that the matrix $\widehat{S}$ in \cref{eq:NonConstEigProbl} has the same
tensor structure as each term of the sum in \cref{eq:GnnBlock2}, but the sum
over $\ell$ does not factorise in general. However, because each term is built
from nested chains of weights, the factorisation holds approximately. To show
this, we note that both $A_{\ell}$ and $B_{\ell}$ are obtained by inserting one
more layer into the chain of weights, that is
\begin{equation}
  A_{\ell+1} = \Omega_{<\ell}^T\left(\Omega^{(\ell)T}\Omega^{(\ell)}\right)\Omega_{<\ell}\,,
  \qquad
  B_{\ell-1} = \Omega_{>\ell}\left(\Omega^{(\ell)}\Omega^{(\ell)T}\right)\Omega_{>\ell}^T\,.
  \label{eq:ABRecursion}
\end{equation}
Since $\Omega^{(\ell)}$ is independent of the chain of weights of the previous layers, conditioning
on the latter yields a recursive structure
\begin{equation}
  \expct\left[ A_{\ell+1} \,\big|\, \Omega_{<\ell}\right]
  = \Omega_{<\ell}^T \expct\left[\Omega^{(\ell)T}\Omega^{(\ell)}\right]\Omega_{<\ell}
  =  n_\ell A_{\ell}\,,
  \label{eq:ExpctARec}
\end{equation}
where in the second equality we used the fact that the entries of $\uwght{\ell}$
are unit normal. The fluctuation around this average,
$\Delta A_{\ell+1} \equiv A_{\ell+1} - n_{\ell}A_{\ell}$, has zero conditional
mean, and its conditional variance follows from Wick contractions, since
$\uwght{\ell}$ is i.i.d.\ Gaussian:
\begin{equation}
  \expct\left[ \left(\Delta A_{\ell+1}\right)_{ab}^2 \,\big|\, \Omega_{<\ell} \right]
  = n_{\ell}
    \left[ (A_{\ell})_{aa}(A_{\ell})_{bb} + (A_{\ell})_{ab}^2 \right]\,.
  \label{eq:AVariance}
\end{equation}
Compared with the average $n_{\ell}A_{\ell}$, the fluctuations are therefore
suppressed by $n_{\ell}^{-1/2}$. The same argument applies to the flavour
factors, conditioning on $\Omega_{>\ell}$: one finds
$\expct[B_{\ell-1}\,|\,\Omega_{>\ell}] = n_{\ell-1}B_{\ell}$, and the entries of
$\Delta B_{\ell-1} \equiv B_{\ell-1} - n_{\ell-1}B_{\ell}$ are suppressed by
$n_{\ell-1}^{-1/2}$. Since $B_{\ell}$ has $\nout^2$ entries, the relative size
of $\Delta B_{\ell-1}$ as a matrix is $(\nout/n_{\ell-1})^{1/2}$.

Iterating the recursions from $A_1 = \mathbb{1}_2$ and
$B_L = \mathbb{1}_{\nout}$, and normalising each factor by its expectation
value, we find
\begin{align}
  &\hat{A}_{\ell} \equiv \frac{A_{\ell}}{a_{\ell}} = \mathbb{1}_2 + \delta A_{\ell}\,,
  \qquad
  \delta A_{\ell} = \sum_{\ell'=2}^{\ell} \frac{\Delta A_{\ell'}}{a_{\ell'}}\,,
  \qquad
  a_{\ell} = \prod_{\ell'=1}^{\ell-1} n_{\ell'}\,,
  \label{eq:AHat}\\
  &\hat{B}_{\ell} \equiv \frac{B_{\ell}}{b_{\ell}} = \mathbb{1}_{\nout} + \delta B_{\ell}\,,
  \qquad
  \delta B_{\ell} = \sum_{\ell'=\ell}^{L-1} \frac{\Delta B_{\ell'}}{b_{\ell'}}\,,
  \qquad
  b_{\ell} = \prod_{\ell'=\ell+1}^{L} n_{\ell'-1}\,.
  \label{eq:BHat}
\end{align}
Each term in these sums has zero mean given the layers already summed, so the
terms are uncorrelated and their variances add: $\delta A_{\ell}$ has relative
size $(\sum_{\ell'<\ell} 1/n_{\ell'})^{1/2}$, and $\delta B_{\ell}$ has relative
size $(\nout\sum_{\ell'\geq\ell} 1/n_{\ell'})^{1/2}$.

Substituting into \cref{eq:GnnBlock2} gives
$\mathcal{G}_{nn} = r^{2\gamma}\sum_{\ell} w_{\ell}\,\hat{A}_{\ell}\otimes\hat{B}_{\ell}$
with $w_{\ell} = a_{\ell}b_{\ell}/\sigma_{\ell}^2$. We define the common factors
as weighted averages over the layers,
\begin{equation}
  \mathcal{G}_2 = \frac{\sum_{\ell} w_{\ell}\,\hat{A}_{\ell}}{\sum_{\ell} w_{\ell}}\,,
  \qquad
  \mathcal{F} = r^{2\gamma}\sum_{\ell} w_{\ell}\,\hat{B}_{\ell}\,.
  \label{eq:G2Fdef}
\end{equation}
When $\mathcal{G}_2\otimes\mathcal{F}$ is expanded in $\delta A_{\ell}$ and
$\delta B_{\ell}$, its terms of zeroth and first order coincide with those of
$\mathcal{G}_{nn}$, so that
\begin{equation}
  \mathcal{G}_{nn} = \mathcal{G}_2 \otimes \mathcal{F} + R\,,
  \qquad
  R = r^{2\gamma}\sum_{\ell} w_{\ell}\,
      \left(\delta A_{\ell} - \overline{\delta A}\right)\otimes
      \left(\delta B_{\ell} - \overline{\delta B}\right)\,,
  \label{eq:GnnBlock3}
\end{equation}
where the bars denote averages with weights $w_{\ell}$. The remainder is a
weighted covariance, across layers, of the input-space and flavour-space
fluctuations, and is therefore quadratic in them. Using
$\Vert X\otimes Y\Vert = \Vert X\Vert\,\Vert Y\Vert$ and the sizes of
$\delta A_{\ell}$ and $\delta B_{\ell}$ given above, its relative size at
initialisation is
\begin{equation}
  \frac{\Vert R \Vert}{\Vert \mathcal{G}_{nn} \Vert}
  = \mathcal{O}\left( \sqrt{\nout}\sum_{\ell=1}^{L-1}\frac{1}{n_{\ell}} \right)\,,
  \label{eq:RScaling}
\end{equation}
\ie\ a few percent for the widths used here. In other words, all terms in the
sum of \cref{eq:GnnBlock2} are proportional to one another up to corrections
that shrink as the hidden layers get wider, and the leading corrections are
absorbed into $\mathcal{G}_2$ and $\mathcal{F}$. \Cref{eq:NonConstEigProbl}
factorises accordingly, so that we can write
\begin{equation}
  \mathcal{G}_{nn}\widehat{S} \simeq
    \left(\mathcal{G}_{2}\widehat{S}_{2}\right)\otimes
    \left(\mathcal{F}\,\mathrm{diag}(m_{i'})_{\nout}\right)\,.
  \label{eq:NonConstKron}
\end{equation}
By the definition of the spectrum of the Kronecker product
\begin{equation}
  \spec{ A\otimes B }
  = \left\{a_i b_j \right\}_{i=1,\dots,n}^{j=1,\dots,m}\,,
  \label{eq:KroneckerSpec}
\end{equation}
where ${a_i}$ and ${b_j}$ are the eigenvalues of $A$ and $B$, the
$2\nout$ eigenvalues can thus be written as
\begin{equation}
  \lambda^{(n)}_{\pm,k} = \mu_{\pm}\,\nu_{k}\,,
  \qquad
  \mu_{\pm}\in\spec{\mathcal{G}_{2}\widehat{S}_{2}}\,,
  \qquad
  \nu_{k}\in\spec{\mathcal{F}\,\mathrm{diag}(m_{i'})_{\nout}}\,,
  \label{eq:NonConstSpec}
\end{equation}
which reproduce the exact eigenvalues to a few percent in the regime of large
$\gamma$ where the separation of scales of \cref{eq:GOrders} holds. The
non-constant sector therefore splits into two tiers of $\nout$ eigenvalues each,
separated by the ratio $\mu_{+}/\mu_{-}$.

The eigenvectors are obtained by projecting back, as in the constant sector.
Using the three column blocks of \cref{eq:Psi}, $z=\Psi\zeta$ reads
\begin{equation}
  z_{i\alpha} = \sum_{i'} \left(\rfk\right)_{i'i}
    \left[\, \zeta_{c,i'} + \zeta_{2,i'}\,x_{\alpha}
             + \zeta_{3,i'}\log x_{\alpha} \,\right]\,,
  \label{eq:ZProjBack}
\end{equation}
and $\zeta_{c}$ is eliminated with \cref{eq:ZetaCSlaved}. The multiplicities drop
out of that step: both $S_{cc}$ and $S_{cn}$ carry the same factor
$\mathrm{diag}(m_{i'})$ by \cref{eq:SMatrixBlocks}, which therefore cancels,
\begin{equation}
  S_{cc}^{-1}S_{cn}
    = \left(\langle x\rangle,\; \langle \log x\rangle\right)\otimes\id_{\nout}\,,
  \qquad
  \zeta_{c,i'} = -\langle x\rangle\,\zeta_{2,i'}
                 -\langle \log x\rangle\,\zeta_{3,i'}\,,
  \label{eq:ZetaCExplicit}
\end{equation}
so that the constant term in \cref{eq:ZProjBack} does nothing but centre the
other two. By \cref{eq:NonConstKron} the eigenvector $\zeta_{n}$ itself
factorises, $\zeta_{n} = \eta_{\pm}\otimes\chi_{k}$ with $\eta_{\pm}$ and
$\chi_{k}$ the eigenvectors of the two factors, and the modes separate into a
flavour term times a term in $x$,
\begin{equation}
  z^{(\pm,k)}_{i\alpha} \propto
    \left[ \sum_{i'} \chi_{k,i'}\left(\rfk\right)_{i'i} \right]
    \left[\, \cos\varphi_{\pm}\left(x_{\alpha}-\langle x\rangle\right)
           + \sin\varphi_{\pm}
             \left(\log x_{\alpha}-\langle \log x\rangle\right) \,\right]\,,
  \label{eq:NonConstModes}
\end{equation}
where $\left(\cos\varphi_{\pm},\,\sin\varphi_{\pm}\right) \equiv \eta_{\pm}$.
The second and the third tiers of the spectrum are therefore separated by
$\mu_{+}/\mu_{-}$, and not by the input variable they carry. We show the modes
of the NTK at different values of $\gamma$ at initialisation in
\cref{fig:QmodesGamma}, which will be discussed later. The figure displays the
PDF modes $q^{(r)}$ of \cref{eq:hamiltonian_eigvec_pdf} (first row of each block
is at initialisation), which are combinations of the NTK eigenvectors $z^{(r)}$
computed here, weighted by $\sqrt{\lambda^{(r)}}$, and not the $z^{(r)}$
themselves. The functional form is nonetheless preserved.

Note that $\widehat{S}_{2}$ is strongly anisotropic because the grid spans
several orders of magnitude while $x$ stays below one, so that the variance of
$\log x$ is far larger than that of $x$. This is why the modes beyond $r=\nout$
are dominated by the $\log x$ term of \cref{eq:NonConstModes}, and look like
straight lines in the logarithmic scale used in \cref{fig:QmodesGamma}.

\subsubsection*{The tier structure of the NTK spectrum}
The derivation above has two consequences for the NTK at initialisation. First,
the dimension of its image cannot exceed three times the number of independent
flavours, $\dim\ntkimg \leq 3\nout = 24$, where 3 counts the dimension of the
tangent subspace $\vlin$. Second, its spectrum is organised into tiers
of $\nout$ modes. We now confront both predictions with the spectrum measured at
different values of $\gamma$.

The first one is readily verified. The median dimension of the image of the NTK
at initialisation, computed for the bare network and at the tolerance of
\cref{fig:DimNTKSpace}, is
\begin{equation}
  \begin{array}{lcccccc}
    \gamma & 0.001 & 1 & 2 & 3 & 4 & 5 \\[2pt]
    \dim\ntkimg & 175 & 104 & 74 & 41 & 31 & 24
  \end{array} \,.
  \label{eq:DimNTKInit}
\end{equation}
The predicted bound is approached monotonically as the initialisation width
shrinks, and is exact at $\gamma=5$, where $\dim\ntkimg = 24 = 3\times8$. We
checked explicitly that this holds replica by replica, with no scatter left over
the ensemble. Note that the values of \cref{eq:DimNTKInit} converge to the
linear bound for the same values of $\gamma$ for which \cref{eq:SaturationInit}
reports a negligible saturation, so that the two diagnostics agree on where the
linear regime sets in. For the full model, preprocessing factors included, the
dimensions reduce to those of \cref{eq:DimNTKInitModel}: the preprocessing
factors shrink the image throughout, consistently with what we already observed
in \cref{fig:DimNTKSpace} for the baseline fit.

\begin{figure}[t!]
  \centering
  \includegraphics[width=\linewidth]{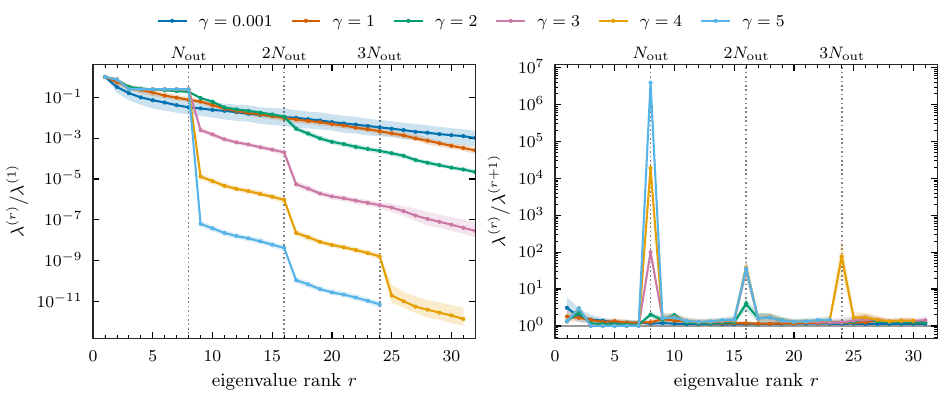}
  \caption{The spectrum of $\ntk$ against the eigenvalue rank $r$ at
  initialisation, for the six initialisation exponents $\gamma$ and computed
  from the neural network alone (no preprocessing). (Left) The
  eigenvalues normalised to the leading one, so that the six fits share an
  ordinate. (Right) The gap between consecutive eigenvalues,
  $\lambda_r/\lambda_{r+1}$, taken replica by replica before the median. Curves
  are medians over the $\nreps = 120$ replicas and bands are $68\%$ intervals;
  there is one marker per eigenvalue. The dotted vertical lines mark the three
  tier boundaries $\nout$, $2\nout$ and $3\nout$ predicted by
  \cref{eq:SpanLinearFeatures}. Note that the curve for each $\gamma$ stops at
  its own $\dim\ntkimg$ of \cref{eq:DimNTKInit}, that is, the last point before
  $\ntkker$.}
  \label{fig:EigValsGammaRank}
\end{figure}
The spectrum of the NTK at initialisation is shown in
\cref{fig:EigValsGammaRank} against the eigenvalue rank $r$ and for the six
initialisation exponents $\gamma$. In the left panel, eigenvalues are normalised
to the leading one, while the right panel shows the gap between consecutive
eigenvalues. This complementary view confirms the group-degenerate structure of
the spectrum predicted above, as far as large values of $\gamma$ are concerned.
The case $\gamma=5$ stands apart for its neatness: the 24 non-zero eigenvalues
of the NTK are arranged in 3 tiers of 8, separated by clear gaps. Within the
first tier, 6 of the 8 eigenvalues are exactly degenerate. Further in the
spectrum the degeneracy is broken, but the 8-fold structure is still visible.
The case $\gamma=4$ is similar, with the same grouping into tiers of 8, though
the gap between the first two tiers is smaller. For $\gamma=3$ the tier
structure is still visible until $r=16$, after which no additional gaps are
visible in the displayed range. For $\gamma=2$ and below, the spectrum follows a
smooth decay and no tiers can be distinguished.

Insofar as the size of the eigenvalues is concerned, we see that the first 8
eigenvalues predicted above are reproduced in \cref{fig:EigValsGammaRank}, that
is \cref{eq:NTKBiasSpecEnd}. This is precisely the $1,\, 0.75,\, 0.25\times6$
pattern displayed by the leading tier of the $\gamma=5$ curve in
\cref{fig:EigValsGammaRank}, and corresponds to the subspace of constant
functions in $\vlin$. On the other hand, the second and third tiers are mixtures
of the $x$ and $\log x$ sectors, as predicted by \cref{eq:NonConstModes}.

%% file: sections/dln_dynamics.tex
\section{NTK dynamics at different initialisation widths}
\label{app:init_dynamics}
In this appendix we complement the discussion of
\cref{sec:init_speed,sec:init_frozen} with a closer look at the NTK during the
training of the six fits with different initialisation widths. Here we focus on
the escape from the linear regime in \cref{app:escape}: we relate the growth of
the image of the NTK to the saturation of the hidden layers, and show how the
escape changes the eigenvectors of the NTK, in their functional form and in
their orientation. Finally, we return to the analytical solution with the frozen
NTK, at initialisation (\cref{app:frozen_ntk_init}) and along the whole training
(\cref{app:frozen_ntk_along_training}).

\subsection{Escape from the linear regime}
\label{app:escape}
We now look at the escape from the linear regime from three complementary
angles: the dimension of the image of the NTK, which we relate to the saturation
of the hidden layers, the functional form of the modes, and the orientation of
the eigenvectors.

\begin{figure}[tbp]
  \centering
  \includegraphics[width=\linewidth]{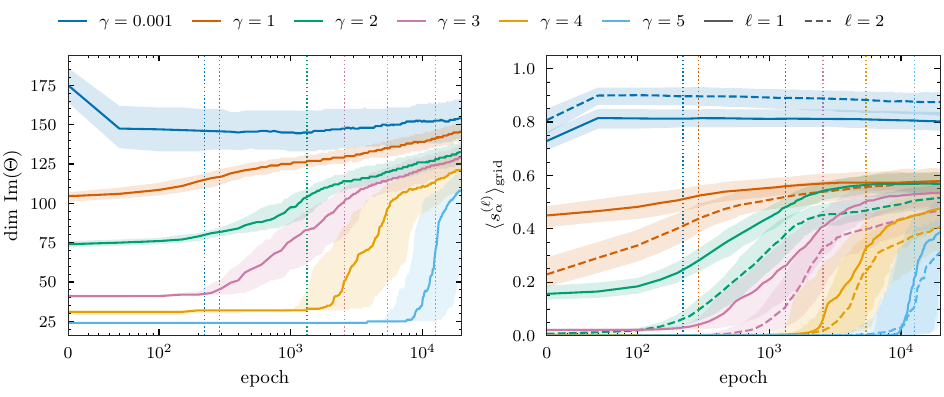}
  \caption{(Left) The dimension of the image of the NTK, $\dim\ntkimg$, as a
  function of the training epoch, for the six values of $\gamma$ considered in
  this analysis. The tolerance is set as in \cref{fig:DimNTKSpace}.
  Preprocessing factors are excluded from the NTK. The curve with $\gamma=1$ is
  the same as \cref{fig:DimNTKSpace} without preprocessing factors. (Right) The saturation
  $s^{(\ell)}_{\alpha}$ of \cref{eq:SaturationMetric}, averaged over the input grid
  in $x$, as a function of the training epoch. Here shown for the first (solid)
  and second (dashed) hidden layer, whose values at $\tau = 0$ are those
  tabulated in \cref{eq:SaturationInit}. In both panels, solid lines are medians
  over the $\nreps=120$ replicas of each fit and the shaded bands are the $68\%$
  intervals; the vertical dotted lines mark the median escape epoch $\tesc$ of
  the corresponding fit. The abscissa is linear below $50$ epochs and
  logarithmic above, so that the initialisation at $\tau=0$ fits on the axis.}
  \label{fig:DimNTKGamma}
\end{figure}

\paragraph{Image of the NTK and saturation.}
The dimension of the image of the NTK is shown in the left panel of
\cref{fig:DimNTKGamma} as a function of the training epoch. When $\gamma=1$,
most of the modes are retained from initialisation, and only a few are
discovered during training (\cfr~\cref{fig:DimNTKSpace}). The case
$\gamma=0.001$ is even more extreme, with the curve remaining almost constant
throughout the training, though it differs from all the other cases in that the
initial dimension is larger and some modes are lost in the first few iterations
-- a similar behaviour is also observed in \cref{fig:EigValsGamma}. The sudden
increase signalling the escape from the linear regime is instead clearly visible
starting from $\gamma=2$, and even more so for $\gamma\geq3$, where new modes
flow into the image of the NTK at about the escape epoch $\tesc$. At $\gamma=5$,
for instance, the median $\dim\ntkimg$ goes from the $24$ of
\cref{eq:DimNTKInit} at initialisation to $71$ at $\tesc$ and $109$ at $\tend$.

Such an increase can be traced back to the saturation of the internal layers,
which sets in at the same epochs (right panel of \cref{fig:DimNTKGamma}). When
this happens, the deep linear network approximation of
\cref{app:dln_ntk} breaks down, and the higher-order terms of the
Taylor expansion of the activations become relevant. This supports the picture
of \cref{sec:init_speed}: with a small initialisation width the network is
confined to the linear regime for the first phase of the training, and it is the
saturation of the hidden layers that allows it to explore a richer functional
space.

\begin{figure}[tbp]
  \centering
  \includegraphics[width=\linewidth]{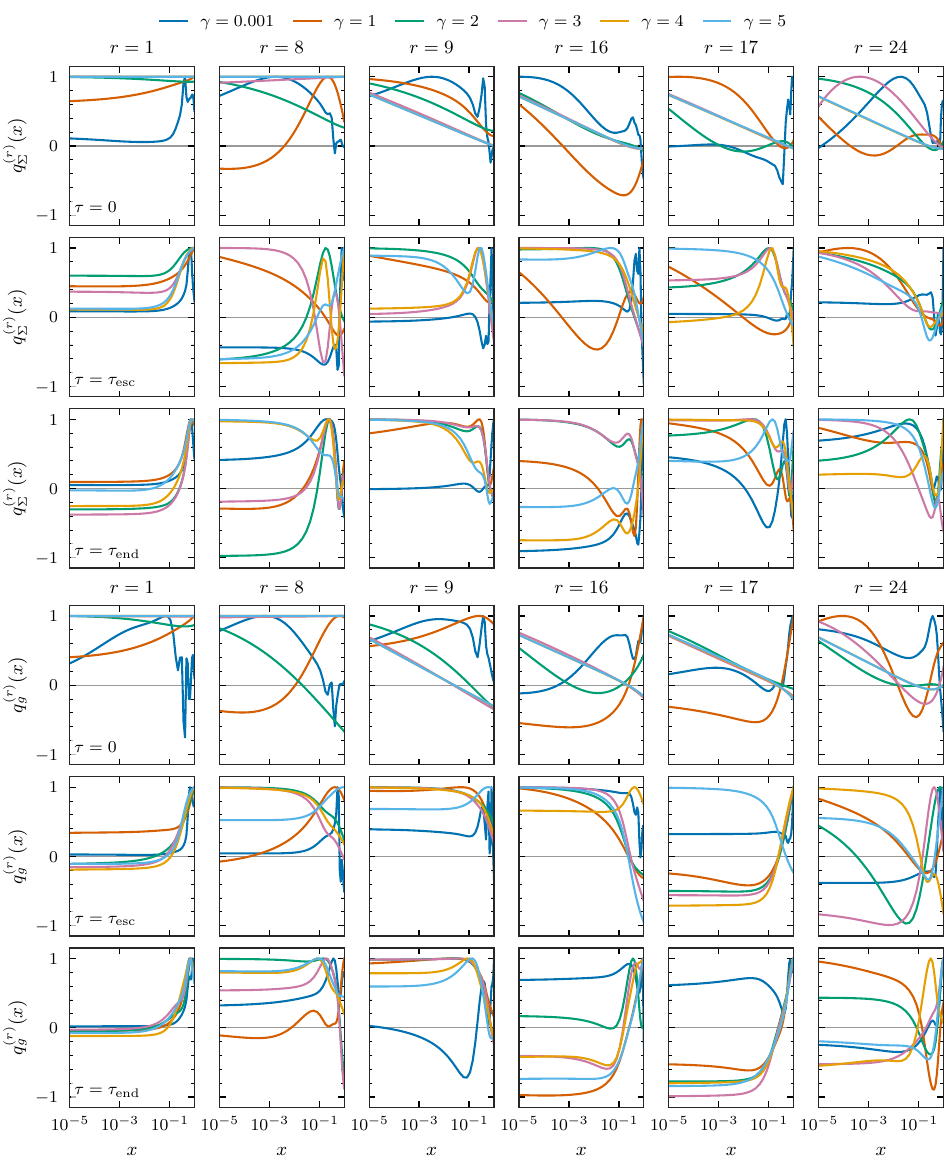}
  \caption{The singlet (upper block) and gluon (lower block) components of six
  modes $q^{(r)}$ of the training Hamiltonian, \cref{eq:hamiltonian_eigvec_pdf},
  as functions of $x$, for the six values of the initialisation exponent
  $\gamma$. Within each block the three rows are the epoch at which the mode is
  evaluated: initialisation, the escape epoch $\tesc$ of each fit, and the last
  stored epoch $\tend = 2\times10^{4}$, well beyond the drop of the loss for
  every $\gamma$. The columns are the mode rank, taken as the first and the last
  of each tier of $\nout$. The modes are computed for the bare network, with the
  preprocessing factors removed, on a single replica common to the six fits.
  Each curve is normalised by its own signed value at the point of largest
  $|q^{(r)}|$, independently for each flavour and each rank, so that the largest
  excursion of every curve is $+1$. Note that in logarithmic scale a straight
  line is a pure $\log x$ dependence.}
  \label{fig:QmodesGamma}
\end{figure}

\paragraph{Functional form of the modes.}
In \cref{fig:QmodesGamma} we show the modes of the training Hamiltonian,
\cref{eq:hamiltonian_eigvec_pdf}, for the singlet and the gluon, at
initialisation (first row of each block), at $\tesc$ (second row) and at $\tend$
(third row). The columns are the first and the last rank of each tier of
$\nout$, so that modes at adjacent tier boundaries are displayed next to each
other.

At initialisation and for $\gamma=4$ and 5, the modes lie in $\vlin$ up to the
small residual quantified below: being linear combinations of the tangent
vectors, they inherit the subspace that the DLN approximation imposes on the
frame (\cref{app:dln_ntk}). The eight modes of the leading tier are constant in
$x$, while every mode beyond $r=\nout$ is a straight line in $\log x$, departing
from it only at $x \gtrsim 0.1$, where the component along $x$ ceases to be
negligible. This is the $3\nout$ structure already visible in the spectrum of
\cref{fig:EigValsGammaRank}, seen now in the functional forms rather than in the
eigenvalues. At $\gamma=3$ the modes of the leading tier are visibly constant,
although the functional form shows some $x$ dependence by the eighth mode. At
$\gamma\lesssim2$ no constant mode is left among the leading eight: the modes of
the $\gamma=1$ fit already curve across the grid, and those of $\gamma=0.001$
oscillate, with several turning points.

Note that the simple functional forms at initialisation with a large $\gamma$
survive only as long as the plateau does. At the escape epoch the modes acquire
more complex functional forms even for large values of $\gamma$, consistently
with the transition in the spectrum of \cref{fig:EigValsGamma}: a network
confined to the linear regime at the beginning of the training can eventually
leave it, provided that the fit is given enough time.

\begin{figure}[tbp]
  \centering
  \includegraphics[width=\linewidth]{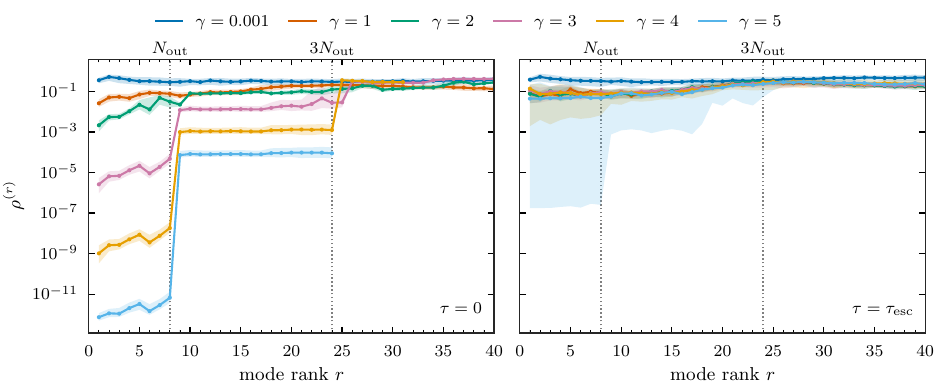}
  \caption{Residual $\rho^{(r)}$ of \cref{eq:ModeResidual} as a function of the
  rank $r$, at initialisation (left) and at the escape epoch $\tesc$ (right).
  Curves are medians over the $\nreps = 120$ replicas of each fit and bands are
  $68\%$ intervals. The curves for $\gamma=4$ and $5$ terminate at the median
  $\dim \ntkimg$ of the corresponding fit (\cfr~\cref{fig:DimNTKGamma}). The
  dotted vertical lines mark the boundaries of the tiers as in
  \cref{fig:EigValsGammaRank}. The wide band on the $\gamma = 5$ curve of the
  right panel is due to the $22$ replicas out of $120$ that never escape within
  the training and therefore remain close to $\vlin$.}
  \label{fig:QmodesGammaLinearity}
\end{figure}
\paragraph{Distance from the linear span.}
We quantify how well a mode $q^{(r)}$ is described by the linear span by means
of the following residual
\begin{equation}
  \rho^{(r)} = \frac{\left\| q^{(r)} - q^{(r)}_{\rm linear} \right\|}
                    {\left\| q^{(r)} \right\|}\,,
  \label{eq:ModeResidual}
\end{equation}
where $q^{(r)}_{\rm linear}$ is the best approximation to $q^{(r)}$ that $\vlin$
can provide.\footnote{Explicitly, and flavour by flavour,
  $
    q^{(r)}_{i,\,{\rm linear}} =
    \operatorname*{arg\,min}_{a_1,\,a_2,\,a_3}
    \left\| q^{(r)}_{i} - a_1 - a_2\,x - a_3 \log x \right\|,
  $
  that is, the projection of the mode onto \cref{eq:SpanLinearFeatures},
  obtained by least squares on the $x$ grid.} It vanishes when the mode lies
  entirely in $\vlin$ and tends to unity when the mode is orthogonal to it.
  \cref{fig:QmodesGammaLinearity} displays it as a function of the rank at
  initialisation (left) and at the escape epoch $\tesc$ (right). Starting from
  the left panel, we see that for large values of $\gamma$ the residuals are
  small and stratify in tiers of $\nout$, similarly to
  \cref{fig:EigValsGammaRank}. As $\gamma$ decreases, the residuals grow and the
  tiers merge, until for $\gamma \lesssim 2$ the modes are far from $\vlin$ and
  no tier structure is left. In the right panel, at the escape epoch, the
  residuals are large for every value of $\gamma$: the modes have left $\vlin$,
  consistently with the growth of the image of the NTK in
  \cref{fig:DimNTKGamma}.

\paragraph{Rotation of the eigenvectors.}
The stationary behaviour of the spectrum on the initial plateau for large
$\gamma$ does not tell us whether the eigenvectors themselves are frozen, or
whether they rotate within the degenerate subspaces of the tiers, thus leaving
the image of the NTK unchanged. To clarify this aspect, we propose the following
measure of the drift of the leading $k$ eigenvectors of the NTK:
\begin{equation}
  d_k(\tau) = 1 - \frac{1}{k} \sum_{i,\,j = 1}^{k}
    \left(z^{(i)}_{\tau},\, z^{(j)}_{0}\right)^2\,.
  \label{eq:SubspaceDrift}
\end{equation}
Here $z^{(i)}_{\tau}$ and $z^{(j)}_{0}$ are the $i$-th and $j$-th eigenvectors
of the NTK at training epoch $\tau$ and at initialisation, respectively, and
$(\cdot,\cdot)$ denotes the Euclidean inner product. This metric compares the
subspaces spanned by the leading $k$ eigenvectors of the two NTKs, and vanishes
when they coincide. Note also that $d_k$ is invariant under any rotation of the
modes within the $k$-dimensional subspace, namely it measures the drift of the
subspace itself rather than of the individual modes.\footnote{This property
guards against two eigenvectors $z^{(i)}_{\tau}$ and $z^{(j)}_{\tau}$ swapping
their labels during training, which would be recorded as a large drift by
individual comparisons.}

\begin{figure}[t]
  \centering
  \includegraphics[width=\linewidth]{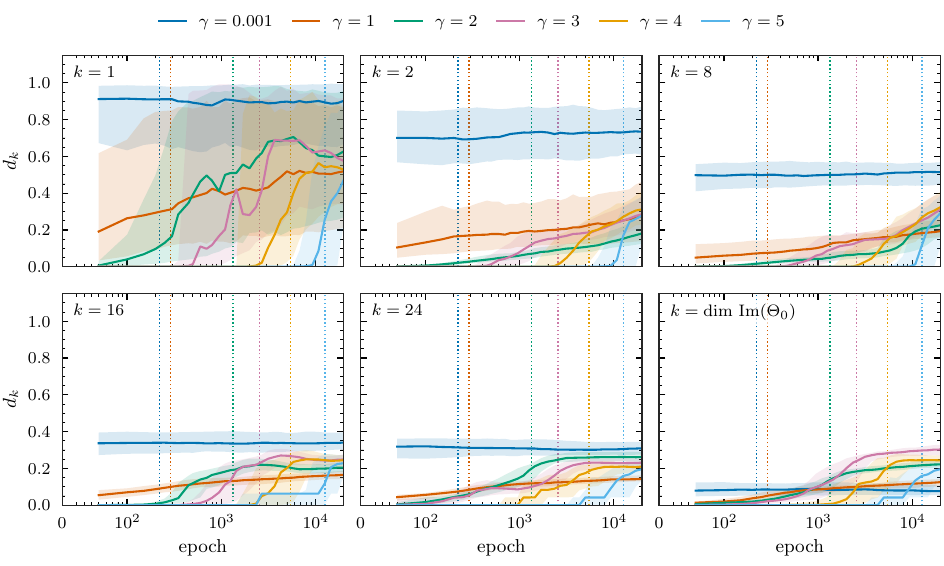}
  \caption{Subspace drift $d_k$ of \cref{eq:SubspaceDrift}, measuring how far
  the $k$ leading eigenvectors of $\ntk_\tau$ have rotated away from the
  subspace they spanned at initialisation, as a function of the training epoch.
  One panel per subspace size $k$, sharing both axes. The first five panels use
  a value of $k$ common to all six fits, chosen at the tier boundaries $8$, $16$
  and $24 = 3\times 8$ together with $k=1$ and $k=2$ within the first tier. The
  last panel sets $k$ to the dimension of the image at initialisation,
  $\dim\mathrm{Im}(\ntk_0)$. Curves are medians over the $\nreps = 120$ replicas
  of each fit and bands are $68\%$ intervals; the dotted vertical lines mark the
  median escape epoch $\tesc$. The NTK is that of \cref{fig:EigValsGamma}, with
  the neural network alone. The initialisation is omitted, $d_k$ vanishing there
  by construction.}
  \label{fig:SubspaceDriftGamma}
\end{figure}

\cref{fig:SubspaceDriftGamma} shows $d_k$ along the training. It exhibits the
same two regimes that the loss and the spectrum have already displayed, now for
the eigenvectors: a step at large $\gamma$, and a smooth, steady rise at small
$\gamma$. At $\gamma=5$ the leading eigenvectors are completely frozen along the
plateau. At $\tau \simeq 8\times10^{3}$ -- thousands of epochs into the
training, but still on the plateau -- the drift of the leading mode is $d_1 =
0.002$ and that of the leading tier is $d_8 = 0.000$ to three decimals, meaning
that the eigenvectors occupy the very subspace they were given at
initialisation. They begin to rotate only at the escape, reaching $d_1 = 0.26$
and $d_8 = 0.11$ at $\tesc$, and keep rotating afterwards. At $\gamma=1$ there
is no such plateau: $d_1$ has already reached $0.28$ within the first few
hundred epochs and grows smoothly from the first epochs onwards, while
$\gamma=3$ sits between the two.

The end-of-training values of $d_k$ carry a further message:
\begin{equation}
  \begin{array}{lcccccc}
    \gamma & 0.001 & 1 & 2 & 3 & 4 & 5 \\[2pt]
    d_1 & 0.90 & 0.52 & 0.63 & 0.58 & 0.53 & 0.47 \\
    d_2 & 0.73 & 0.27 & 0.18 & 0.29 & 0.31 & 0.28 \\
    d_8 & 0.51 & 0.19 & 0.23 & 0.31 & 0.32 & 0.29 \\
    d_{16} & 0.34 & 0.17 & 0.20 & 0.25 & 0.24 & 0.23 \\
    d_{24} & 0.31 & 0.14 & 0.26 & 0.23 & 0.21 & 0.20 \\
    d_{\dim\mathrm{Im}(\ntk_0)} & 0.076 & 0.13 & 0.22 & 0.30 & 0.24 & 0.20
  \end{array}
  \label{eq:SubspaceDriftEnd}
\end{equation}
Apart from the $\gamma=0.001$ fit, whose saturated activations at initialisation
leave it drifting more than the others at every $k$ common to the six fits --
its leading mode rotates almost completely, $d_1 = 0.90$ -- the drift
accumulated by the end of the training is largely independent of $\gamma$:
$d_{24}$ lies between $0.14$ and $0.26$ across the remaining five initialisation
widths. This suggests that $\gamma$ controls \emph{when} the kernel moves, not
\emph{where} it ends up. The final orientation of the NTK is a property of the
data, and the initialisation only decides how long the network waits before
going to find it. It also shows that the rotation is more pronounced for the leading
eigenvectors, while the sub-leading ones rotate less.
\par\bigskip\noindent Taken together, these results complete the picture of the
linear regime drawn in \cref{fig:EigValsGamma,fig:EigValsGammaRank}, thus
proving that the NTK is stationary when confined to this regime and not just
slowly varying. Its eigenvalues are pinned at the values fixed at
initialisation, its eigenvectors do not move, and its image does not grow beyond
the dimension of \cref{eq:DimNTKInit} -- exactly $3\nout$ at $\gamma=5$, where
the degeneracy of \cref{eq:NTKBiasSpecEnd} is also exact.

\subsection{Timescales of the frozen NTK at initialisation}
\label{app:frozen_ntk_init}
\begin{figure}[tbp]
  \centering
  \includegraphics[width=0.7\linewidth]{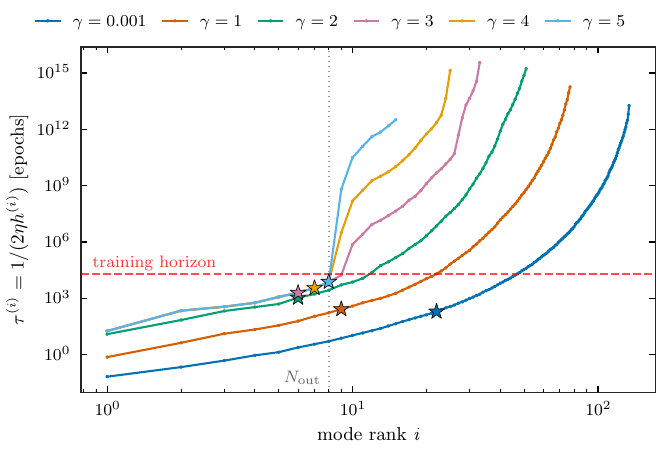}
  \caption{Characteristic resolution times $\tau^{(i)} = 1/(2\eta h^{(i)})$ of
  \cref{eq:tau}, in epochs, of every mode in $\ntkimg$ with the NTK frozen at
  initialisation, against the mode rank $i$ and for the six initialisation
  exponents $\gamma$. The kernel is that of the full model, as in
  \cref{fig:ResolutionTimes}, so that each curve terminates at the corresponding
  entry of \cref{eq:DimNTKInitModel}. Points are medians over the $\nreps = 120$
  replicas of each fit. The red dashed line marks the training horizon
  $\tend = 2\times10^{4}$ epochs, the last epoch of the numerical
  training: a mode whose resolution time lies above it has not had the time to
  be learned. The dotted vertical line marks the $\nout = 8$ modes of the
  constant sector. The star on each curve is the slowest mode that the
  corresponding fit could have resolved by its own escape epoch $\tesc$.}
  \label{fig:ResolutionTimesGamma}
\end{figure}

In \cref{sec:init_frozen} we showed that, for large values of $\gamma$, the
analytical solution with the NTK frozen at initialisation follows the trained
network along the plateau, but not past it. Here we ask whether this is
because the frozen solution is simply too slow, i.e.\ whether the directions
needed to leave the plateau are present in $\ntkimg$ at initialisation but are
not resolved within the training horizon. To this end, we compute the
characteristic resolution times $\tau^{(i)}$ of \cref{eq:tau} for every mode
in $\ntkimg$ at $\tntk = 0$, shown in \cref{fig:ResolutionTimesGamma} against
the training horizon $\tend$ for the six values of $\gamma$.

Within the training horizon, the frozen kernel of the $\gamma=4$ and $\gamma=5$
fits resolves exactly eight modes -- those of the constant sector, which are
generated by the biases of the last layer and do not depend on $\gamma$,
\cref{eq:NTKBiasSpecEnd}. Every direction carrying a dependence on $x$ or $\log
x$ is instead pushed to $\tau^{(i)} \gtrsim 3\times10^6$ epochs, and remains
unresolved. We then conclude that the escape is not a slow linear learning
running its course, but a genuinely non-linear event driven by the growth of the
weights, which saturates the hidden layers and takes the network out of $\vlin$,
rather than by the spectrum of the initial kernel.

\subsection{The frozen NTK along the training}
\label{app:frozen_ntk_along_training}

\begin{figure}[t!]
  \centering
  \includegraphics[width=\linewidth]{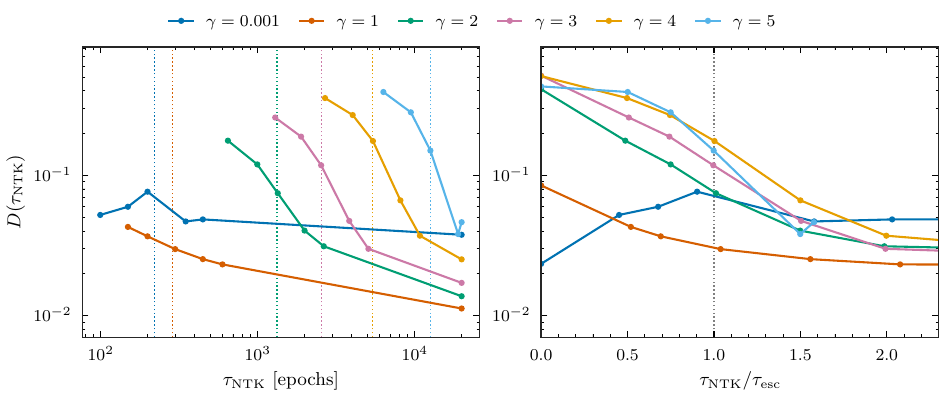}
  \caption{The discrepancy $D$ of \cref{eq:AnltSolDiscrepancy} between the
  analytical and the trained solution at the end of the training, as a function
  of the epoch $\tntk$ at which the NTK is frozen, for the six values of the
  initialisation exponent $\gamma$. (Left) Against $\tntk$ in epochs, with the
  median escape epoch $\tesc$ of each fit marked by a dotted vertical line of
  the same colour. The abscissa is logarithmic, so that the point at $\tntk = 0$
  is shown only in the right panel. (Right) The same curves against $\tntk$ in
  units of the escape epoch of each fit, on a linear abscissa truncated at
  $2.3$. The dashed grey line marks $\tntk = \tesc$. Markers are the sampled
  freezing epochs, and lines only interpolate between them. $D$ is computed from
  the ensemble means over the $\nreps = 120$ replicas of each fit.}
  \label{fig:AnltSolDiscrepancyGamma}
\end{figure}

Here we supplement the discussion of \cref{sec:init_frozen} by assessing the
frozen-NTK description at intermediate freezing epochs. Specifically, we
consider $\tntk = 0.5,\,0.75,\,1,\,1.5$ and $2$ times the median escape epoch
$\tesc$ for each value of $\gamma$, in addition to the two ends $\tntk = 0$ and
$\tntk = \tend$ considered in \cref{sec:init_frozen}.\footnote{Each epoch is
rounded to the closest stored one, the epochs being stored every $50$. For
$\gamma = 5$ the last of them exceeds $\tend$, and only six epochs are
available.} For each $\tntk$, the analytical solution $\hat{f}$ is evolved from
the same initial condition $f_0$ as the trained network, as in
\cref{sec:init_frozen}, up to the end of the training, where the two solutions
are compared. The distance between the analytical solution $\hat{f}$ and the
numerically trained network is measured using the following figure of merit
\begin{equation}
  D(\tntk) = \frac{\left\| \left\langle \hat{f}_{\tend} \right\rangle
                         - \left\langle f_{\tend} \right\rangle \right\|_M}
                  {\left\| \left\langle f_{\tend} \right\rangle \right\|_M}\,,
  \label{eq:AnltSolDiscrepancy}
\end{equation}
where $\langle\cdot\rangle$ denotes the mean over the ensemble of replicas and
$\|v\|_M = (v^T M v)^{1/2}$ is the norm induced by the metric $M$ of
\cref{eq:M_operator}. Here $M$ is built from the whole dataset, rather than from
the training split of each replica, so that it is the same for all replicas, and
the discrepancy is weighted by the experimental uncertainties after projecting
the PDF onto the data space.

\cref{fig:AnltSolDiscrepancyGamma} shows $D$ as a function of $\tntk$, in epochs
(left) and in units of the escape epoch of each fit (right). Two groups of
curves can be clearly distinguished. For $\gamma \lesssim 1$ the discrepancy
remains below $0.1$ for every choice of $\tntk$. With $\gamma = 1$ it decreases
steadily, from $D \simeq 0.08$ at $\tntk = 0$ to $D \simeq 0.01$ at $\tntk =
\tend$. As we already observed in \cref{fig:AnltSolInit,fig:AnltSolSigma}, later
kernels give better approximations, but the kernel at initialisation is already
a reasonable one. With $\gamma = 0.001$, instead, $D$ stays between $0.02$ and
$0.08$ without a clear trend, the smallest value being obtained with the NTK
frozen at initialisation. In both cases the frozen-NTK description holds to a
good approximation all along the training, as expected in the lazy regime of
\cref{sec:init_speed} and in agreement with Ref.~\cite{jacot2022sts}.

The picture is different from $\gamma = 2$ upwards. With the NTK frozen at
initialisation the discrepancy is $D \simeq 0.4$--$0.5$, five to twenty times
larger than for $\gamma \lesssim 1$, in agreement with the failure of the
analytical solution to follow the escape discussed in \cref{sec:init_frozen}. As
$\tntk$ approaches the escape epoch, $D$ decreases steadily, dropping by a
factor of four to ten between $\tntk = 0.5\,\tesc$ and $\tntk = 1.5\,\tesc$, and
between $1.5$ and $2\,\tesc$ it reaches $D \simeq 0.03$--$0.07$ for all four
fits, comparable with the values of the lazy fits. The right panel shows that,
once $\tntk$ is measured in units of $\tesc$, the four curves share the same
shape: whatever the value of $\gamma$, the escape sets the timescale over which
the NTK becomes an accurate description of the training. A residual dependence
on $\gamma$ remains before the escape, where $D$ tends to be larger for larger
$\gamma$ at the same fraction of $\tesc$ -- at $\tntk = \tesc$ we find $D \simeq
0.08,\,0.12,\,0.18$ and $0.15$ for $\gamma = 2,\,3,\,4$ and $5$, respectively --
but it is largely washed out after it. 

Together, these results confirm the picture drawn by the subspace drift,
\cref{eq:SubspaceDriftEnd}: the initialisation controls \emph{when} the
frozen-NTK description becomes accurate, but not how accurate it eventually is.

\FloatBarrier